\documentclass[galaxies,article,accept,pdftex,moreauthors]{Definitions/mdpi}

\firstpage{1} 
\pubvolume{14}
\issuenum{2}
\articlenumber{14}
\pubyear{2026}
\copyrightyear{2026}
\externaleditor{Francesco Calura} %
\datereceived{21 December 2025} 
\daterevised{17 February 2026} %
\dateaccepted{24 February 2026} 
\datepublished{27 February 2026} 

\usepackage{bm}
\usepackage{xspace}

\newcommand{\mnras}{Mon. Not. R. Astron. Soc.}
\newcommand{\apj}{Astrophys. J.}
\newcommand{\apjs}{Astrophys. J. Suppl.}
\newcommand{\apjl}{Astrophys. J. Lett.}
\newcommand{\aap}{Astron. Astrophys.}
\newcommand{\araa}{Annu. Rev. Astron. Astrophys.}
\newcommand{\pasa}{Publ. Astron. Soc. Aust.}

\usepackage{xcolor}
\newcommand{\rev}[1]{\textcolor{black}{#1}} %

\def\gthree{\textsc{gadget3-osaka}\xspace}
\def\gfour{\textsc{gadget4-osaka}\xspace}

\Title{Validating the CROCODILE Model Within the AGORA Galaxy Simulation Framework}

\Author{{Pablo Granizo} %
 $^{1,2,}$*\orcidA{}, Yuri Oku $^{3,4}$\orcidB{} and Kentaro Nagamine $^{1,5,6,7,8}$\orcidC{}}

\address{%
$^{1}$ \quad Theoretical Astrophysics, Department of Earth and Space Science, Graduate School of Science, The University of Osaka, {Toyonaka} %
 560-0043, Japan; {kn@astro-osaka.jp} %
\\
$^{2}$ \quad Facultad de Ciencias, {Universidad Aut\'{o}noma de Madrid}, Ciudad Universitaria de Cantoblanco, %
\mbox{28049 Madrid, Spain}\\
$^{3}$ \quad Center for Cosmology and Computational Astrophysics, Institute for Advanced Study in Physics, Zhejiang University, Hangzhou 310058, China; {yuri.oku.astro@gmail.com}\\
$^{4}$ \quad Institute of Astronomy, School of Physics, Zhejiang University, Hangzhou 310058, China\\
$^{5}$ \quad Theoretical Joint Research, Forefront Research Center, Graduate School of Science, The University of Osaka, {Toyonaka} 560-0043, Japan\\
$^{6}$ \quad Kavli IPMU (WPI), University of Tokyo, 5-1-5 Kashiwanoha, {Kashiwa} 277-8583, Japan\\
$^{7}$ \quad Department of Physics \& Astronomy, University of Nevada{,} %
 Las Vegas, NV 89154, USA\\
$^{8}$ \quad Nevada Center for Astrophysics, University of Nevada, %
Las Vegas, \mbox{NV {89154}%
,} USA}

\corres{Correspondence: pablo.granizo@estudiante.uam.es}

\abstract{Numerical galaxy formation simulations are sensitive to numerical methods and sub-grid physics models, making code comparison projects essential for quantifying uncertainties. Here, we {evaluate} %
 \gfour within the AGORA project framework by conducting a systematic comparison with its predecessor. We perform an isolated disk galaxy and a cosmological zoom-in run of a Milky Way-mass halo, following the multi-step AGORA calibration procedure. By systematically deconstructing the updated stellar feedback model, we demonstrate that mechanical momentum injection is necessary to suppress unphysical gas fragmentation and regulate star formation, yielding agreement with the \mbox{Kennicutt--Schmidt} relation. Meanwhile, stochastic thermal heating is essential for driving a hot metal-enriched gaseous halo, thereby creating a multiphase circumgalactic medium that is absent in the predecessor code. In the cosmological context, we calibrate the simulation to match the stellar mass growth history targeted by the AGORA collaboration. The validated \gfour simulation has been contributed to the AGORA CosmoRun suite, providing a new data point for understanding the impact of numerical and physical modeling choices on galaxy evolution.}


\begin{document}

\section{Introduction}

The origin and evolution of galaxies are central questions in modern astrophysics. Galaxy formation spans a vast range in both space and time, from~parsec-scale star-forming regions to the megaparsec scales of the cosmic web, with~timescales ranging from the evolution of supernova remnants ($\sim$$10^4$ years) to the age of the universe ($\sim$$10^{10}$ years). The~current cosmological paradigm, the $\Lambda$ cold dark matter ($\Lambda$CDM) model, provides a remarkably successful theoretical framework for understanding the formation of galaxies. Within~this framework, current models favor hierarchical structure formation: small dark matter halos collapse under gravity, then grow by accretion and merging with other halos to form the massive halos that host today's observed galaxies~\cite{White1978}.

Due to the nonlinear nature of this problem, accurately modeling galaxy evolution requires numerical simulations. The~field has progressed from early dark-matter-only calculations, which were instrumental in establishing the large-scale predictions of \mbox{$\Lambda$CDM \citep{Springel_2005_1,Klypin_2011}}, to~modern simulations that include hydrodynamics and an expanding suite of physical processes. These simulations typically evolve $10^7$--$10^9$ elements with spatial resolutions reaching sub-kiloparsec scales, enabling detailed studies of galactic internal structure and evolution. Exascale systems and GPU-powered algorithms will soon enable simulations with $10^{12}$ resolution elements and beyond \citep{hirashima2023,Cavelan_2020,2025arXiv251023330H}.

Over the last decade, large-volume cosmological hydrodynamical simulations have achieved remarkable success in reproducing present-day statistics of the universe \citep{Springel_2017,Schaye_2014,Kaviraj_2017,Oku2024,Dav__2019,dolag2025emagneticum,Feldmann_2023,schaye2025colibre}. However, key processes remain poorly understood, particularly stellar and AGN feedback. Different codes implement substantially different physical models, producing qualitatively and quantitatively different results even when calibrated to similar observational \mbox{targets \citep{Scannapieco_2012}}. This sensitivity stems from the challenge of modeling baryonic physics across a vast range of scales. Current codes cannot resolve the physics of individual supernova remnants or accretion disks, necessitating sub-grid models that approximate physical processes at unresolved scales. %

The diversity of numerical methods (for hydrodynamics, SPH \citep{Wadsley_2004,Springel_2005}, AMR \citep{Bryan_2014,Kravtsov_1997,Teyssier_2002}, moving-mesh \citep{Weinberger_2020}, and~meshless \citep{Hopkins_2012}) combined with different sub-grid physics implementations has produced a rich ecosystem of zoom-in galaxy formation simulations \citep{Sawala_2016,Wetzel_2016,Grand_2017,Guedes_2011,Wang_2015,Agertz_2021,Ceverino_2014}. Each major simulation typically employs different combinations of gravitational solvers, hydrodynamic schemes, and~sub-grid models, making it difficult to isolate the effects of specific modeling choices on~outcomes.

This complexity has motivated the development of controlled code comparison projects that systematically investigate the origins of differences between simulation codes and improve their robustness. Several efforts have targeted specific aspects of this problem, including (but not limited to) the Santa Barbara Cluster Comparison Project \citep{Frenk_1999}, the~nIFTy series \citep{Sembolini_2016}, and~the Three Hundred \citep{Cui_2018} project. The~Santa Barbara Cluster Comparison Project pioneered this approach, revealing systematic differences in gas thermodynamics between particle-based codes and mesh-based codes. At~cluster scales, the~Santa Barbara comparison first exposed systematic differences in gas thermodynamics between particle-based and mesh-based codes;~later projects such as nIFTy and The Three Hundred have extended these comparisons to modern codes and full-physics treatments. For~Milky Way-like simulations, the~Aquila project \citep{Scannapieco_2012} compared multiple zoom-in simulations of the same halo among different codes with varying hydrodynamics, cooling, and feedback implementations, which produced differing morphologies and stellar masses. Other \mbox{studies \citep{Hu_2023Lag,Smith_2018,Stewart_2017}} have compared different codes and feedback prescriptions in the isolated or cosmological zoom-in regimes. The~FIRE collaboration \citep{Fire2_2018} has performed extensive comparisons with the \textsc{GIZMO} code across different numerical implementations in both~regimes.

While these targeted studies have proven valuable,~a more comprehensive approach is needed in order to disentangle the coupled effects of numerical methods and sub-grid physics. This is the context in which the AGORA collaboration was~established.

AGORA (Assembling Galaxies of Resolved Anatomy) is an international effort aimed at systematically comparing the results of state-of-the-art cosmological hydrodynamics codes. The~primary goals of the project are to quantify the systematic uncertainties inherent in galaxy simulations, understand the interplay between numerical methods and sub-grid physics, and~provide a standardized platform for testing and calibrating new models and codes, thereby improving reproducibility and robustness across the field. To~enable this, the~AGORA infrastructure is built on three pillars: common initial conditions (generated with Makedisk \citep{Springel_2005fdb} and MUSIC \citep{Hahn_2011}), common physics assumptions (e.g., cooling and star formation), and~a common analysis toolkit (\texttt{yt} \citep{Turk_2010}).

The comparison has progressed through several key phases, each building on the previous one to isolate distinct physical and numerical effects. Paper I \citep{Kim_2013} focused on dark matter-only simulations to establish a baseline by comparing the performance of different gravity solvers and code architectures. Paper II \citep{Kim_2016} introduced hydrodynamics in the controlled environment of an isolated disk galaxy, using a common set of physics to compare how different codes handle gas dynamics, star formation, and~stellar feedback in a controlled~setting.

More recently, in~Papers III and IV \citep{Roca_F_brega_2021,rocafabrega2024} the AGORA collaboration has applied these lessons to full cosmological zoom-in simulations, following a meticulous multi-step calibration procedure. This process is designed to incrementally add physical complexity, from gravity and hydrodynamics alone to the full preferred feedback models, along with~tests to verify cooling and star formation procedures performed along the way. This approach ensures that any differences in the final calibrated simulations can be confidently attributed to the distinct stellar feedback~implementations.

After a decade-long calibration effort, the~collaboration produced a fully calibrated suite of simulations from codes widely used across the community. More than five subsequent papers \citep{Roca_F_brega_2021,rocafabrega2024,Jung_2024,Clayton2024,Rodr_guez_Cardoso_2025,jung2025} have analyzed these results, significantly advancing our understanding of how numerical schemes and feedback models shape galaxy evolution. The~AGORA project can now expand in two ways: adding more codes and physical models, or~performing simulations with different \rev{initial conditions (ICs)}, more detailed schemes, and higher resolution. This work contributes directly to the first~goal.

We introduce \gfour, a~new implementation based on the GADGET-4 \mbox{code \citep{Springel_2021}} with a distinct set of sub-grid physics, as~a new participant in the AGORA project. Our primary objective in this paper is to elucidate the factors that drive differences between \gthree and \gfour within the AGORA framework. While a comprehensive calibration of \gfour against all AGORA targets has been performed and contributed to AGORA Paper VIII \citep{jung2025}, this work focuses on a detailed comparison between our predecessor (\gthree) and current (\gfour) codes in both isolated and cosmological contexts. By~systematically varying feedback components and sub-grid parameters, we can quantify which differences arise from numerical scheme updates versus feedback-physics~changes.

This question is particularly relevant given the transition from GADGET-3 to GADGET-4 numerics and the simultaneous evolution of the Osaka Feedback Model from thermal-dominated in \gthree to momentum-and-wind-inclusive schemes in \gfour. Previous work has demonstrated the importance of feedback implementation in isolated galaxies \citep{Shimizu_2019} and introduced the current momentum-based model \citep{Oku_2022}, but~a systematic comparison across both idealized and cosmological contexts within a standardized framework has not been~performed.

The structure of this paper is as follows: Section~\ref{sec:methods} describes the initial conditions, numerical methods, and~sub-grid physics for both \gthree and \gfour, emphasizing points of divergence; Section~\ref{sec:setup} categorizes all simulations performed, including our systematic feedback component variations and robustness tests; Section~\ref{sec:results} presents results in order of increasing physical complexity: dark matter, hydrodynamics and cooling, inclusion of star formation, and~a systematic dissection of the feedback model component; finally, Section~\ref{sec:conclusion} summarizes our conclusions and outlines future~work.

\section{Numerical~Methods}
\label{sec:methods}

The simulations presented here were performed with \gfour, a~modified version of the publicly available GADGET-4 code \citep{Springel_2021}. GADGET-4 is the successor to the widely-used GADGET-2 \citep{Springel_2005} and GADGET-3 codes, and is redesigned for improved performance and scalability on modern supercomputing architectures. Our specific version incorporates SPH variants and a suite of sub-grid physics models developed by the Theoretical Astrophysics group at the University of Osaka, with~the current model described in \citet{Oku2024}.

For comparison, we analyze simulations performed with \gthree, our predecessor code. The~\gthree runs analyzed are identical to those contributed to Paper II (isolated) and Paper III (cosmological) of the AGORA project. While the base code used was the same for those two papers, some implementation details differ, so we will briefly describe each in the following~subsections.

The AGORA project specifies a series of common parameters and models described in detail in Papers II and III. Here, we briefly cover them for context; full motivation can be found in the original papers. Our new \gfour runs also employ these parameters, as~demonstrated in Section~\ref{sec:setup}.

\subsection{Initial~Conditions}
\label{sec:ics}

\textls[-15]{The simulations utilize common \rev{ICs} established by the AGORA project for \mbox{two regimes:}}

\textit{{Isolated disk IC:}} %
We employ the Milky Way-mass disk IC from Paper II, consisting of a dark matter halo of mass $M_{200}=1.074 \times 10^{12} M_{\odot}$, radius $R_{200}=205.5$~kpc, \rev{and circular velocity $v_{c,200} = 150$~km s$^{-1}$, following a NFW \rev{\citep{1997NFW}} profile with concentration $c = 10$ and spin parameter $\lambda = 0.04$,} generated with \textsc{Makedisk} \citep{Springel_2005fdb}. We use \mbox{$3.125 \times 10^{5}$ particles} across dark matter, gas, and stars, separated in bulge and disk components ({{both} %
 components use the same particle mass; for~more details about this IC, please refer to Table~1 in AGORA Paper II \citep{Kim_2016}}), with~masses $M_{\rm DM,IC}=1.254 \times 10^{7} M_{\odot}$, $M_{\rm gas,IC}=8.593 \times 10^{4} M_{\odot}$, and $M_{\rm star,IC}=3.437 \times 10^{5} M_{\odot}$, respectively. \rev{The disk has total mass $M_{d}=4.297\times10^{10} M_{\odot}$, following an exponential profile with scale length $r_{d}=3.432$~kpc and scale height \mbox{$z_{d}=0.1$ $r_{d}$}, and~is 80\% stars/20\% gas by mass. The~stellar bulge has $M_{b,\star}=4.297\times10^{9} M_{\odot}$ and follows a Hernquist profile ($M_{b,\star}/M_{d} = 0.1$) \citep{1990hernquist}}.

\textit{Cosmological zoom-in IC}: We adopt the public AGORA zoom-in ICs ({cosmological} and isolated ICs available {at} %
 \url{https://sites.google.com/site/santacruzcomparisonproject/data}, accessed on 25 February 2026) (Paper III) targeting a $M_{\rm vir}\approx10^{12}M_{\odot}$ halo at $z=0$, generated with the MUSIC IC generator \citep{Hahn_2011} using \mbox{five nested} refinement levels ($128^3$ to effective $4096^3$ resolution) in a (\mbox{60 \text{ comoving } h$^{-1}\mathrm{Mpc})^{3}$} parent volume. The~adopted cosmological parameters follow a flat $\Lambda$CDM model: $\Omega_{\rm m}=0.272$, $\Omega_{\Lambda}=0.728$, $\sigma_{8}=0.807$, $n_{\rm s}=0.961$, and~$H_{0}=70.2 \text{ km s$^{-1}$ Mpc$^{-1}$}$ (based on WMAP7/9 + SNe + BAO) \citep{komatsuic, Hinshaw_2013}. Particle masses in the highest-resolution region correspond to $M_{\rm DM,IC}=2.8 \times 10^{5} M_{\odot}$ and \mbox{$M_{\rm gas,IC}=5.65 \times 10^{4} M_{\odot}$}.

\subsection{Gravity~Treatment}
\label{sec:grav}

Both codes employ a tree algorithm in isolated simulations, computed to third order in \gfour and second order in \gthree. In~cosmological runs, \gfour uses the novel Fast Multipole Method--Particle Mesh (\mbox{FMM--PM}) approach up to fourth order, while \gthree uses a Tree--Particle Mesh (TreePM) method. The \mbox{FMM--PM} solver in GADGET-4 includes a hybrid MPI+shared memory parallelization scheme and sophisticated domain decomposition that simultaneously balances computational load and memory usage, yielding significantly improved strong and weak scaling for highly clustered zoom-in simulations \citep{Springel_2021}. \gfour also makes use of hierarchical time integration, which conserves momentum despite the use of multiple timestep bins for~particles.

All methods use the force softening law from \citet{Springel_2001G1}, with~a gravitational softening length of $80$~pc for isolated runs. For~cosmological runs, we use $800$~comoving~pc until $z = 9$ and $80$~proper~pc thereafter. In~both cases, the~minimum SPH smoothing length is set to $0.2 \times \varepsilon_{\rm grav}$. These choices result in an effective spatial resolution of $80$~pc for our simulations at late~times.

\subsection{SPH~Formulation}
\label{sec:sph}

Both codes solve hydrodynamics using Smoothed Particle Hydrodynamics (SPH), a~Lagrangian approach in which the fluid is discretized into particles. The~\gfour implementation incorporates several advancements designed to improve the accuracy and robustness of capturing shocks and fluid~instabilities.

\textls[-15]{\textit{In \gfour:} We use the Wendland $C^4$ kernel \citep{Dehnen_2012} with \mbox{$N_{\rm nbg} = 128 \pm 8$ neighbors}} (isolated) and $128 \pm 4$ (cosmological). The~pressure-energy SPH formulation \citep{Hopkins_2012, Saitoh_2013} enables direct, non-iterative updates of energy from sub-grid physics. Artificial viscosity \citep{Frontiere_2017,Rosswog_2020} and artificial thermal conduction \citep{Price_2008,Borrow_2021} are included. The~SPH time step limiter uses an mutual signal velocity formulation \citep{Oku2024}, improving upon the one-sided limiter in base GADGET-4 to maintain accuracy when neighboring particles experience feedback~effects.

\textit{In \gthree:} isolated runs employ the pressure--entropy formulation, time-dependent artificial viscosity \citep{Morris_1997}, and a quintic spline smoothing kernel \citep{Morris1996} with \mbox{$N_{\rm nbg} = 64 \pm 2$}. Cosmological runs additionally include the time-step limiter \citep{Saitoh_2009} and a modified neighbor number $N_{\rm nbg} = 128 \pm 8$.

\subsection{Common AGORA~Physics}
\label{sec:commonphys}

Since AGORA requires standardized “common physics”, differences arise mainly from hydrodynamics and primary sub-grid models. Below, we summarize the implementation in both~codes.

\textit{Gas cooling and chemistry:} Thermal evolution is governed by the \textsc{GRACKLE} v3.3.1 library \citep{Smith_2016}. Although~previous AGORA runs used older versions (v2.1--v3.1), we found no significant differences in our validation tests. We use the tabulated equilibrium cooling mode, where pre-computed cooling and heating rates are read from a lookup table as a function of gas density, temperature, metallicity, and redshift. These tables were generated using the photoionization code CLOUDY \citep{ferland20132013releasecloudy} and include primordial (H, He) and metal species. A~redshift-dependent UV background radiation field from Haardt \& Madau \citep{Haardtable} is included in the heating rates for both isolated and cosmological cases. Cosmological simulations additionally employ hydrogen self-shielding and a redshift-dependent temperature floor due to CMB~heating.

\textsc{GRACKLE}'s chemistry network parameter controls the number of chemical species tracked when computing thermal evolution. The~fiducial AGORA prescription uses mode 0, which interpolates directly from pre-computed equilibrium tables. In~order to clarify the physical inaccuracies introduced by this approach (Section~\ref{sec:cal3coolvar}), we also employ non-equilibrium chemistry networks: mode 1 (H, H$^+$, He, He$^+$, He$^{++}$, e$^-$), mode 2 (adds H$_2$, H$^-$, H$_2^+$), and~mode 3 (adds D, D$^+$, HD). These networks enable molecular hydrogen and deuterium cooling but increase computational~costs.

\textit{Star formation:} Gas particles are eligible to form stars once they exceed a critical hydrogen number density threshold of $n_{\rm H, thres} = 10$ cm$^{-3}$ for isolated and \mbox{$n_{\rm H, thres} = 1$ cm$^{-3}$} for cosmological simulations. \rev{This lower threshold was found to better reproduce the $M_{\star}/M_{\rm halo}$ relation in the Paper III calibration, and~we follow it now for better comparison with other cosmological runs from this suite}. Eligible particles form stars stochastically following a local Schmidt law \citep{Schmidt1959}:
\begin{equation}
\dot{\rho}_\star = \epsilon_\star \frac{\rho_{\rm gas}}{t_{\rm ff}}
\end{equation}
{where} %
 $t_{\rm ff} = \sqrt{3\pi/(32G\rho_{\rm gas})}$ is the local freefall time and~$\epsilon_\star = 0.01$ is the star formation efficiency for both isolated and cosmological runs. Each star particle represents a Simple Stellar Population (SSP) with age and metallicity recorded at formation (inherited from the parent gas particle). We adopt a Chabrier \citep{Chabrier_2003} initial mass function (IMF), with~metal yields computed using the \textsc{CELib} library \citep{Saitoh_2017}.

A uniform metal floor is applied to all gas particles: $Z_{\rm floor} = Z_{\odot} = 0.02041$ in isolated simulations and $Z_{\rm floor} = 10^{-4} Z_{\odot}$ in cosmological simulations, consistent with the AGORA~standard.

\textit{Jeans pressure floor:} To prevent artificial numerical fragmentation in dense gas and to model a minimum level of the unresolved turbulent pressure in the interstellar medium (ISM), we implement a Jeans pressure floor \citep{truelove}:
\begin{equation}
P_{\rm Jeans} = \frac{1}{\gamma\pi} G N_{\rm Jeans}^2 \rho_{\rm gas}^2 \Delta x^2
\end{equation}
where $\gamma=5/3$ is the adiabatic index, $N_{\rm Jeans}$ = 4, and $\Delta x$ is the local spatial resolution (proportional to the smoothing length). In~\gfour this floor is implemented via a minimum internal energy for gas, while \gthree sets a minimum pressure. This is due to the different SPH schemes used, since~setting a minimum pressure in the \mbox{pressure--energy} formulation leads to catastrophic heating and cooling failures, as~demonstrated in early test~runs.

\subsection{Feedback~Models}
\label{sec:feedback}

We now describe the stellar feedback implementations, which constitute the primary difference between \gthree and \gfour. Tables~\ref{tab:iso_runs} and \ref{tab:cosmo_runs} provide a side-by-side comparison, with all our runs~included.

\begin{table}[H]
\small
\caption{{Isolated} %
 disk galaxy simulation suite. All runs use the AGORA Milky Way-mass disk initial condition (Section~\ref{sec:ics}) with $3.125 \times 10^5$ particles and $80$~pc spatial resolution. Fiducial metallicity floor $Z_{\rm floor} = 0.02041$ and star formation efficiency $\epsilon_\star = 0.01$ for all runs. All runs evolved to \mbox{$t=500$ Myr}.}
\label{tab:iso_runs}

\begin{adjustwidth}{-\extralength}{0cm}
\begin{tabularx}{\fulllength}{cCcCcCc}
\toprule
\multirow{2}{*}{\textbf{Name}} & \multirow{2}{*}{\textbf{Code} \boldmath{{$^a$}}} %
 & \multirow{2}{*}{\textbf{Physics} \boldmath{$^b$}} & \multirow{2}{*}{\textbf{Cooling} \boldmath{$^c$}} & \multirow{2}{*}{\textbf{FB Model} \boldmath{$^d$}} & \textbf{E$_{\rm \textbf{SN}}$} & \multirow{2}{*}{\textbf{Thermal/Kinetic} \boldmath{$^e$}} \\
 & & & & & \textbf{(\boldmath{$10^{51}$} erg)} & \\
\midrule
\multicolumn{7}{c}{{Fiducial/Baseline Runs}} \\
\midrule
G3-NSFF & G3 & Hydro + Cooling & 0 & -- & -- & -- \\
G4-NSFF & G4 & Hydro + Cooling & 0 & -- & -- & -- \\
G3-SFF & G3 & Full & 0 & ST Thermal & 1.0 & 100/0 \\
G4-SFF & G4 & Full & 0 & Full $^f$ & 1.0 & 72/28 \\
\midrule
\multicolumn{7}{c}{Feedback Component Variations (\gfour)} \\
\midrule
G4-FB-Thermal & G4 & Full & 0 & ST Velocity & 1.0 & 100/0 $^g$ \\
G4-FB-Kinetic & G4 & Full & 0 & ST Velocity & 1.0 & 0/100 $^g$ \\
G4-FB-Mix & G4 & Full & 0 & ST Velocity & 1.0 & 50/50 $^g$ \\
G4-FB-Oku-Rshock & G4 & Full & 0 & ST Velocity + Oku $^h$ & 1.0 & 50/50 $^g$ \\
G4-FB-Momentum & G4 & Full & 0 & Momentum + Oku $^h$ & 1.0 & -- \\
G4-FB-Stochastic & G4 & Full & 0 & Stochastic + Oku $^h$ & 1.0 & 50/50 $^i$ \\

\bottomrule
\end{tabularx}
\end{adjustwidth}
\end{table}

\begin{table}[H]\ContinuedFloat
\small
\caption{{\em Cont.}}
\begin{adjustwidth}{-\extralength}{0cm}
\begin{tabularx}{\fulllength}{cCcCcCc}
\toprule
\multirow{2}{*}{\textbf{Name}} & \multirow{2}{*}{\textbf{Code} \boldmath{{$^a$}}}  & \multirow{2}{*}{\textbf{Physics} \boldmath{$^b$}} & \multirow{2}{*}{\textbf{Cooling} \boldmath{$^c$}} & \multirow{2}{*}{\textbf{FB Model} \boldmath{$^d$}} & \textbf{E$_{\rm \textbf{SN}}$} & \multirow{2}{*}{\textbf{Thermal/Kinetic} \boldmath{$^e$}} \\
 & & & & & \textbf{(\boldmath{$10^{51}$} erg)} & \\
\midrule

\multicolumn{7}{c}{Feedback Strength Variations (\gfour)} \\
\midrule
G4-FB-Low & G4 & Full & 0 & Full $^f$ & 3.0 & -- \\
G4-FB-Mid & G4 & Full & 0 & Full $^f$ & 5.5 & -- \\
G4-FB-High & G4 & Full & 0 & Full $^f$ & 8.0 & -- \\
G4-FB-VHigh & G4 & Full & 0 & Full $^f$ & 10.0 & -- \\
\bottomrule
\end{tabularx}
\end{adjustwidth}
\noindent{\footnotesize{$^a$ G3 corresponds to \gthree and G4 to \gfour; $^b$ Full: Hydrodynamics, Cooling, Star Formation and Feedback (Section~\ref{sec:methods}); $^c$ 0 corresponds to tabulated cooling (Section~\ref{sec:commonphys}); $^d$ ST: Sedov-Taylor; Full: Momentum-driven + TIGRESS + Metal Winds (Section~\ref{sec:fb_variations}); $^e$ Thermal/Kinetic energy partition (percentages); \mbox{$^f$ Full} \gfour feedback includes mechanical momentum injection, TIGRESS hot winds, stochastic thermal heating, metal wind enhancement and dust production/destruction (Section~\ref{sec:g4feedback}); $^g$ Note: Velocity-kick models may not rigorously conserve total energy; see Section~\ref{sec:fb_variations}; $^h$ Oku: Uses \citet{Oku_2022} shock radius formula instead of {Chevalier} %
 \citep{Chevalier74}; $^i$ Thermal component uses discrete stochastic heating with entropy floor \mbox{$S_{\rm OF} = k_B \times 10^8$ K cm$^2$}.}}
\end{table}

\vspace{-9pt}

\begin{table}[H]
\small
\caption{{Cosmological} zoom-in simulation suite. All runs use the AGORA MW-mass halo initial condition (Section~\ref{sec:ics}) targeting {$M_{\rm vir} \approx 10^{12} M_\odot$} %
 at $z=0$. Spatial resolution down to $80$~pc in the high-resolution region. Metallicity floor $Z_{\rm floor} = 10^{-4}\,Z_\odot$ for all baryonic runs with cooling. Star formation efficiency $\epsilon_\star = 0.01$ for Cal-3 and Cal-4.}
\label{tab:cosmo_runs}
\begin{adjustwidth}{-\extralength}{0cm}
\begin{tabularx}{\fulllength}{cCcCcCc}
\toprule
\multirow{2}{*}{\textbf{Name}} & \multirow{2}{*}{\textbf{Code} \boldmath{$^a$}}  & \multirow{2}{*}{\textbf{Physics} \boldmath{$^b$}} & \multirow{2}{*}{\textbf{Cooling} \boldmath{$^c$}} & \multirow{2}{*}{\textbf{FB Model} \boldmath{$^d$}} & \textbf{E$_{\rm \textbf{SN}}$} & \multirow{2}{*}{\textbf{Thermal/Kinetic} \boldmath{$^e$}} \\
 & & & & & \textbf{(\boldmath{$10^{51}$} erg)} & \\
\midrule
\multicolumn{7}{c}{Fiducial Calibration Sequence} \\
\midrule
G3 DM Only & G3 & -- & No & -- & -- & -- \\
G4 DM Only & G4 & -- & No & -- & -- & -- \\
G3 {Cal-1} %
 & G3 & Hydro & No & -- & -- & -- \\
G4 Cal-1 & G4 & Hydro & No & -- & -- & -- \\
G3 Cal-2 & G3 & Hydro + Cooling & 0 & -- & -- & -- \\
G4 Cal-2 & G4 & Hydro + Cooling & 0 & -- & -- & -- \\
G3 Cal-3 & G3 & Hydro + Cooling + SF & 0 & -- & -- & -- \\
G4 Cal-3 & G4 & Hydro + Cooling + SF & 0 & -- & -- & -- \\
G3 Cal-4 & G3 & Full & 0 & T + K & 5.5 & 50/50 \\
G4 Cal-4 & G4 & Full & 0 & Full $^f$ & 10.0 & -- \\
\midrule
\multicolumn{7}{c}{Cooling Network Variations (\gfour, Cal-3)} \\
\midrule
G4 Cal-3 gr1 & G4 & Hydro + Cooling + SF & 1 & -- & -- & -- \\
G4 Cal-3 gr2 & G4 & Hydro + Cooling + SF & 2 & -- & -- & -- \\
G4 Cal-3 gr3 & G4 & Hydro + Cooling + SF & 3 & -- & -- & -- \\
\midrule
\multicolumn{7}{c}{Feedback Strength Variations (\gfour, Cal-4)} \\
\midrule
G4 Cal-4 FB Low & G4 & Full & 0 & Full $^f$ & 5.5 & -- \\
G4 Cal-4 FB Mid & G4 & Full & 0 & Full $^f$ & 8.0 & -- \\
\bottomrule
\end{tabularx}
\end{adjustwidth}
\noindent{\footnotesize{$^a$ G3 corresponds to \gthree and G4 to \gfour; $^b$ SF: Star Formation, Full: Hydrodynamics, Cooling, Star Formation and Feedback (Section~\ref{sec:methods}); $^c$ Cooling modes: 0 = Tabulated; 1 = H, H$^+$, He, He$^+$, He$^{++}$, e$^-$ non-equilibrium; 2 = adds H$_2$, H$^-$, H$_2^+$; 3 = adds D, D$^+$, HD (Section~\ref{sec:commonphys}); $^d$ T: Thermal, K: Kinetic; $^e$ Thermal/Kinetic energy partition (percentages); $^f$ Full \gfour feedback includes: momentum-driven mechanical feedback with terminal momentum from \citet{Oku_2022}, TIGRESS-calibrated hot wind model \citep{Kim_2020}, stochastic thermal heating, metal wind enrichment, angle-weighted Voronoi injection and dust production/destruction (Section~\ref{sec:g4feedback}).}}
\end{table}

\subsubsection{\gthree Feedback~Model}
\label{sec:g3feedback}

The \gthree feedback implementations differ substantially between isolated and cosmological runs owing to the evolution of the Osaka feedback model over the~timeline of the AGORA project.

\textit{Isolated runs (Paper II, 2016):} When a star particle ``explodes'', we compute the shock radius and find the gas particles within it. Then, the~energy and metal yields are injected into the identified gas particles, weighted by the SPH spline kernel. Cooling is turned off for gas particles only during the hot phase (defined as slightly larger than the adiabatic phase) in order to avoid radiating away most of the energy via overcooling. While \mbox{\citet{Aoyama_2016}} separates the SN energy into kinetic and thermal components, in~this case all of the energy is injected thermally at a constant rate until $t = 4$ Myr has~passed. 

\textit{Cosmological runs (Paper III, 2021):} An updated model \citep{Shimizu_2019} includes \rev{Early Stellar Feedback} (ESFB, which also considers photons from the {\sc Hii} region) and \rev{Asymptotic Giant Branch} (AGB) feedback, with~kinetic and thermal energy efficiencies set to $0.5$ each. Energy, momentum, and~metals are now deposited gradually over multiple events, with~SN duration dependent on the stellar metallicity obtained from \textsc{CELib} tables. In~order to reach the AGORA target stellar mass, SN energy is boosted to $E_{\rm SN} = 5.5 \times 10^{51}$ erg per~event.

In both of these cases, Sedov--Taylor shock radius and velocity kicks are computed similarly to what we will show in the next Section~\ref{sec:fb_variations} for the first runs. This predates the \citet{Oku_2022} model, in which terminal momentum injection is~considered.

\subsubsection{\gfour Feedback~Model}
\label{sec:g4feedback}

The \gfour Osaka model \citep{Oku_2022, Oku2024} tracks mass, energy, momentum, and~metal return from evolving stellar populations, including Type II SNe (\mbox{SN-II}), Type Ia SNe (\mbox{SN-Ia}), \rev{AGB} stars, and~\rev{ESFB} from young stars in cosmological runs. Chemical yields for \mbox{12 elements} (H, He, C, N, O, Ne, Mg, Si, S, Ca, Fe, and~Ni) are deposited by these events, and their abundance is tracked individually in each gas particle. Dust and metals in the gas are diffused using a Smagorinsky--Lilly model \citep{Smagorinsky} with a diffusion parameter of $C = 0.002$ in cosmological and $C = 0.0002$ for isolated simulations. {We} note that in isolated runs this explicit diffusion is only enabled for the fiducial run, since it increased the runtime significantly. We compared runs with different diffusion parameters and found no major differences with the analysis done in this paper. This is consistent with results from \mbox{FIRE-2 \citep{Fire2_2018}}.

The feedback implementation distinguishes between a local mechanical feedback model and a stochastic thermal feedback model that generates hot galactic winds. The~mechanical feedback injects the terminal momentum from unresolved SN remnants, calibrated using high-resolution simulations of SN superbubbles. The~stochastic thermal model increases the internal energy of gas particles around SN-II events, using a modified \citet{Dalla_Vecchia_2012} model with the outflow entropy $S_{\rm OF}$ set as a free parameter instead of the temperature increase $\Delta T$. 
In our cosmological simulations, momentum and energy injection are solid angle-weighted using a Voronoi tessellation around each particle, thereby ensuring momentum conservation. 
If the constant entropy model is not used, then the galactic wind properties (mass, velocity and metallicity) are informed by the TIGRESS framework \citep{Kim_2020}, %
with launched wind particles temporarily decoupled from hydrodynamic forces to mimic free-streaming escape from dense star-forming regions. 
These two schemes enable simulations to~effectively resolve the multiphase ISM in galaxies.

Metal yields are computed using the \textsc{CELib} library~\cite{Saitoh_2017}, rather than the metallicity- and redshift-dependent top-heavy IMF employed in the fiducial CROCODILE \mbox{simulations \citep{Oku2024}}. Specifically, for~SN-II feedback, we combine yield tables from \citet{Portinari} and \citet{Nomoto}. This choice yields a relatively high effective metal yield of $0.042$, compared with $0.025$ in \gthree (Table 1 in Paper III). The~discrepancy disappears if we instead use only the yield tables of \citet{Nomoto}, which also give an effective yield of $0.025$. This is a consequence of the broader SN progenitor mass coverage achieved by combining the two tables, which also yields more SN per stellar~population.

For the fiducial AGORA Cal-4 run, the~total energy from Type II and Type Ia supernovae is boosted to $E_{\rm SN} = 10.0 \times 10^{51}$ ergs/SN in~order to reach the required stellar mass at $z = 4$ (see Section~\ref{sec:fbstresults}). This parameter sets the overall feedback strength and is key for calibrating the code to produce a total stellar mass that fits in the AGORA calibration range, specifically, a target mass within 1.0--5.0 $\times 10^{9} M_{\odot}$ at $z = 4$. Additionally, the~production and destruction of dust follow the prescriptions in~\citet{Aoyama2020,Romano_2022}.

\subsection{Testing Feedback Model Variations with Isolated~Galaxy}
\label{sec:fb_variations}

To systematically bridge the gap between the \gthree and \gfour feedback implementations and isolate the effects of individual modeling choices, we perform a suite of \gfour isolated disk simulations with progressively increasing feedback model complexity. These variations span four key aspects of supernova feedback: energy partition between thermal and kinetic channels, shock radius prescriptions, terminal momentum injection and stochastic thermal injection. Each variation disables or modifies specific components of the full \gfour model, allowing us to trace how galaxy properties change as we transition from the simpler \gthree approach to the full momentum-driven stochastic \gfour implementation. Due to computational constraints, these systematic tests are performed only for the isolated disk setup (Table~\ref{tab:iso_runs}); cosmological simulations employ the full fiducial models (Table~\ref{tab:cosmo_runs}).

The first three tests explore how the split between thermal heating and bulk kinetic motion affects feedback coupling to the ISM. They use the \citet{Chevalier74} shock radius and Sedov--Taylor velocity kicks:
\begin{equation}
v_{\rm kick} = 151.41 \, ({\rm km \, s^{-1}}) \times \xi_{\rm cool}^{-0.2} \times n_{0}^{0.14} 
\end{equation}
\rev{where $n_0 = n_{\rm H}/(1 \, {\rm cm^{-3}})$ is the local ambient hydrogen number density and~$\xi_{\rm cool} = 0.5$ is a dimensionless parameter that calibrates the cooling timescale at the shell \mbox{formation radius}.}

\begin{itemize}
    \item `\textit{G4-FB-Thermal}' deposits 100\% of supernova energy as thermal heating with no explicit kinetic kick. This most closely resembles the \gthree isolated model.
    \item `\textit{G4-FB-Kinetic}' inverts this, injecting 100\% kinetic energy with zero direct thermal heating. Both runs are unphysical, since during the Sedov--Taylor phase approximately 72\% of the SN energy is thermal, while the remaining 28\% is in the kinetic motion of the expanding shell.
    \item `\textit{G4-FB-Mix}' splits energy equally (50\% thermal, 50\% kinetic) to probe whether a balanced split can approximate the effects of the full model. This is also the same partition used for \gthree cosmological runs, but~we note that the feedback model in that simulation is more complete than this simple test case (see Section~\ref{sec:g3feedback}).
    \item `\textit{G4-FB-Oku-Rshock}' retains the 50/50 energy split and velocity-kick formulation from \textit{G4-FB-Mix} but replaces the \citet{Chevalier74} shock radius with the \citet{Oku_2022} prescription. The~remaining tests use Oku's calculation.
\end{itemize}

We note that these velocity-kick implementations do not rigorously conserve energy, as~was noted in \citet{Shimizu_2019}. Now, we separately activate two different feedback variants (one kinetic, the~other thermal) that can both help alleviate overcooling and regulate star formation more efficiently than the previous treatment:

\begin{itemize}
    \item `\textit{G4-FB-Momentum}' implements full momentum-driven mechanical feedback with superbubble momentum $\hat{p}(n_0,Z)$ calculated from resolved simulations of SN remnants in \citet{Oku_2022}:
\begin{equation}
    \hat{p}(n_0,Z) = 1.75 \times 10^5 \,( M_\odot \, {\rm km \, s^{-1}}) n_0^{-0.05}\Lambda_{6,-22}^{-0.17} 
    \end{equation}
    \rev{where $\Lambda_{6,-22} = \Lambda(10^6\, {\rm K})/(10^{-22} {\rm erg \, cm^3 \, s^{-1}})$ represents the cooling function from \citet{1993Sutherland} at $T = 10^6 \, \rm K$.}
    The remaining energy, $E_{\rm thermal} = E_{\rm SN} - \Delta E_{\rm kinetic}$, is deposited as heat. This ensures exact energy~conservation.

    \item `\textit{G4-FB-Stochastic}' heats particles to high entropy stochastically rather than distributing energy uniformly. Terminal momentum is not calculated in this case; instead, we use parameters from \textit{G4-FB-Oku-Rshock} and additionally enable the stochastic thermal model. We use the entropy of hot outflows $S_{\rm OF}$ as a free parameter rather than a fixed increase in temperature. The~thermal energy required to heat the $i$-th gas particle to target entropy is
\begin{equation}
    \Delta E_{\rm req,i} = \frac{1}{\gamma - 1} \frac{m_i}{\mu m_p} n_i^{2/3} S_{\rm OF},
    \end{equation}
    where $n_i$ is the number density and we adopt $S_{\rm OF} = 10^8\, k_B \, {\rm K\,cm^2}$ as the fiducial value\rev{, with~$k_B$ the Boltzmann constant}. The~solid-angle-weighted thermal energy available from the supernova event is
\begin{equation}
    \Delta E_{\rm th,i} = \epsilon_{\rm th} E_{\rm SN} \frac{\Omega_i}{4\pi},
    \end{equation}
    where $\epsilon_{\rm th} = 1 - \epsilon_{\rm kin}$ is the thermal energy fraction (we adopt $\epsilon_{\rm th} = 0.5$) and $\Omega_i$ is the solid angle subtended by particle $i$ in the Voronoi tessellation around the star particle. The~probability of heating particle $i$ is then
\begin{equation}
    P_i = \frac{\Delta E_{\rm th,i}}{\Delta E_{\rm req,i}}.
    \end{equation}
    
    A random number $0 < \theta_i < 1$ is drawn for each gas particle; if $\theta_i < P_i$, then particle $i$ receives thermal energy $\Delta E_{\rm req,i}$. When $P_i > 1$, thermal energy $\Delta E_{\rm th,i}$ is injected~directly.

\end{itemize}

\section{Simulation~Setup}
\label{sec:setup}

We perform two categories of simulations, namely, isolated disk galaxies and cosmological zoom-ins, each with a hierarchy of physical complexity designed to isolate the effects of specific numerical and physical modeling choices. All simulations employ the common AGORA initial conditions (Section~\ref{sec:methods}) and shared physics modules (Section~\ref{sec:commonphys}), with~differences arising from the gravity solvers (Section~\ref{sec:grav}), hydrodynamic schemes (Section~\ref{sec:sph}), and~feedback implementations (Section~\ref{sec:feedback}) detailed in the previous~section.

\subsection{Isolated Galaxy~Simulations}
\label{sec:iso_setup}

Isolated galaxy simulations provide an idealized testbed for code validation and physics exploration. Their computational efficiency (hours rather than days or weeks) enables rapid iteration and parameter studies, while their simplified environment consisting of a~single rotating disk free from cosmological accretion and mergers facilitates clean attribution of results to specific model or parameter~choices. 

We run two baseline configurations: NSFF (No Star Formation or Feedback), hydrodynamics + cooling only, and~SFF (Star Formation and Feedback) with full physics and all sub-grid modules active. Both configurations are run to $t=500$~Myr, consistent with Paper~II. 

To isolate which feedback components drive differences between \gthree and \gfour, we systematically disable or modify elements of the \gfour feedback model, with~the simplest model being the most similar to the one in \gthree. Six variations are performed, detailed in Table~\ref{tab:iso_runs}. Similarly, we vary the feedback strength in the fiducial model to match our cosmological calibration variations and deduce its effects, with~runs also detailed in the previous~table.

\subsection{Cosmological Zoom-In~Simulations}

\begin{figure}[H]
    \includegraphics[width=.99\linewidth]{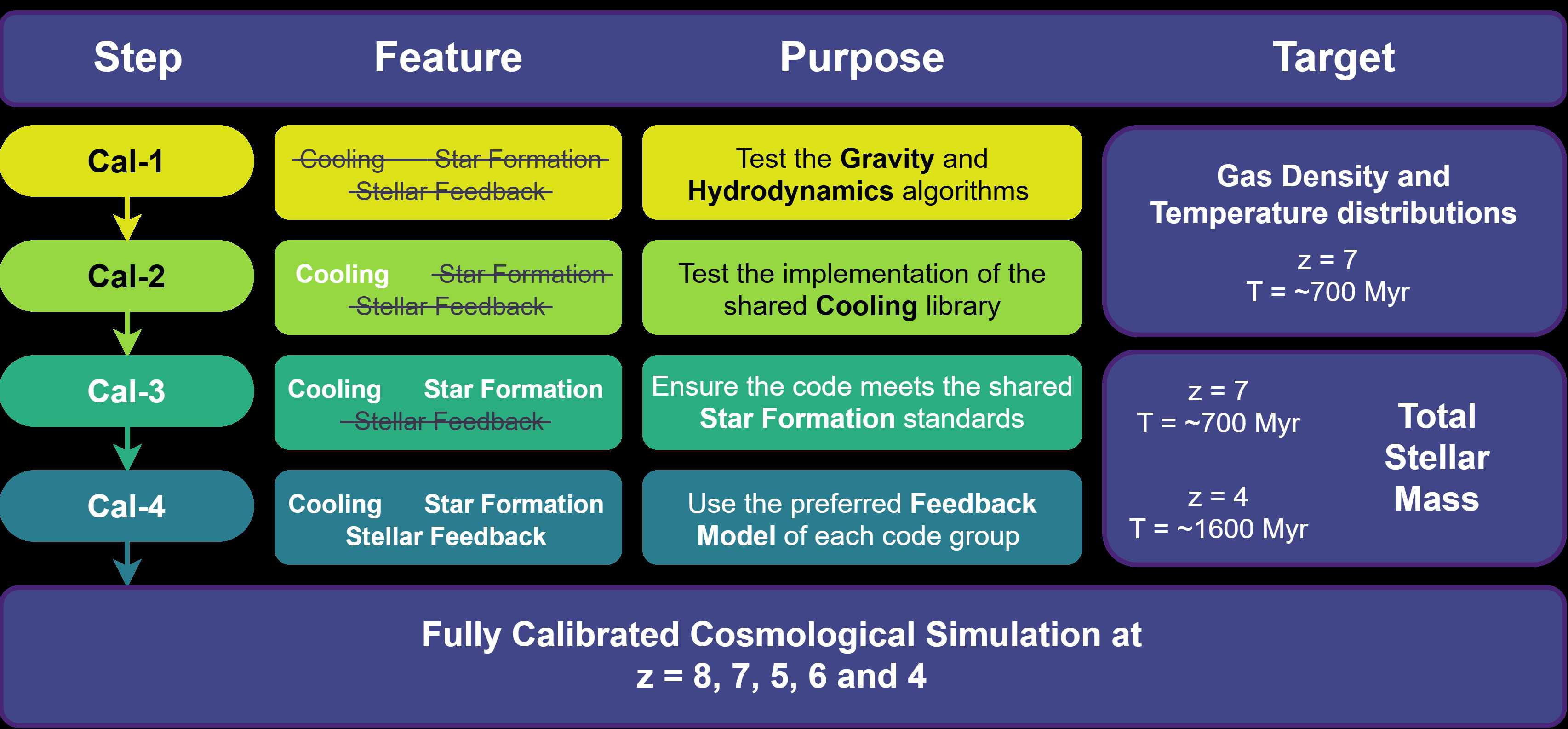}
    \caption{{AGORA} %
 cosmological calibration procedure. Physical complexity increases from top to bottom: Cal-1 (gravity and adiabatic hydrodynamics), Cal-2 (radiative cooling), Cal-3 (star formation), and Cal-4 (stellar feedback). Each step has a specified target redshift ($z=7$ for Cal-1/2/3, $z=4$ for Cal-4) for inter-code comparison and a diagnostic focus (e.g., gas and temperature projections in Cal-1/2, stellar mass in Cal-3/4). This staged approach ensures differences in final calibrated runs can be confidently attributed to feedback implementations rather than accumulated numerical artifacts. Adapted and modified from Figure~1 of \citet{Roca_F_brega_2021}.}
    \label{fig:calsteps}
\end{figure}

Cosmological simulations subject our models to the full dynamical range and environmental complexity of hierarchical structure formation. Following the AGORA calibration protocol (Paper III \citep{Roca_F_brega_2021}), we proceed through four calibration steps, incrementally activating physical processes to isolate their effects. \rev{This procedure is outlined in Figure~\ref{fig:calsteps}.}

We begin with Cal-1, a~no cooling, star formation, or feedback run used to verify that the gravity and hydrodynamic solvers converge properly. We then turn on radiative cooling (via \textsc{GRACKLE}) for Cal-2. For~both runs, the~main objective is to reproduce the overall gas density and temperature distributions of the other codes, which should also reflect an agreement in the gas phase. In~the next step, Cal-3, we turn on star formation. Now, the~target shifts to a consistent stellar mass between the codes, testing the star formation models to ensure that they are well calibrated. The~pressure floor is also turned on in this step. Finally, Cal-4 turns on the ``favorite'' stellar feedback implementation of each group. The~only requirement set here is for the stellar mass to be between \mbox{$10^9 M_{\odot} < M < 5 \times 10^9 M_{\odot}$}, which is in the range predicted by semi-analytical galaxy formation models. This final run serves as the \gfour entry into the \textit{CosmoRun} suite of galaxy formation simulations. A~detailed account of the calibration process, results, and~comparison with other simulations in the \textit{CosmoRun} suite is given in Appendix~E of \citet{jung2025}.

We also run a dark-matter-only simulation (no baryons), which provides a baseline for validating gravity solvers independent of hydrodynamics (Section~\ref{sec:dmo}). It is run to $z=7$ for comparison with~Cal-1. 

In addition to this fiducial set of simulations, we perform a series of additional runs with varying feedback strengths and \textsc{GRACKLE}. These runs pave the way for a fully AGORA-consistent CosmoRun and also demonstrate the effects of minor variations in sub-grid physics. Table~\ref{tab:cosmo_runs} summarizes all \rev{fifteen} cosmological runs: five \gthree runs (DM-only + Cal-1/2/3/4) and \rev{ten} \gfour runs (DM-only + Cal-1/2/3/4 + 3 Cal-3 cooling modes + \rev{2} Cal-4 feedback strengths).

\subsection{Analysis~Methodology}
\label{sec:analysis}

All analysis is performed using the \texttt{yt} python library \citep{Turk_2010}. While Papers II and III used \texttt{yt} v3, this work uses \texttt{yt} v4, which introduces changes that affect visualization: SPH field projections now compute kernel-weighted contributions at each pixel directly rather than using an octree deposition. This produces smoother and more accurate visualizations, but yields minor differences in projected quantities relative to earlier AGORA papers. Consequently, our \gthree visualizations differ slightly in appearance from those in Papers II and III, though~the underlying physical quantities remain~identical.

Halo centering follows the iterative procedure described in Paper III, in which we compute the center of mass within progressively smaller spheres until convergence. The~galaxy's face-on and edge-on axes are determined from the total angular momentum of star particles. The~virial radius $R_{\rm vir}$ is defined as the radius enclosing a mean density $\bar{\rho} = \Delta_{\rm vir}(z) \rho_{\rm c}(z)$, where $\rho_{\rm c}(z)$ is the critical density of the universe and $\Delta_{\rm vir}(z)$ is the redshift-dependent density contrast from \citet{Bryan_1998}.

\subsection{Computational~Resources}
\label{sec:compute}

Isolated galaxy simulations were performed on a local linux cluster \textsc{Virgo} (AMD EPYC) at U. Osaka, using single nodes with 64--128 CPU cores and 64--128 GB RAM. MPI parallelization across multiple nodes degraded performance because communication overhead dominated computation. NSFF runs required 0.5--2 h to reach 500\,Myr, while SFF runs required 2--12 h, depending on the feedback model and~parameters.

Cosmological zoom-in simulations were performed on the SQUID supercomputer (Intel Xeon Platinum 8368 [Icelake]/2.40\,GHz, 38-cores per CPU\,$\times$\,2-CPUs/256\,GB memory per node) at the D3 Center of the University of Osaka, using four nodes with 64 cores each (256 cores total). Cal-1, Cal-2, and~Cal-3 run required $\sim$2 days to reach $z=7$. Cal-4 runs required $\sim$4 days to $z=4$, $\sim$8 days to $z=1$, and~$\sim$24 days to $z=0$. These timings are for the fiducial \gfour model; variations with different feedback strengths or cooling networks had similar runtimes. These runtime numbers can be highly~machine-dependent.

\section{Results and Discussion}
\label{sec:results}

We present our systematic comparison of \gthree and \gfour across isolated and cosmological contexts, organized by increasing physical complexity. We begin with purely gravitational runs (Section~\ref{sec:dmo}), then include hydrodynamics, radiative cooling (Section~\ref{sec:hydro_cool}), star formation (Section~\ref{sec:cal3coolvar}), and~feedback (Section~\ref{sec:fbresults}). This hierarchy mirrors the calibration philosophy in AGORA, allowing us to isolate sources of divergence at each level of complexity. All figures employ techniques described in Section~\ref{sec:analysis}, with~projections computed via SPH kernel-weighted deposition in \texttt{yt} v4. Where appropriate, we reference results from other AGORA codes and other comparison studies to contextualize our findings within the broader galaxy formation~literature.

\subsection{Baseline: Dark Matter-Only~Runs} 
\label{sec:dmo}

\begin{figure}[H]
\begin{adjustwidth}{-\extralength}{0cm}
        \centering
        \includegraphics[width=.99\linewidth]{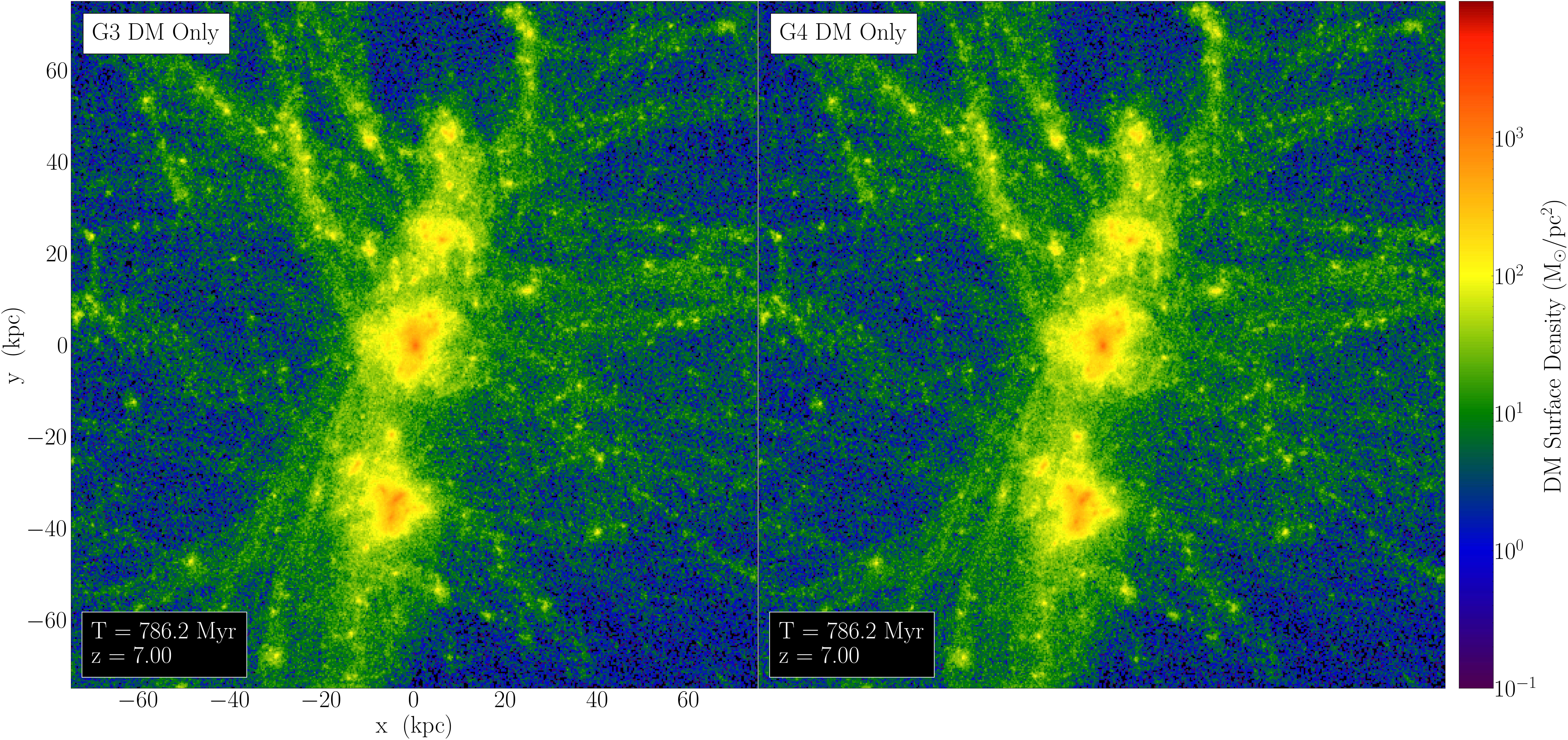}
        \end{adjustwidth}
        \caption{Dark matter density projections at $z=7$ for (\textbf{left}) \gthree and (\textbf{right}) \gfour. Particles are deposited via a cloud-in-cell (CIC) scheme on a 300\,pc grid, and~surface densities are then calculated for a slice of $150$\,kpc thickness. Both codes produce visually indistinguishable dark matter distributions, demonstrating convergence of gravity solvers (TreePM vs. FMM-PM).}
        \label{fig:proj_DM}
\end{figure}

We begin by verifying that our gravity solvers, TreePM (2nd order) in \gthree versus FMM-PM (4th order) in \gfour for cosmological runs, produce consistent dark matter distributions independent of baryonic physics. Figure~\ref{fig:proj_DM} shows the dark matter density projections at $z=7$ in the zoom-in cosmological simulation. The~dark matter distributions are visually indistinguishable. Both codes reproduce the filamentary cosmic web structure feeding the central halo, with~satellite halos positioned identically along the filaments. The~central halo exhibits the same stretched morphology in both projections, and~the density contrast between filaments and voids is~comparable.

\begin{figure}[H]\begin{adjustwidth}{-\extralength}{0cm}
        \centering
        \includegraphics[width=.99\linewidth]{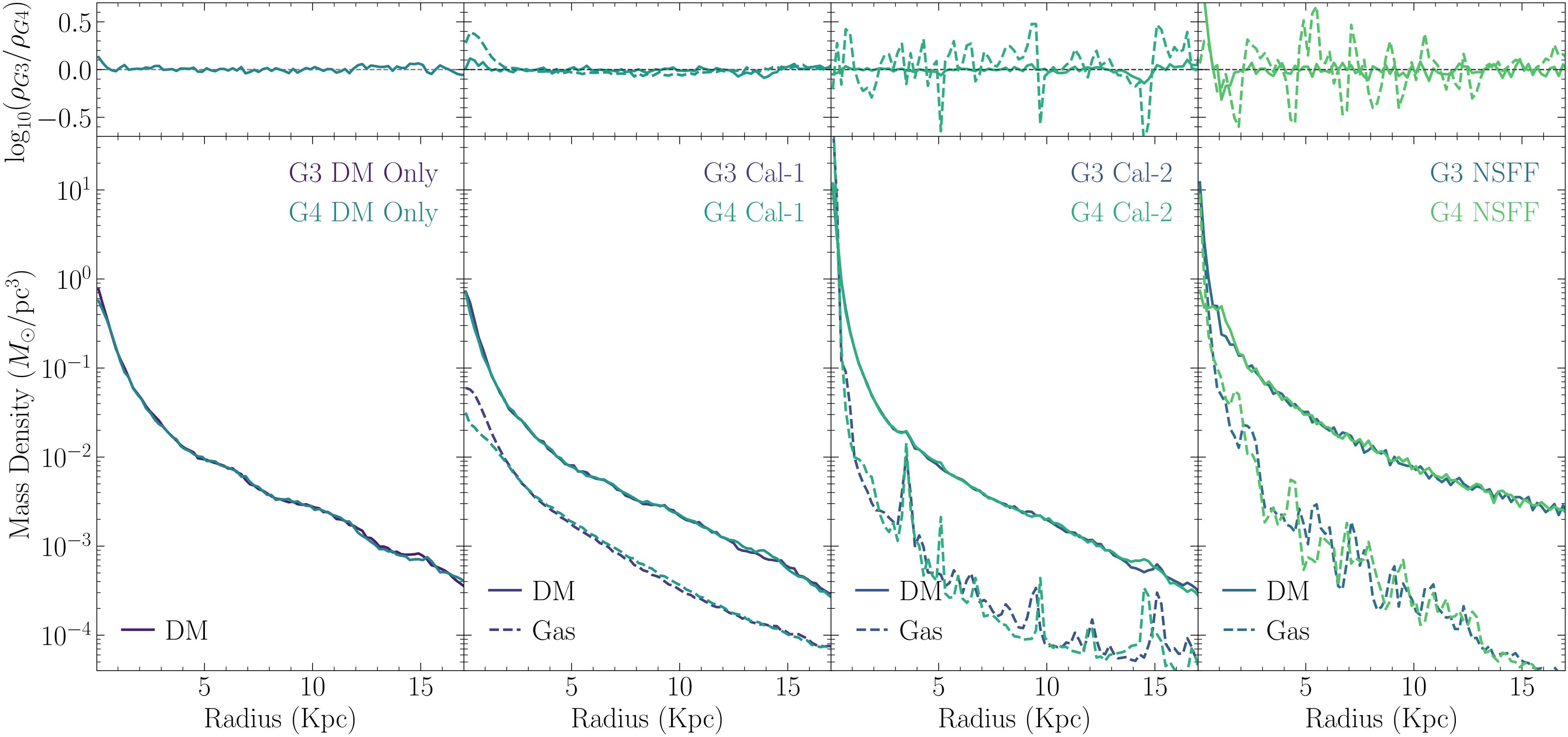}\end{adjustwidth}
        \caption{Spherically averaged density profiles for dark matter (solid line) and gas (dashed line, when applicable) at $z=7$, shown for \textit{DM-Only} (col 1), Cal-1 (col 2), Cal-2 (col 3), and~\textit{NSFF} isolated runs at $t=500$ Myr (col 4). The~upper subplot in each panel displays the logarithmic density ratio $\log_{10}(\rho_{\rm G3}/\rho_{\rm G4})$ to highlight~deviations.}
        \label{fig:surf_12}
\end{figure}

Quantitative analysis confirms this visual agreement. The~spherical density profiles ({Figure}%
~\ref{fig:surf_12}, leftmost panel) show dark matter profiles converging to within $<$0.05 dex at all radii except the innermost 500 pc, where a deviation of $\sim$$0.1$ dex ($\sim$$25\%$) appears. This deviation is consistent with other comparisons of dark matter-only density profiles. The~nIFTy galaxy cluster comparison \citep{Sembolini_2016} found $\sim$$20\%$ discrepancies in inner and outer dark matter profiles even between different flavors of the same base code (GADGET-2 vs. GADGET-3). Similarly, AGORA Paper I \citep{Kim_2013} reported central ($r < 1$ kpc) deviations of $\sim$$ 20\%$ between codes for a slightly smaller halo ($M_{\rm vir} = 1.7 \times 10^{11} M_\odot$ at $z=0$), also between different flavors of the same~code.

We attribute these small differences to the low particle counts at the inner kpc radius, with~a noisy distribution that makes fluctuations in the profile common. This is also suggested by the dark matter profiles in our other runs, which do not always manifest a divergence at $r < 1$~kpc (see {Figures} \ref{fig:surf_12}, \ref{9}a and \ref{13}a). When including baryonic physics, the~deviations persist and are dominated by hydrodynamic and feedback processes rather than gravity-solver~artifacts.

Therefore, we conclude that the TreePM and FMM-PM algorithms produce numerically convergent dark matter distributions at our $80$~pc spatial resolution. This ensures that all subsequent divergences in baryonic runs arise primarily from gas physics and feedback~models.

\subsection{Hydrodynamics and~Cooling}
\label{sec:hydro_cool}

Having established gravity solver convergence, we now activate the hydrodynamic modules: first adiabatically (Cal-1), then with radiative cooling (Cal-2 and NSFF) to test the consistency of the SPH formulation and the implementation of the cooling~library.

\vspace{-3pt}

\subsubsection{{Adiabatic} %
 Hydrodynamics (Cal-1)}

\begin{figure}[H]\begin{adjustwidth}{-\extralength}{0cm}
        \centering
        \includegraphics[width=.99\linewidth]{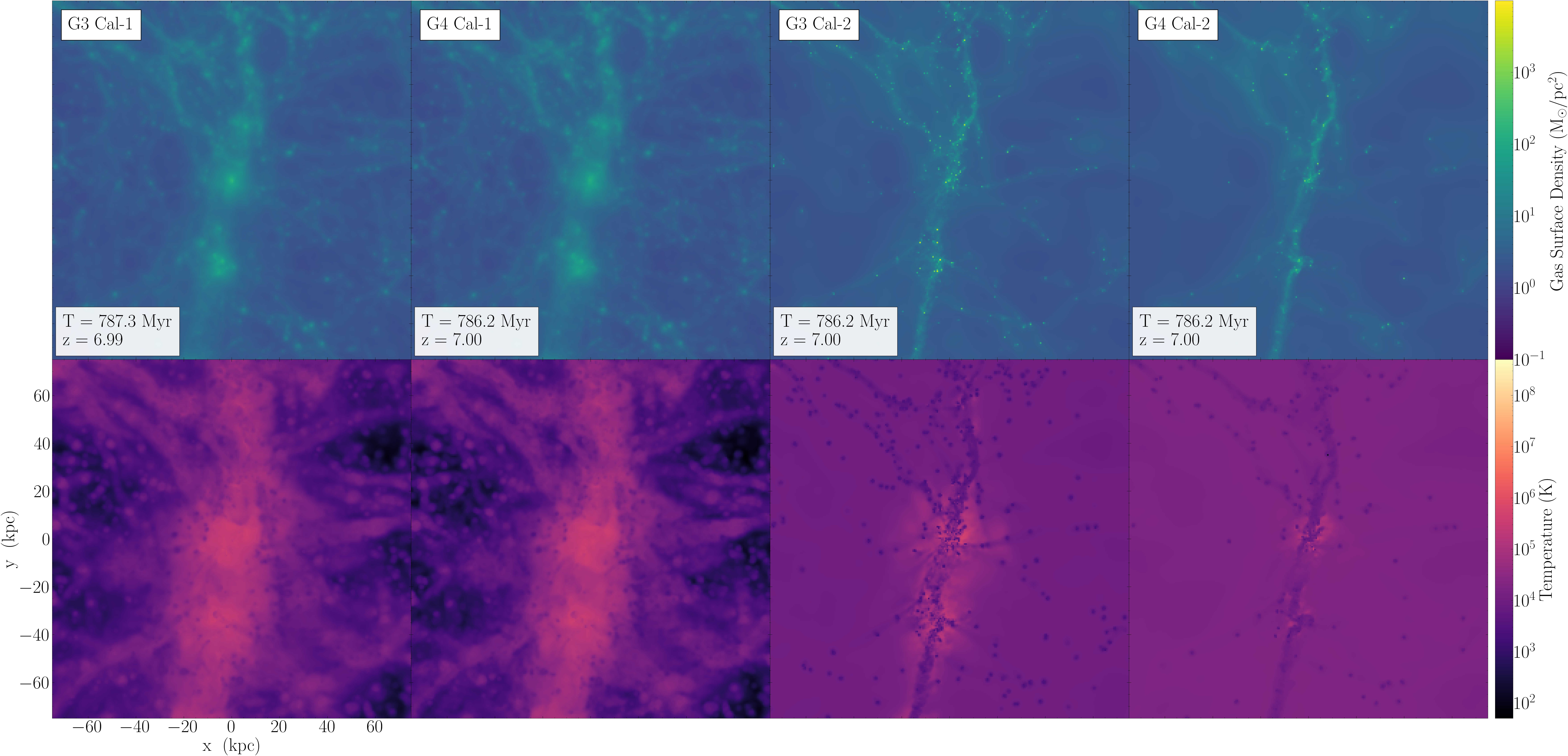}\end{adjustwidth}
        \caption{Gas density projection (\textbf{top}) and density-weighted temperature projection (\textbf{bottom}), each projected through a slab of thickness $150$ kpc at $z=7$, for~Cal-1 (adiabatic, cols 1--2) and Cal-2 (with cooling, cols 3--4). While both \gthree and \gfour converge in Cal-1, minor variations in the temperature distribution are present in Cal-2.}
        \label{fig:proj_12}
\end{figure}

Figure~\ref{fig:proj_12} (columns 1--2) demonstrates that \gfour successfully reproduces the large-scale gas density and temperature distributions of \gthree in Cal-1, with~agreement extending down to individual kpc-scale features. These results are also consistent with other AGORA~codes. 

However, quantitative analysis reveals a subtle but persistent discrepancy. The~radial gas density profiles ({Figure}~\ref{fig:surf_12}, column 2) agree to within <0.1 dex for $r > 2$ kpc, but~exhibit a $\sim$$ 0.4$ dex ($\times 2.5$) difference in the inner core. This manifests as a lack of $\rho \gtrsim 10^{-24}$ g cm$^{-3}$ dense gas in \gfour, which \gthree readily produces. We confirmed that this is not a centering artifact; the denser gas is at the center of the halo, and is correlated with the higher surface~density.

\begin{figure}[H]\begin{adjustwidth}{-\extralength}{0cm}
        \centering
        \includegraphics[width=.95\linewidth]{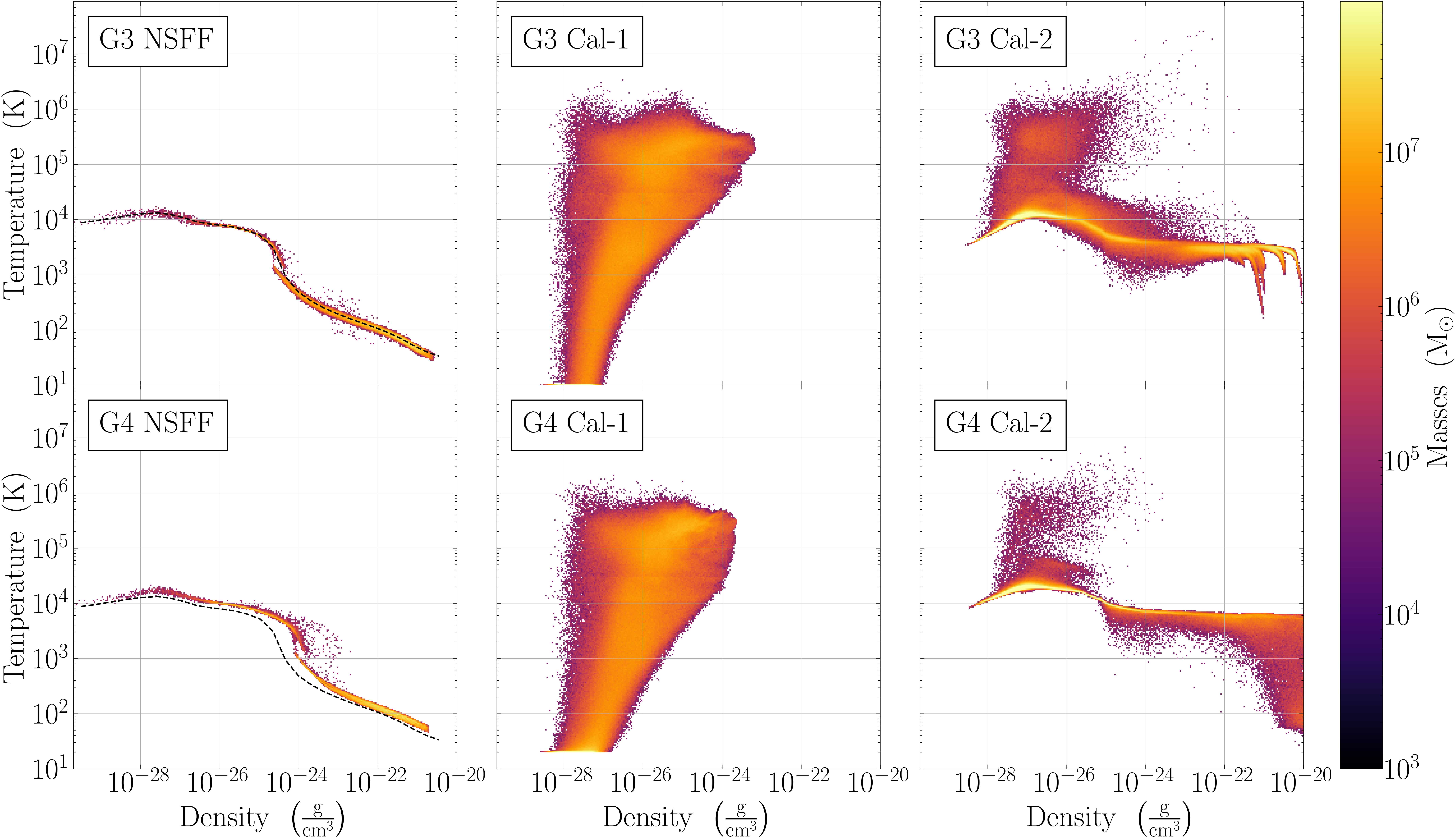}\end{adjustwidth}
        \caption{{Mass-weighted} phase diagrams of gas density vs. temperature for the gas within $100$\,kpc from the center of the galaxy in the \textit{NSFF} (\textbf{left}), Cal-1 (\textbf{middle}), and~Cal-2 (\textbf{right}) runs, at~$500$ Myr for the first and $z = 7$ for the latter runs. To~guide the eye, we use a thick dashed line in the \textit{NSFF} panel to plot the mean temperature in each density bin for \gthree. Colors represent the total gas mass in each 2-dimensional~bin.}
        \label{fig:phase_12}
\end{figure}

The gas phase diagrams ({Figure}~\ref{fig:phase_12}, middle panels) corroborate this observation, with~a lack of the densest gas in \gfour. This difference disappears when cooling is enabled in Cal-2 and NSFF runs, which is because~radiative energy loss allows the gas to collapse further and form a much denser cusp in both~codes.

\vspace{-3pt}

This discrepancy arises from differences among SPH schemes. In~particular, by~running tests in which we systematically vary SPH modules to more closely imitate \gthree, we find that the most significant contribution comes from disabling artificial conduction. Using a quintic kernel or time-dependent viscosity in \gfour, like in \gthree, does not affect the formation of a denser cusp. This is because artificial conduction can promote gas mixing from infalling clouds, thereby forming a constant entropy core at the halo center, as~observed~here.

Other results in the literature have shown that for~purely adiabatic hydrodynamics, this gas density cusp can be very sensitive to the scheme used. In~Paper III, GIZMO, which uses mesh-free particle hydrodynamics \citep{Hopkins_2012}, presents the same lack of high-density gas in Cal-1 that we noted in \gfour. AREPO, compared in a Paper IV appendix and with a comparable moving-mesh implementation, also gives very similar results. In~the nIFTy cluster comparisons \citep{Sembolini_2016}, differences of up to 1 dex in the gas density profiles at $r < 100$ kpc were found, primarily between mesh-based and particle-based codes but~also between GADGET-2 and GADGET-3~variants.

Critically, this adiabatic disagreement has no lasting impact on our conclusions. Once cooling is activated (Section \ref{sec:rcid}), both codes converge in their central density structure, and~the effect becomes negligible compared to feedback-driven variations (Section~\ref{sec:fbresults}).

\subsubsection{{Radiative} Cooling: Isolated Disks (NSFF)}
\label{sec:rcid}

\begin{figure}[H]
\begin{adjustwidth}{-\extralength}{0cm}
  \centering
    \subfloat[\centering\label{fig:proj_NSFF_a}]{
    \includegraphics[width=0.48\linewidth]{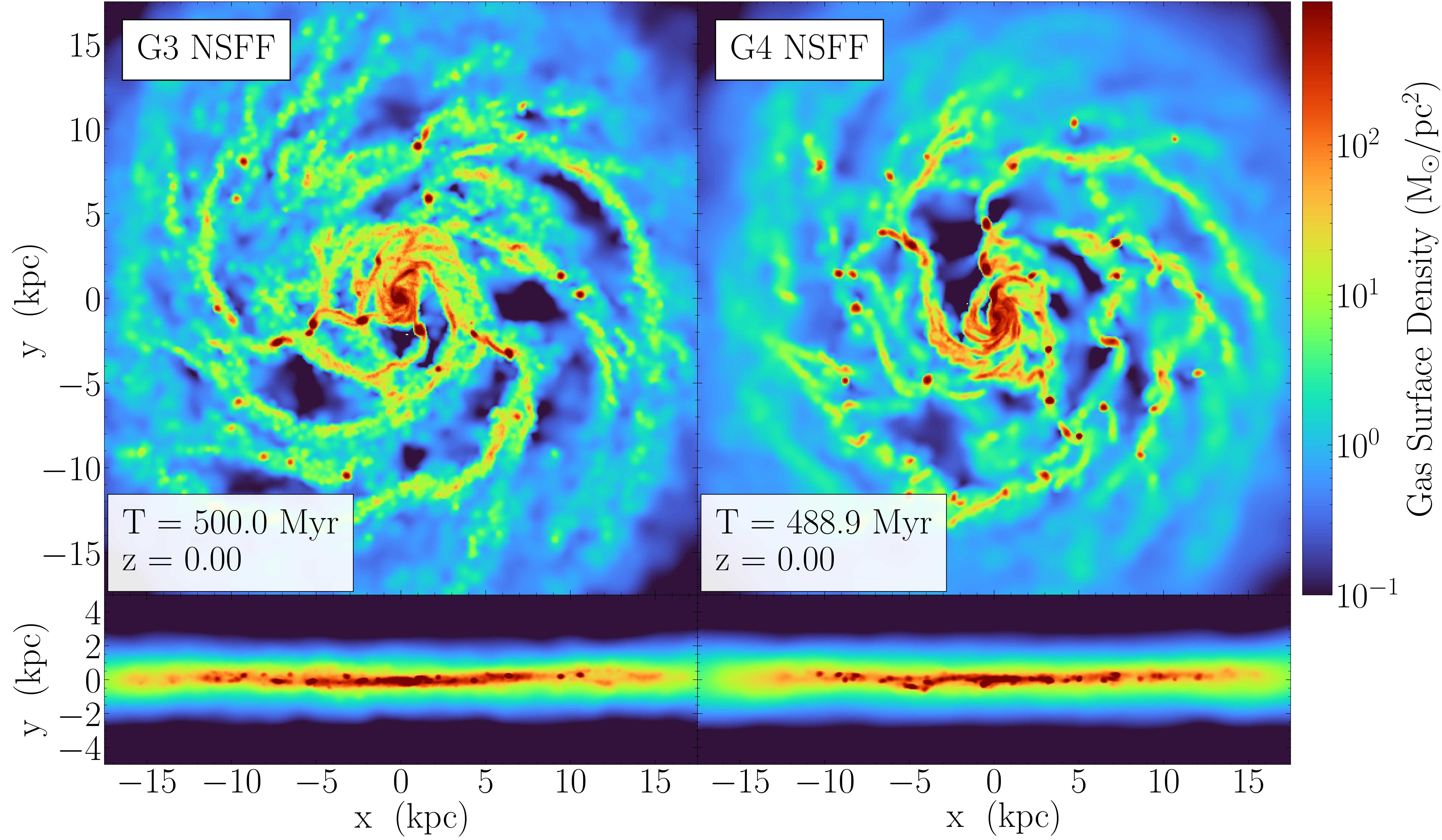}
    }~
    \subfloat[\centering\label{fig:proj_NSFF_b}]{
    \includegraphics[width=0.48\linewidth]{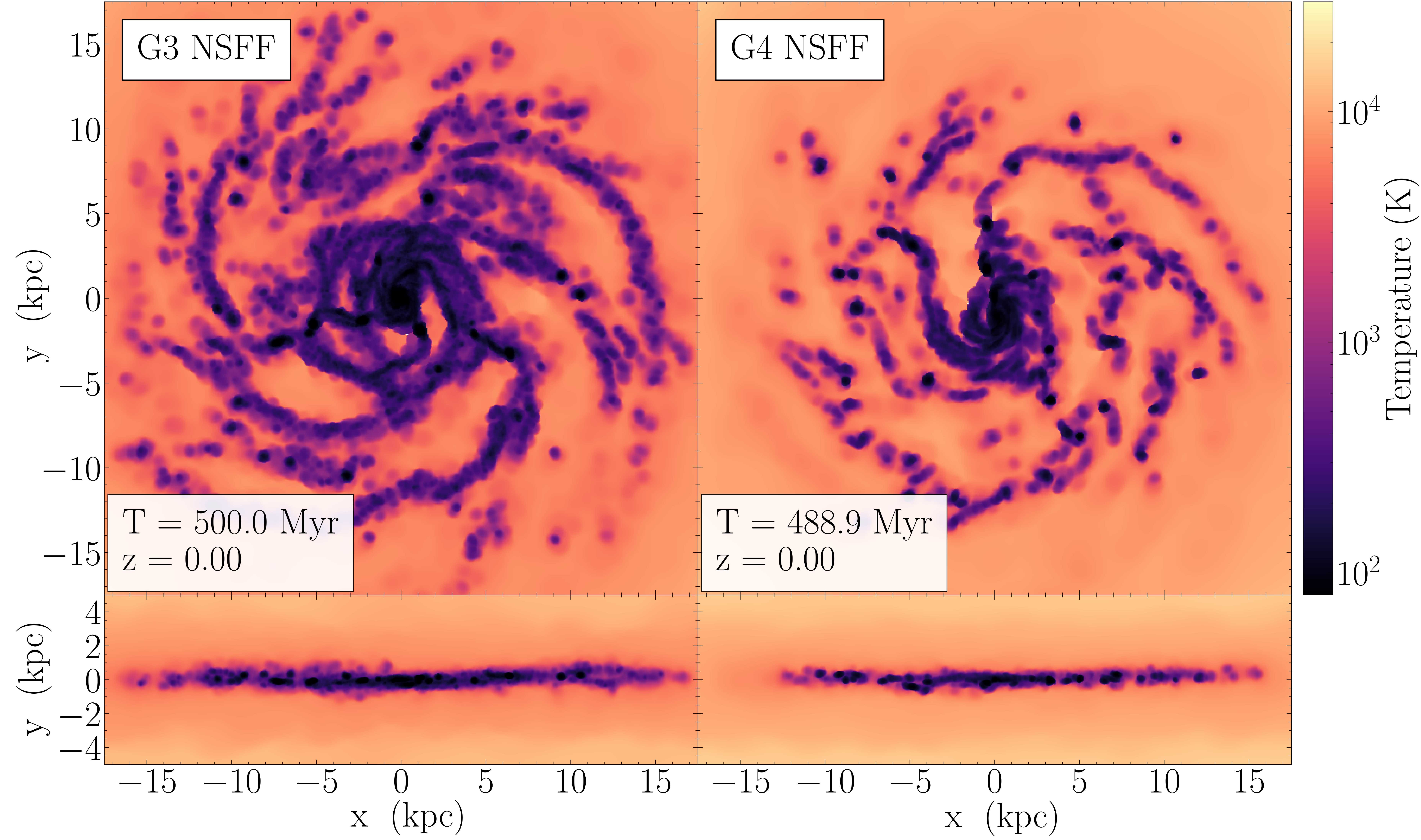}
    }
\end{adjustwidth}
    \caption{{(\textbf{a})} %
 Gas density projection projected through a slab of $35$\,kpc thickness at $500$\,Myr for~\textit{NSFF} runs. The~left panel shows \gthree, while the right panel shows \gfour. There is overall agreement between the two, with~minor differences in the gas clumpiness and the distribution of cold~gas.  (\textbf{b}) \rev{Same as the left panel but~for a density-weighted temperature projection.}}\label{4}
\end{figure}

We now examine \textit{NSFF} (No Star Formation or Feedback), where radiative cooling via \textsc{GRACKLE} is enabled in the isolated disk setup. Both codes produce gravitationally unstable clumpy disks (Figure~\ref{4}a), but~with \rev{slightly} different morphologies. \gthree exhibits point-like dense clumps with sharp boundaries, while \gfour produces a more diffuse filamentary structure with smoother density~gradients.

\vspace{-6pt}

This morphological difference is \textit{not} primarily driven by variations in the SPH scheme, a~conclusion we reached through various numerical experiments. We tested \gfour with: (1) pressure--entropy formulation instead of pressure--energy, (2) disabled time-step limiters, (3) disabled artificial conduction, (4) quintic spline kernel (matching \gthree), and~(5) non-equilibrium chemistry networks. None of these changes reproduced \gthree's point-like clump structure. Conversely, running \gthree with modern SPH enhancements (pressure--energy formulation, time-step limiters, artificial conduction) \textit{also preserved} the point-like clumps. {These} \gthree tests are not included in this paper due to code archaeology difficulties; the~original Paper II runs were performed over a decade ago with a code version that has since evolved significantly, making exact reproduction of the isolated galaxy run challenging despite our \mbox{best efforts.}

Therefore, we think that these features are instead caused by the \textsc{GRACKLE} interface of each code. In~contrast, \citet{Hobbs_2013, Hu_2014} demonstrated that standard SPH without artificial conductivity can produce a similar point-like structure due to a lack of fluid mixing. However, they also showed that improved SPH with conductivity resolves these into filaments that only fragment on smaller scales. Our \gfour implementation includes both improved SPH and artificial conduction \citep{Price_2008}, which enhances mixing and suppresses spurious fragmentation. When running \gfour \textit{NSFF} tests with artificial conduction disabled, we did not find these structures; however, this is likely because a density-independent scheme was used, which has also been found to improve fluid mixing. Curiously, some tests with no UV background and grackle non-equilibrium cooling (mode 3) produced morphologies most similar to \gthree. 

Radial profiles ({Figure}~\ref{fig:surf_12}, rightmost panel) show dark matter agreement to within $\sim$$ 0.1$ dex, with~the exception of the central 1 kpc. This time, the~deviation arises from a centering difference: \gfour selects a gas clump slightly offset from the dark matter density peak (Figure~\ref{4}a, left panel), leading to an apparent $\sim$$ 0.3$ dex offset in the dark matter profile. Gas profiles exhibit a scatter of >0.5 dex due to the stochastic clump distribution, but~the overall radial trend remains~consistent.

Thermodynamic properties \rev{present some minor but systematic differences}. Temperature projections (Figure~\ref{4}b) show that both disks have cooled and settled into thin spiral structures with characteristic $T$ $\sim$$10^4$ K, as~expected from primordial cooling equilibrium in the \textsc{CLOUDY} tables \citep{ferland20132013releasecloudy}. The~phase diagrams (Figure~\ref{fig:phase_12}, left panels) confirm that >95\% of gas mass resides on a tight equilibrium curve. However, careful inspection reveals a subtle offset: \gfour's plateau sits $\sim$$ 0.1$ dex ($\sim$$ 25\%$) higher in temperature than \gthree. This was also observed in Paper II between different codes implementing the same \textsc{GRACKLE} library (see their {Figure~16} \citep{Kim_2016} %
, where \textsc{GEAR}/\textsc{SWIFT} exhibits a similar offset). We hypothesize that this stems from either: (1) subtle \textsc{GRACKLE} interface differences (e.g., iteration convergence criteria, interpolation order), or~(2) different internal energy $\leftrightarrow$ temperature conversion routines (\gfour {explicitly} stores both internal energy and temperature for all particles, while \gthree stores only internal energy in isolated runs, but also both in cosmological simulations). Using internal energy rather than temperature in the phase diagram does not resolve the offset, confirming that it reflects actual physical state differences rather than post-processing artefacts. This, along with the prevalence of this slight shift in essentially all codes in AGORA and in calibrations with cooling enabled, suggests that either the first option is more likely or that there is an unknown coupling between hydrodynamics and cooling that we did not~consider.

We also note that a key parameter to reproduce the \textit{NSFF} \gthree runs was the metal background. Paper II runs use a uniform background with $Z = 0.02041$, which significantly increases cooling and star formation rates. \gfour only attached metallicity values to SPH particles when turning the feedback modules on. Thus, \textsc{GRACKLE} did not read these values and read cooling table values for very low metallicity (essentially only taking into account primordial cooling). Attaching metallicity to fluid particles and fixing their value to the AGORA standard of $Z_{\rm floor} = Z_{\odot} = 0.02041$ allowed gas to cool down to $10^2$ K properly. In~star-forming simulations, slightly changing the background can significantly affect the SFR; this is because increased metal cooling rates, which increase the cold gas fraction, allow more stars to form. However, this is an ultimately unphysical treatment, included solely to more closely resemble present-day galaxy properties in isolated~simulations.

\subsubsection{{Radiative} Cooling: Cosmological Halos (Cal-2)}

In the cosmological Cal-2 runs, radiative cooling is activated for the hierarchically assembled halo. Field projections (Figure~\ref{fig:proj_12}, columns 3--4) show that gas density and temperature distributions broadly trace the same filaments and satellite halos in both codes. However, \gfour's gas appears more diffuse and smoother than \gthree, consistent with our \textit{NSFF} findings. More strikingly, \gfour exhibits a $\sim$$ 5\times$ higher temperature background in the diffuse~IGM.

This temperature offset persists in the phase diagrams (Figure~\ref{fig:phase_12}, right panels). \gfour's equilibrium plateau at $n$ $\sim$$10^{-26}$--$10^{-20}$ g cm$^{-3}$ sits $\sim$$ 0.3$ dex higher than \gthree, mirroring the \textit{NSFF} offset. Additionally, \gfour shows less shock-heated gas at $T > 10^5$ K, a feature more similar to \textsc{GIZMO}'s and \textsc{Changa}'s behavior in Paper III. We attribute this to artificial thermal conduction smoothing out temperature discontinuities across shocks, reducing the hot gas mass fraction compared to \gthree. \rev{Additionally, \gfour produces a larger temperature spread} in the cold, dense gas at $T < 10^4$ K, $n > 10^{-22}$ g cm$^{-3}$, again aligning better with \textsc{GIZMO} and \textsc{CHANGA} results (see Paper III, Figure~4).

The ``tail'' structures in \gthree's phase diagram, consisting of high-density gas offset from the main equilibrium track, are discussed in detail in Paper III, Section~5.2.2. These arise from \textsc{GRACKLE} table interpolation artifacts: when gas density falls between tabulated bins, linear interpolation can produce temperatures inconsistent with local heating/cooling balance, creating transient tails before the next cooling timestep corrects them. The~tail spacing matches the density resolution of the \citet{Haardtable} tables, confirming this~interpretation.

Radial gas profiles (Figure~\ref{fig:surf_12}, third column) show similar behavior to \textit{NSFF}, with~larger scatter in some regions due to different clump distributions. Importantly, the~Cal-1 adiabatic central density deficit in \gfour has now disappeared; both codes produce comparable central gas densities, with~the scatter now arising from the noisy particle distributions. This confirms that radiative cooling, which allows gas to efficiently lose thermal support, overrides the adiabatic compression differences we identified~earlier.

Minor disagreements in \textit{NSFF} and Cal-2 runs stem primarily from the coupling of the \textsc{GRACKLE} interface with each code's specific hydrodynamic implementation. While both codes use the identical \textsc{GRACKLE} v3 library, differences in internal energy update schemes, timestep criteria, and~artificial conduction treatments produce $\sim$$ 0.1$--$0.3$ dex temperature offsets and altered clump morphologies. These effects are subdominant to feedback-driven variations (Section~\ref{sec:fbresults}), but are important for understanding baseline systematic uncertainties. Our additional cooling network experiments (Section~\ref{sec:cal3coolvar}) further disentangle these~effects.

\subsection{Cooling Variations and Star~Formation}
\label{sec:cal3coolvar}

We now activate star formation (Cal-3), turning on the Schmidt law prescription (Section~\ref{sec:commonphys}) and the Jeans pressure floor. Additionally, we perform \gfour reruns with progressively more sophisticated \textsc{GRACKLE} chemistry networks (modes 0/1/2/3) to assess sensitivity to non-equilibrium~chemistry.

\begin{figure}[H]\begin{adjustwidth}{-\extralength}{0cm}
        \centering
        \includegraphics[width=.99\linewidth]{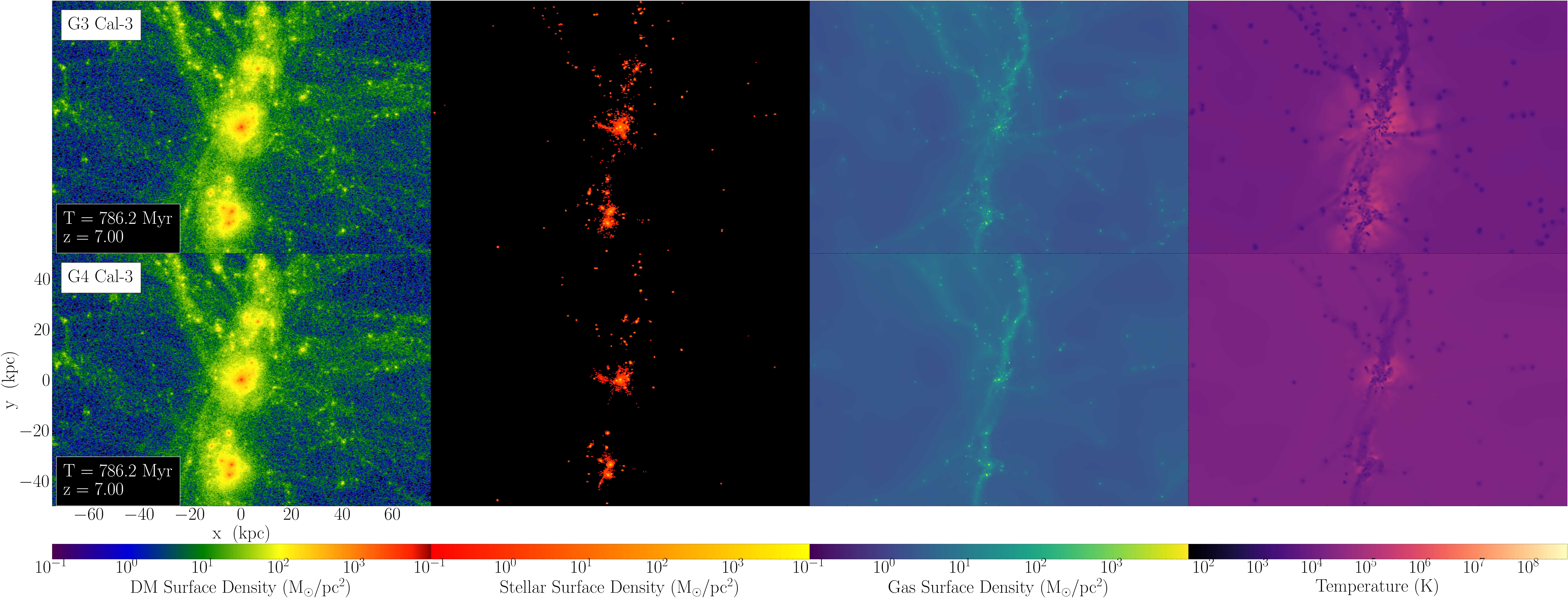}\end{adjustwidth}
        \caption{Projected quantities at $z=7$ for Cal-3 runs. The~top row shows \gthree, the~bottom row shows \gfour. From~left to right: dark matter surface density, stellar surface density, gas surface density, and~density-weighted temperature. Each panel is projected through a 150~kpc-thick~slab. }
        \label{fig:proj_3}
\end{figure}

\begin{figure}[H]\begin{adjustwidth}{-\extralength}{0cm}
\centering
    \subfloat[\centering\label{fig:surf_3}]{
    \includegraphics[width=0.23\linewidth]{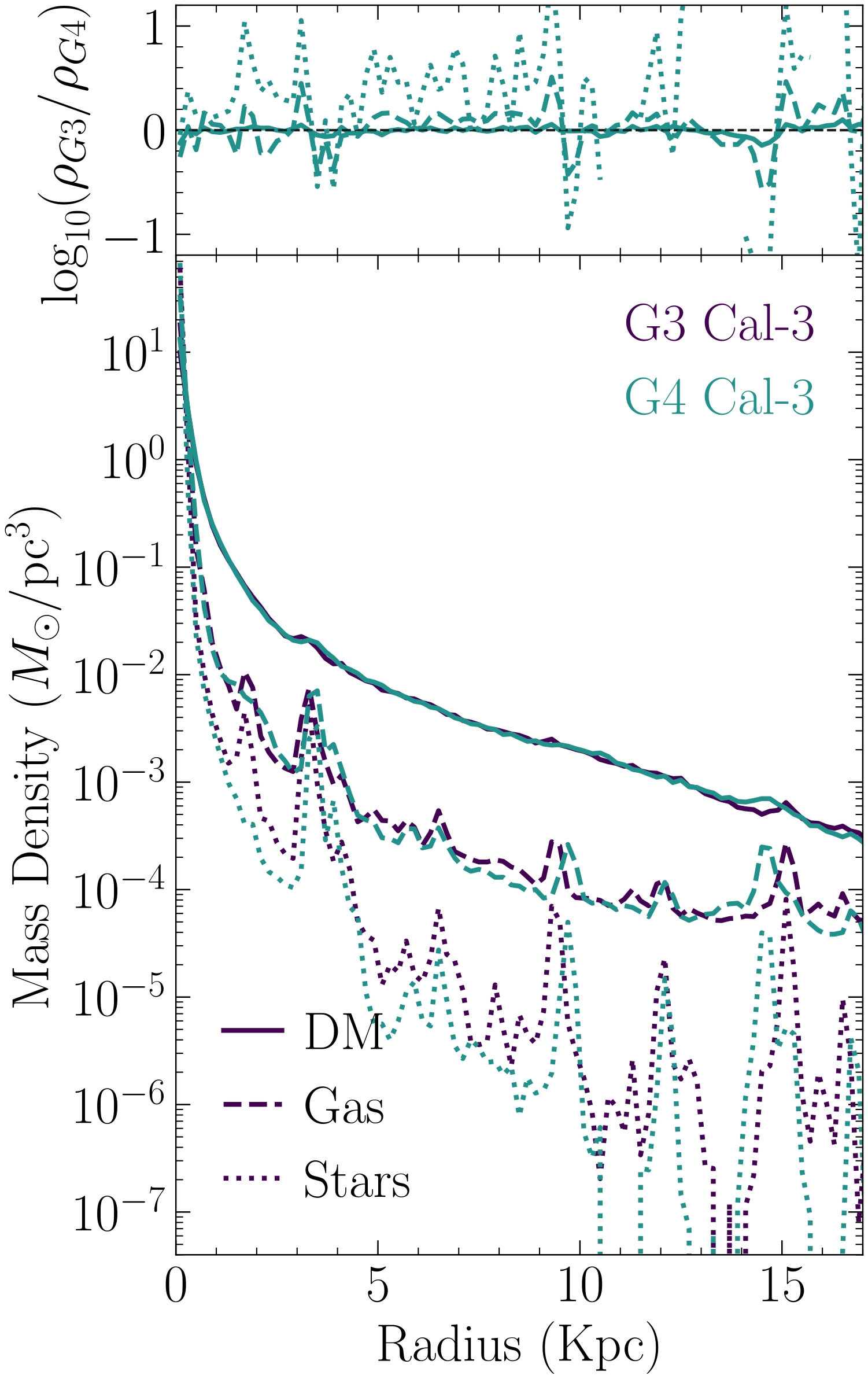}
    }~
    \subfloat[\centering\label{fig:surf_3gr}]{
    \includegraphics[width=0.21\linewidth]{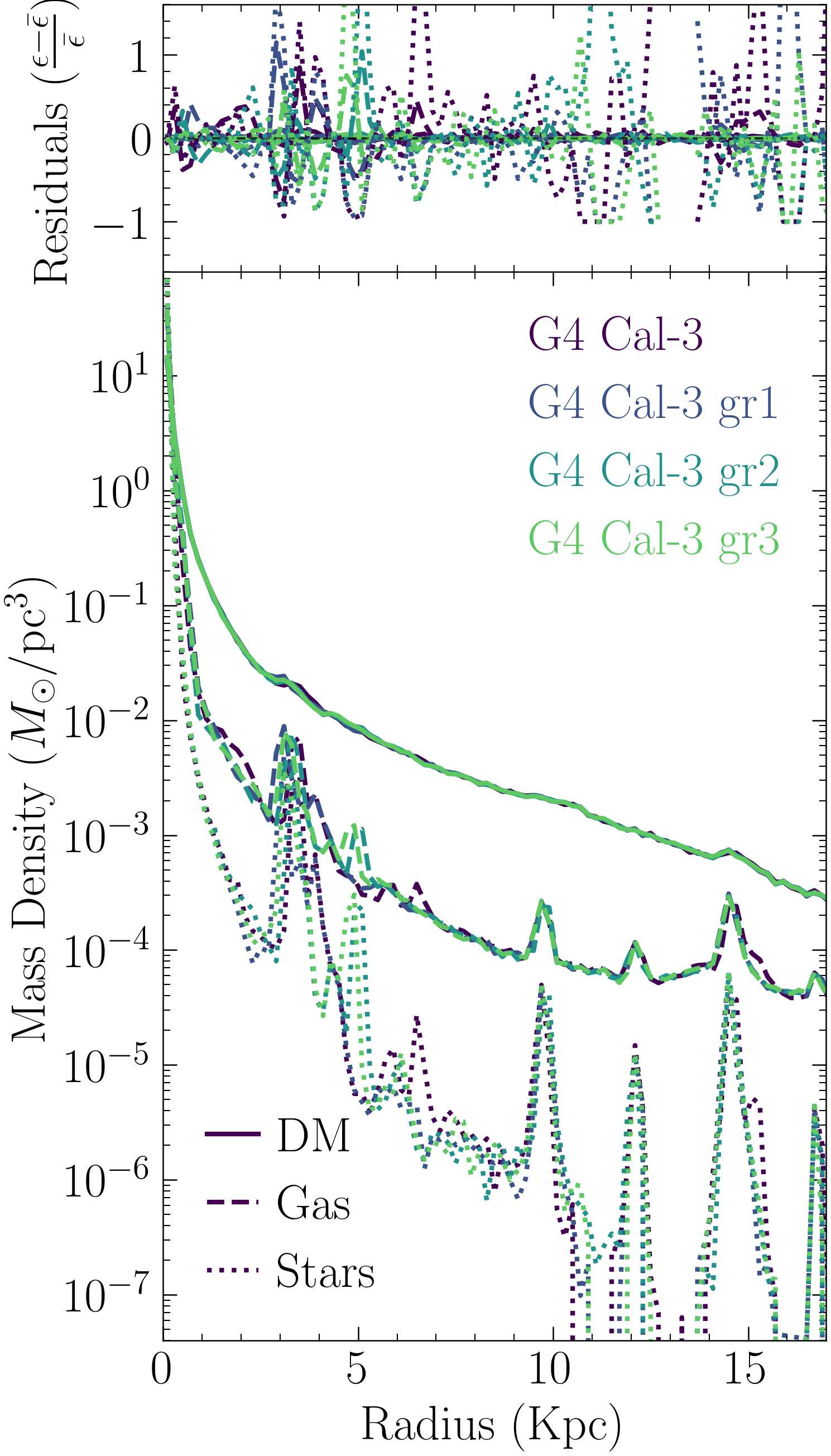}
    }~
    \subfloat[\centering\label{fig:sfr_3}]{
    \includegraphics[width=0.50\linewidth]{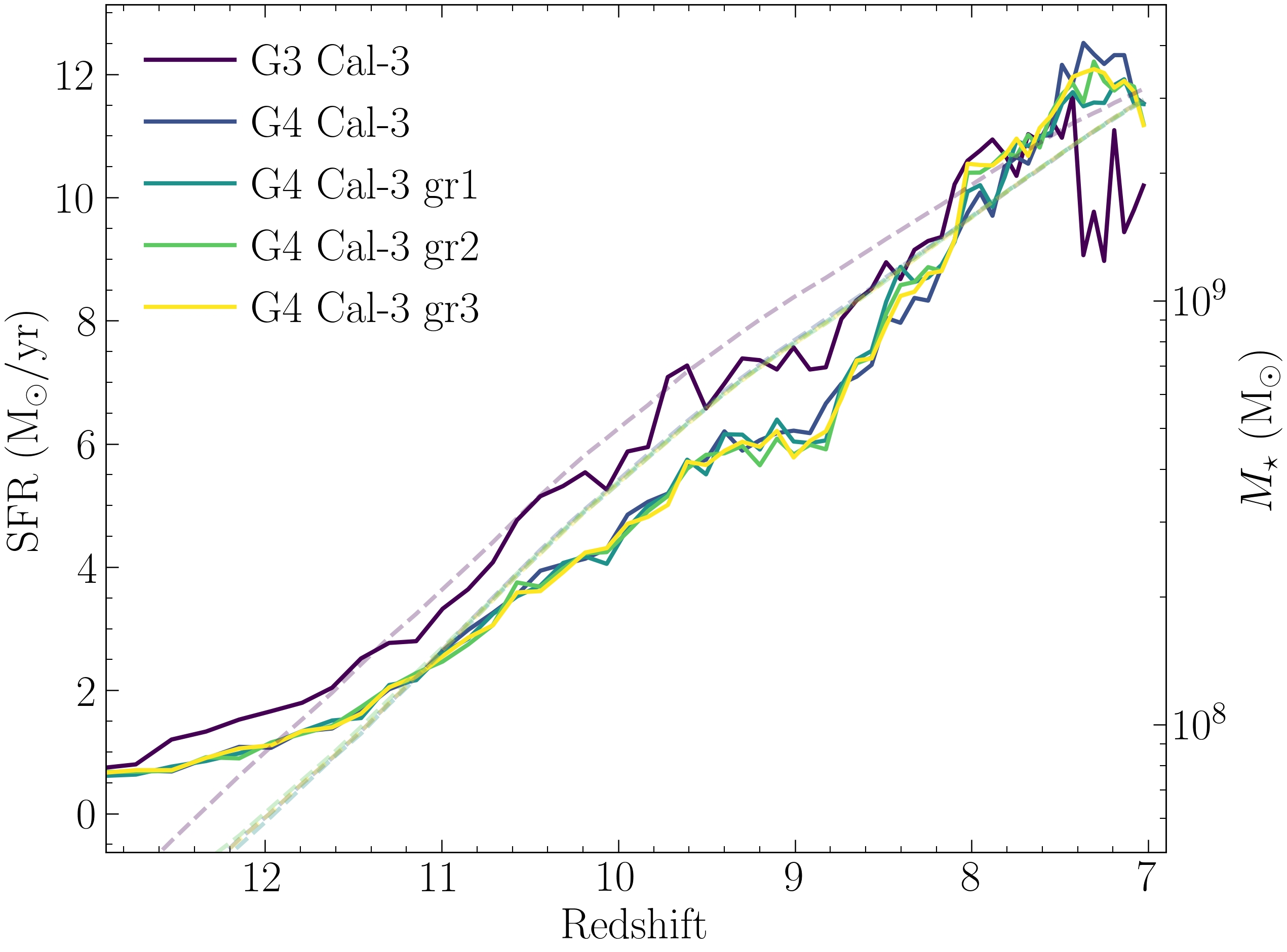}
    }
    \end{adjustwidth}
    \caption{(\textbf{a}) Spherically averaged density profiles for Cal-3 at $z=7$. Dark matter (solid) and baryonic components (dashed: gas, dotted: stars) are shown for \gthree and \gfour. Residual ratios $\log_{10}(\rho_{\rm G3}/\rho_{\rm G4})$ are displayed above each~panel. 
    (\textbf{b}) Same as (\textbf{a}) but for \gfour Cal-3 chemistry network variations (Grackle modes 0/1/2/3, color-coded, see Section~\ref{sec:commonphys} for an explanation of each mode). 
    (\textbf{c}) Star formation rate (solid lines, left axis) and integrated stellar mass (dashed lines, right axis) versus redshift for Cal-3 runs from $z=13$ to $z=7$.}\label{9}
\end{figure}

Figure~\ref{fig:proj_3} shows projections for dark matter, stars, gas, and~temperature at $z=7$. The~patterns from previous calibrations persist: dark matter distributions are visually identical, gas in \gfour appears slightly more diffuse and less clumpy than \gthree, and~the temperature background remains $\sim$$ 2\times$ higher in \gfour. The~new stellar component largely traces the distribution of cold dense gas; this is as~expected, since stars form directly from gas above the density threshold (\mbox{$n_{\rm H, thresh} > 1$ cm$^{-3}$}). The~lower clumpiness of \gfour now manifests in having fewer star-forming halos at $z=7$, but~this difference does not appear significant, as the total stellar mass formed remains consistent between the two codes ({Figure}~\ref{9}c, dashed lines).

Radial profiles ({Figure}~\ref{9}a) quantify these trends. Gas and dark matter profiles agree to $\sim$$ 0.4$ dex for all radii, confirming that star formation alone (without feedback) does not drastically alter the overall density structure. Stellar profiles show central ($r < 2$ kpc) convergence to $\sim$$ 0.2$ dex, but~diverge at larger radii where satellite positions differ. This is the result of the different gas clump positions observed in previous calibration~steps.

\vspace{-15pt}

{Figure}~\ref{9}b compares \gfour runs with different \textsc{GRACKLE} chemistry modes. Dark matter profiles are highly degenerate, with~scatter < 10\%, as~expected. Gas profiles show modest (<40\%) variations, whereas stellar profiles can go up to 160\%, primarily near clumps. We attribute this to the stochastic nature of our star formation algorithm. These differences in turn drive the slight variations in gas and density profiles (most present at radii with high stellar density).

In {Figure}~\ref{9}c, the~star formation rate and total stellar mass are shown. Different cooling schemes show very high agreement, likely due to the stochasticity inherent in star formation, as~observed in the radial profiles. The~\gthree run, while not as convergent, shows similar SFR rates and an almost equal stellar mass at $z = 7$. Further evolution of the \gfour Cal-3 runs shows that the SFR and total mass remain essentially equal up to, at~least, $z = 1$. SFR eventually peaks at $z$ $\sim$4.5, as~most of the cold gas is consumed without regulation by~feedback. 

\begin{figure}[H]\begin{adjustwidth}{-\extralength}{0cm}
        \centering
        \includegraphics[width=.99\linewidth]{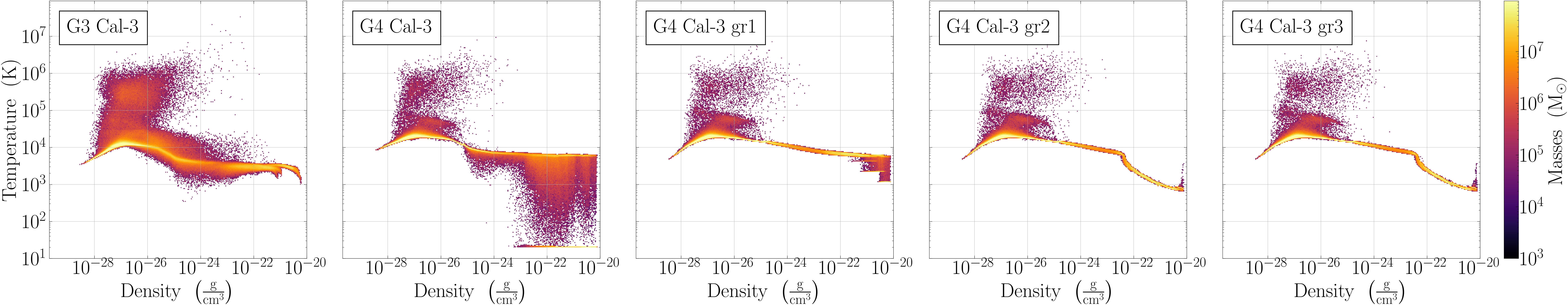}\end{adjustwidth}
        \caption{Gas density vs. temperature phase diagrams for Cal-3 runs at $z=7$, showing gas within 100 kpc of the halo center. Colors represent total gas mass in each 2D bin. The~leftmost panel shows \gthree (Grackle mode 0, tabulated equilibrium cooling), while the four \gfour panels show progressively more detailed chemistry networks: mode 0 (tabulated, fiducial, and~equivalent to \gthree) and modes 1, 2, and~3 (see Section~\ref{sec:commonphys} for an explanation of the chemistry in each).}
        \label{fig:phase_3}
\end{figure}

However, we observe several notable differences in the gas-phase diagrams (Figure~\ref{fig:phase_3}) across all our Cal-3 runs. In~the \gthree case, the~phase does not differ significantly from that observed in Cal-2 results, with~cooling tails and more shock-heated gas than \gfour. However, in~the latter code we see a new low-temperature, high-density structure. The~same structure has been observed in the \textsc{GIZMO} and \textsc{Changa} codes, as~in the previous Cal-2 artifacts in the same phase-diagram region. As~Section~5.3.1 of Paper III discusses, these features arise from a stochastic star-formation scheme and from the implementation of the pressure floor in each code. In~particular, we find that for \gfour this is due to the pressure floor being set by the energy in the SPH pressure--energy implementation. Changing to a pressure--entropy scheme and a pure pressure floor erases this low-density gas and recovers a \gthree-like phase diagram (but with no cooling tails, less shock-heating, and~slightly higher $T$ $\sim$$10^4$ K equilibrium). Another way to demonstrate that this phase-diagram feature is unphysical is to compare it with our cooling-variation runs, in~which the feature completely disappears and~the effect of $H_2$ cooling is observed. Since metallicity is very low in this case, contrary to \textit{NSFF}, $H_2$ cooling dominates over metal line cooling;~a slight shift in $\rho \gtrsim 10^{-22}$ g\,cm$^{-3}$ gas can be seen for modes 2 and 3, which include this~effect.

We also performed \textit{NSFF} cooling variations, but~the results were essentially equal across modes for the same reasons explained earlier. Similarly, performing the same experiment with \textit{SFF} or Cal-4 simulations presented no variations in density profiles, kinematics, or overall stellar mass ({merger} timings, SFR, and stellar streams were different, but~we believe these differences to be stochastic in nature; see merger timing appendix in Paper IV of the AGORA collaboration and the AURIGA variability study \citep{Pakmor_2025}). These results also mirror findings in FIRE-2 \citep{Fire2_2018}, which found that galaxy properties are broadly invariant of specific cooling schemes used. This is because the cooling times are much shorter than the typical dynamical times, especially for cooler gas, which presents the most differences in our tests. Therefore, we warn that the temperature distribution of dense gas is very sensitive to the cooling routine used. This can become especially important in low-metallicity regimes, where metal cooling is nearly absent and other channels~dominate.

\subsection{Feedback Model~Comparison}
\label{sec:fbresults}

Having established broad convergence across gravity, hydrodynamics, cooling, and~star formation, with remaining differences quantified in the previous sections, we now examine simulations that include stellar feedback. This produces the most dramatic divergence between \gfour and \gthree, as~the feedback implementations differ fundamentally (Section~\ref{sec:feedback}). To~systematically explore this divergence, we perform isolated disk simulations with progressively more sophisticated \gfour feedback schemes (Section~\ref{sec:fbvaresults}), revealing that momentum-driven mechanical feedback and stochastic thermal heating are the dominant regulators. We also examine feedback strength sensitivity (Section~\ref{sec:fbstresults}) to distinguish parameter effects from algorithmic~choices.

\subsubsection{{Isolated} Disk Galaxy: Fiducial Feedback (SFF)}

\begin{figure}[H]\begin{adjustwidth}{-\extralength}{0cm}
        \centering
        \includegraphics[width=.99\linewidth]{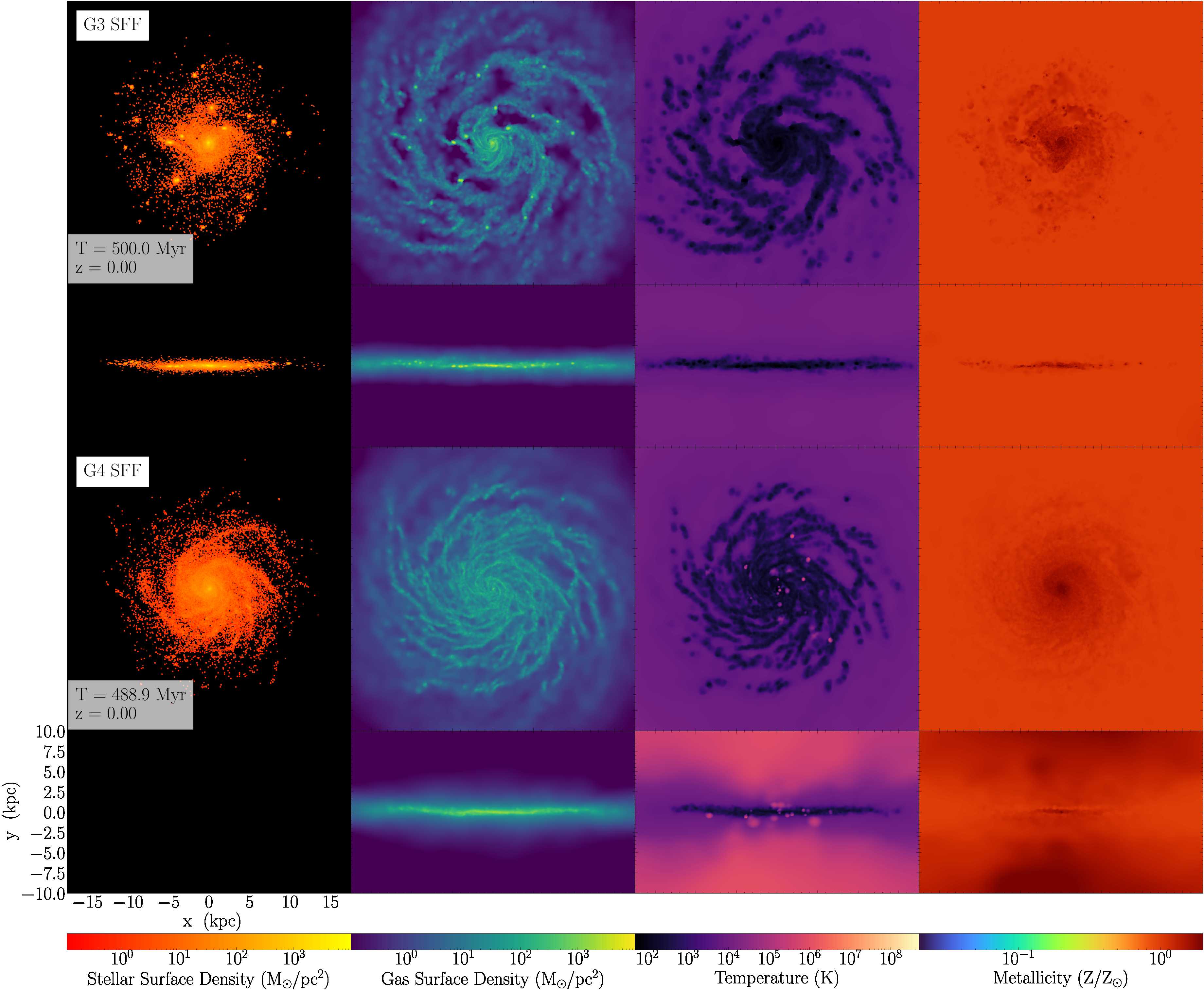}\end{adjustwidth}
        \caption{Face-on (rows 1 and 3) and edge-on (rows 2 and 4) projections at $t=500$ Myr for isolated \textit{SFF} runs. Columns show (left to right) stellar surface density, gas surface density, density-weighted temperature, and~\rev{gas} metallicity. \gthree is displayed in rows 1 and 2, \gfour in rows 3 and 4. Each panel is projected through a 35 kpc thick slab. See Section~\ref{sec:g4feedback}.}
        \label{fig:proj_fid}
\end{figure}

Figure~\ref{fig:proj_fid} shows face-on and edge-on projections of stellar density, gas density, temperature, and~\rev{gas} metallicity at $t=500$ Myr for the \textit{SFF} runs. The~morphological contrast is~stark.

\begin{figure}[H]\begin{adjustwidth}{-\extralength}{0cm}
 \centering
    \subfloat[\centering\label{fig:clumps}]{
    \includegraphics[width=0.48\linewidth]{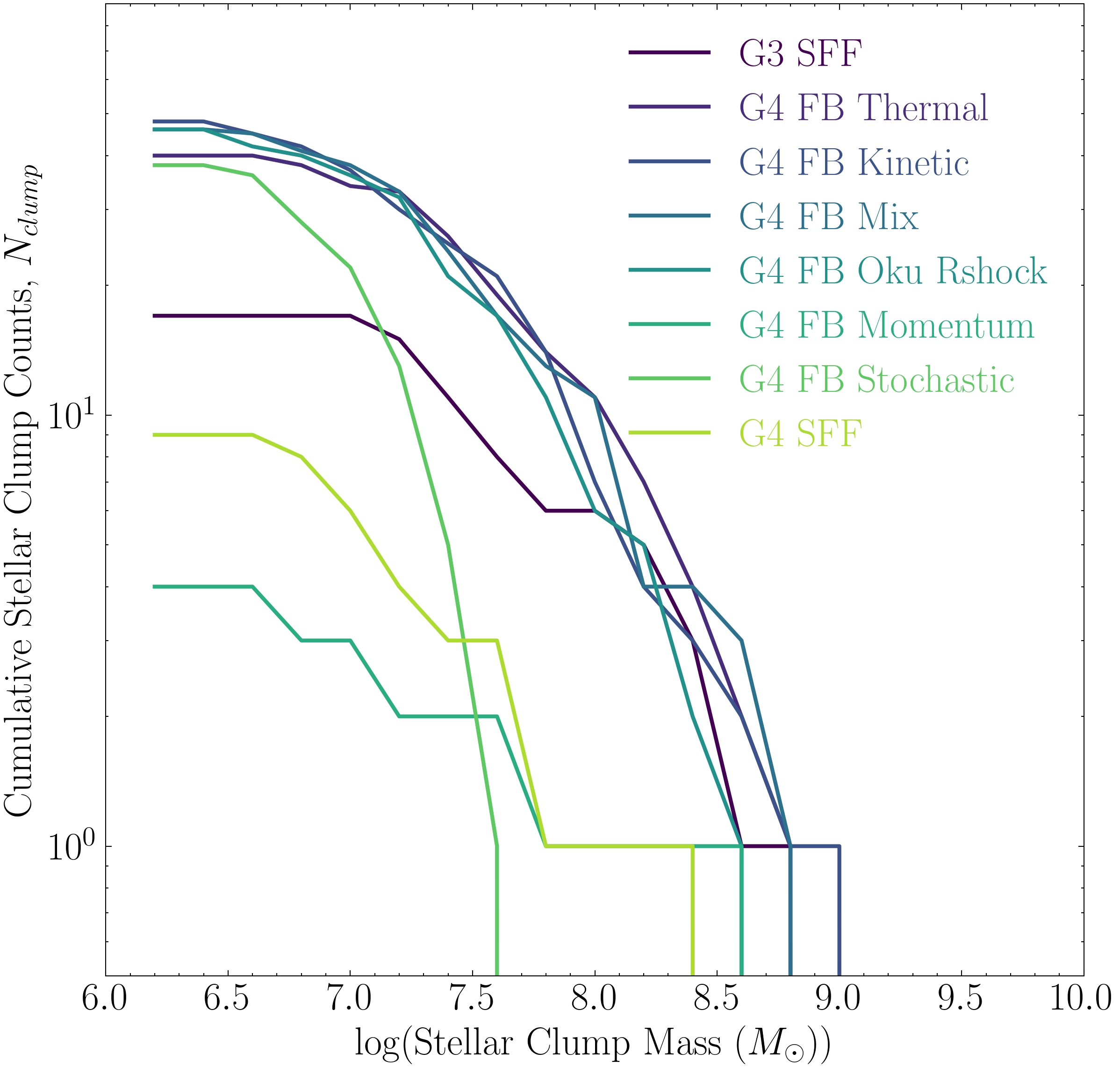}
    }~
    \subfloat[\centering\label{fig:KS}]{
    \includegraphics[width=0.49\linewidth]{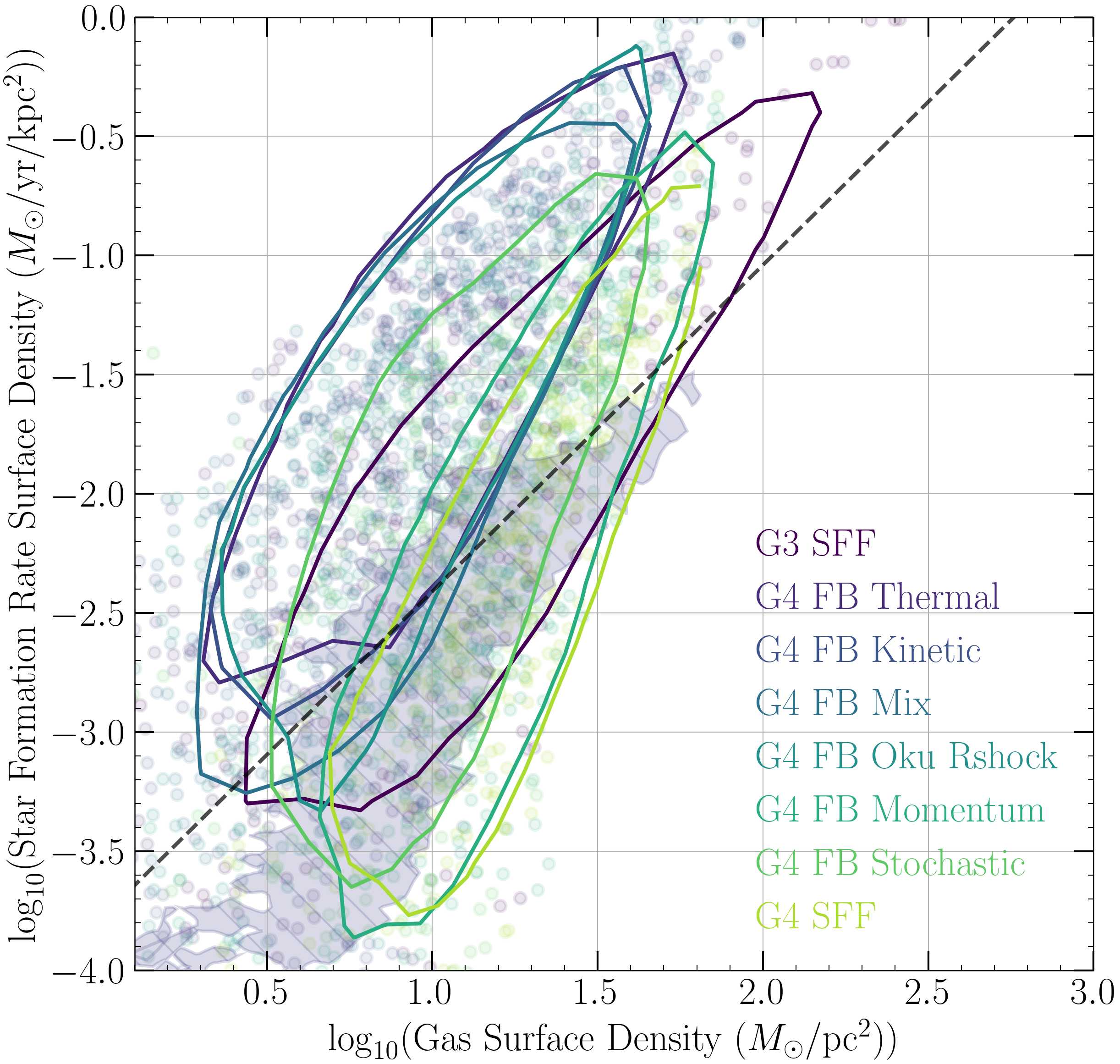}
    }
    \end{adjustwidth}
    \caption{(\textbf{a}) Cumulative clump mass histograms at $t=500$ Myr for feedback model variations. Clumps are identified using the HOP algorithm \citep{Eisenstein_1998}. The~exact clump number or mass can be sensitive to the halo finder (FoF also tested) and parameters used, but~the simulation order remains the same. The~most massive clump ($10^8$--$10^9 M_{\odot}$) corresponds to the central stellar~cusp. \textls[10]{(\textbf{b}) Spatially resolved Kennicutt--Schmidt relation at $t=500$ Myr for feedback model variations. Mock observations use 750 pc pixels (matching \citet{Kim_2016} AGORA Paper II and \citet{Bigiel_2008} observation resolutions) with $\Sigma_{\rm SFR}$ computed from stellar age $< 20$ Myr. Data points are colored by model (see legend), with~observational fits from \citet{Kn2007} (dashed black line) and \citet{Bigiel_2008} local galaxy observations (contour, multiplied by 1.36 to match total gas density) overlaid. Corresponding mock observation images are shown in Figure~\ref{fig:mock_obs}.}}\label{17}
\end{figure}

\begin{figure}[H]\begin{adjustwidth}{-\extralength}{0cm}
    \centering
    \includegraphics[width=1\linewidth]{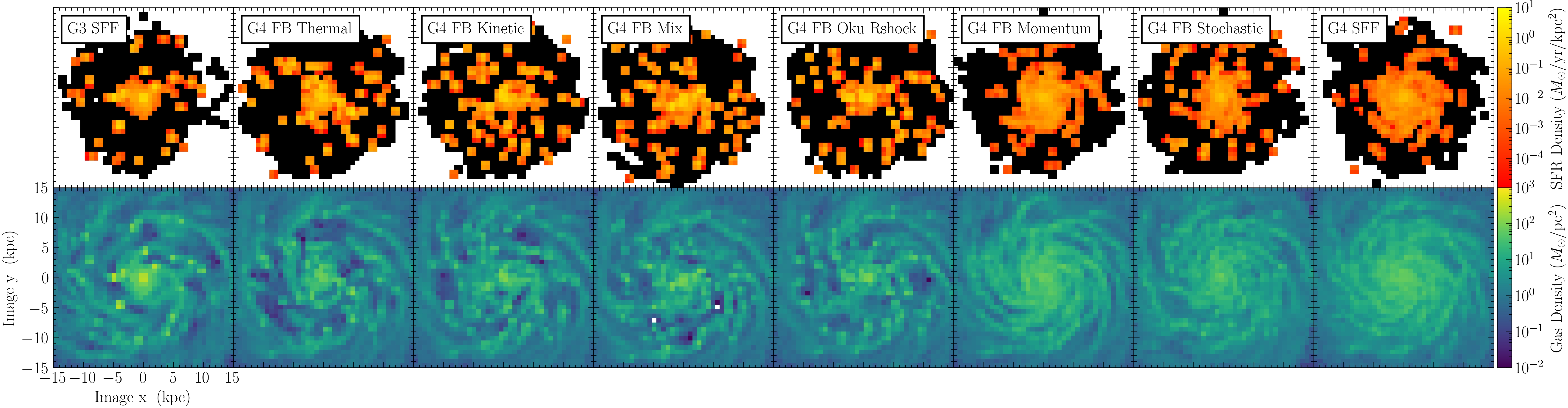}\end{adjustwidth}
    \caption{Mock observation images used to construct the Kennicutt--Schmidt relation in Figure~\ref{17}b. Each $30 \times 30$ kpc field is pixelated at 750 pc resolution. Top row shows star formation rate surface density $\Sigma_{\rm SFR}$ (stars with age $< 20$ Myr), while bottom row shows gas surface density $\Sigma_{\rm gas}$.}
    \label{fig:mock_obs}
\end{figure}

\gthree produces a highly fragmented disk containing 10--20 massive clumps with masses $M_{\rm clump}\sim10^7$--$10^9 M_\odot$ ({Figure}~\ref{17}a). These clumps are cold ($T$ $\sim$$10^4$ K), gravitationally bound, and~serve as the primary sites of ongoing star formation ({Figure}~\ref{fig:mock_obs}). \rev{We note that these structures are more massive and larger (>300 pc) than Giant Molecular Clouds (GMCs). Given our spatial resolution of $\sim$80 pc, we do not resolve the internal physics of GMCs; thus, this fragmentation represents a numerical artifact driven by unchecked cooling rather than physical substructure}. The~disk is geometrically thin, with~a fainter spiral structure arising from clumps rather than a filamentary structure. Critically, there are \textit{no temperature or metallicity outflows}: the edge-on temperature projection shows that the disk remains uniformly $T$ $\sim$$10^4$ K, with no hot ($T > 10^6$ K) halo and~\rev{gas} metallicity confined to the disk plane. This demonstrates that the purely thermal feedback in \gthree fails to (1) prevent runaway gravitational collapse into clumps, and \mbox{(2) drive} galactic-scale~outflows.

In~contrast, \gfour  produces a thick, smooth disk with well-defined spiral arms and essentially no gravitationally bound clumps ({Figure}~\ref{17}a). The~stellar distribution is more extended and has a clear spiral. Temperature projections reveal individual supernova remnants as $\sim$$ 100$ pc-scale hot bubbles ($T$ $\sim$$10^{6}$ K) scattered throughout the disk, a~direct signature of stochastic thermal feedback that heats localized regions around supernovae (Section~\ref{sec:fb_variations}). More importantly, prominent thermal outflows extend above and below the disk, with~similar temperatures. These outflows are metal-enriched, with~the metallicity projection showing \rev{$Z$ $\sim$$1.4$--$2 \, Z_\odot$} gas extending far beyond the stellar disk. It is apparent that the fiducial feedback scheme creates a hot, low-density, and~high-metallicity CGM outside the cold disk, in~stark contrast to \gthree results. This hot halo has actually been observed and linked to stellar feedback outflows in the Milky Way \citep{Nakashima2018}.

\begin{figure}[H]\begin{adjustwidth}{-\extralength}{0cm}
        \centering
        \includegraphics[width=.99\linewidth]{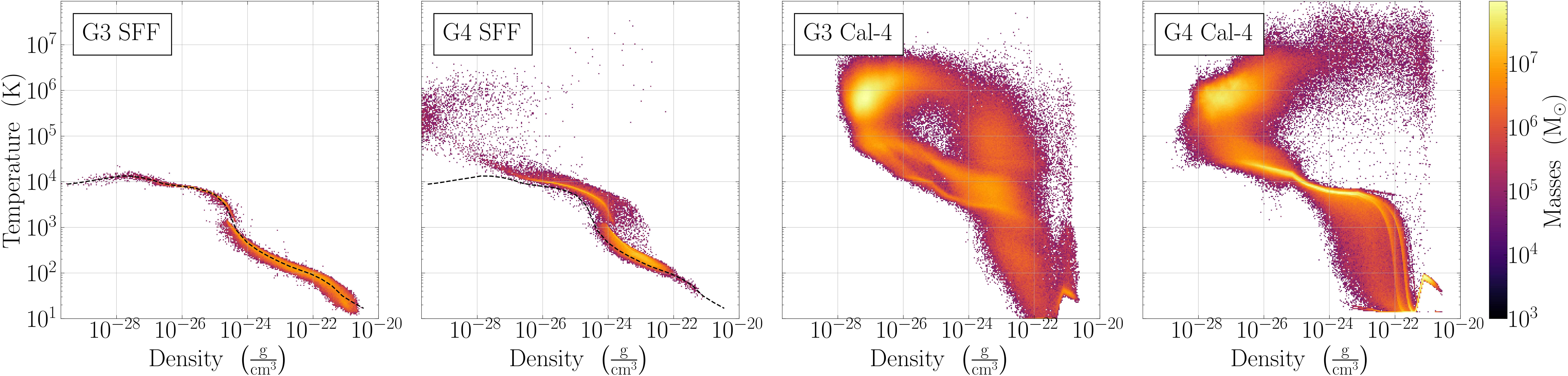}\end{adjustwidth}
        \caption{Gas density--temperature phase diagrams for fiducial feedback runs, showing gas within 30\,kpc (isolated) of the galaxy center. Colors represent the total gas mass in each 2D bin. From~left to right: \textit{G3 SFF} (isolated, $t=500$\,Myr), \textit{G4 SFF} (isolated, $t=500$\,Myr), \textit{G3 Cal-4} (cosmological, $z=1$), \textit{G4 Cal-4} (cosmological, $z=1$). To~guide the eye, we use a thick dashed line in the \textit{SFF} panel to plot the mean temperature in each density bin for \gthree.}
        \label{fig:phase_fid}
\end{figure}

The phase diagrams ({Figure}~\ref{fig:phase_fid}, panels 1 and 2), corroborate this finding. \gthree presents almost no differences with respect to \textit{NSFF} runs, with~just some broadening of the low-density equilibrium track in the temperature direction. Meanwhile, \gfour, in~addition to a similar broadening, contains $T > 10^{4}$\,K gas, which is what forms the hot halo in the projection maps. Additionally, \gfour shows a deficit of the densest gas ($n > 10^{-22}$ g cm$^{-3}$) compared to \gthree; the material that would have formed clumps in the thermal-only case has instead been disrupted by momentum-driven feedback and heated by stochastic thermal feedback, being redistributed into the diffuse~CGM.

\begin{figure}[H]\begin{adjustwidth}{-\extralength}{0cm}
 \centering
    \subfloat[\centering\label{fig:surf_fid}]{
    \includegraphics[width=0.22\linewidth]{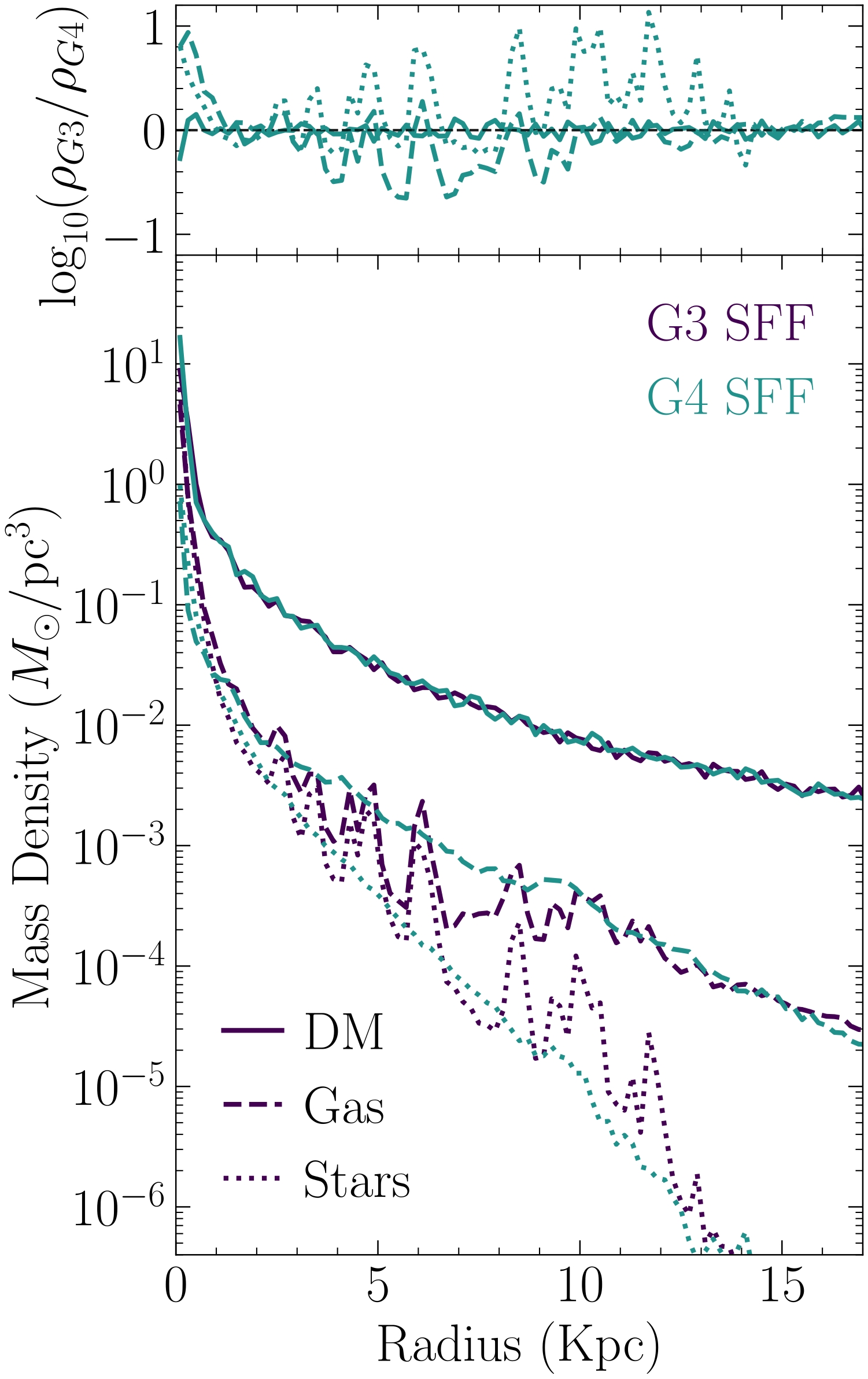}
    }~
    \subfloat[\centering\label{fig:surf_evol}]{
    \includegraphics[width=0.75\linewidth]{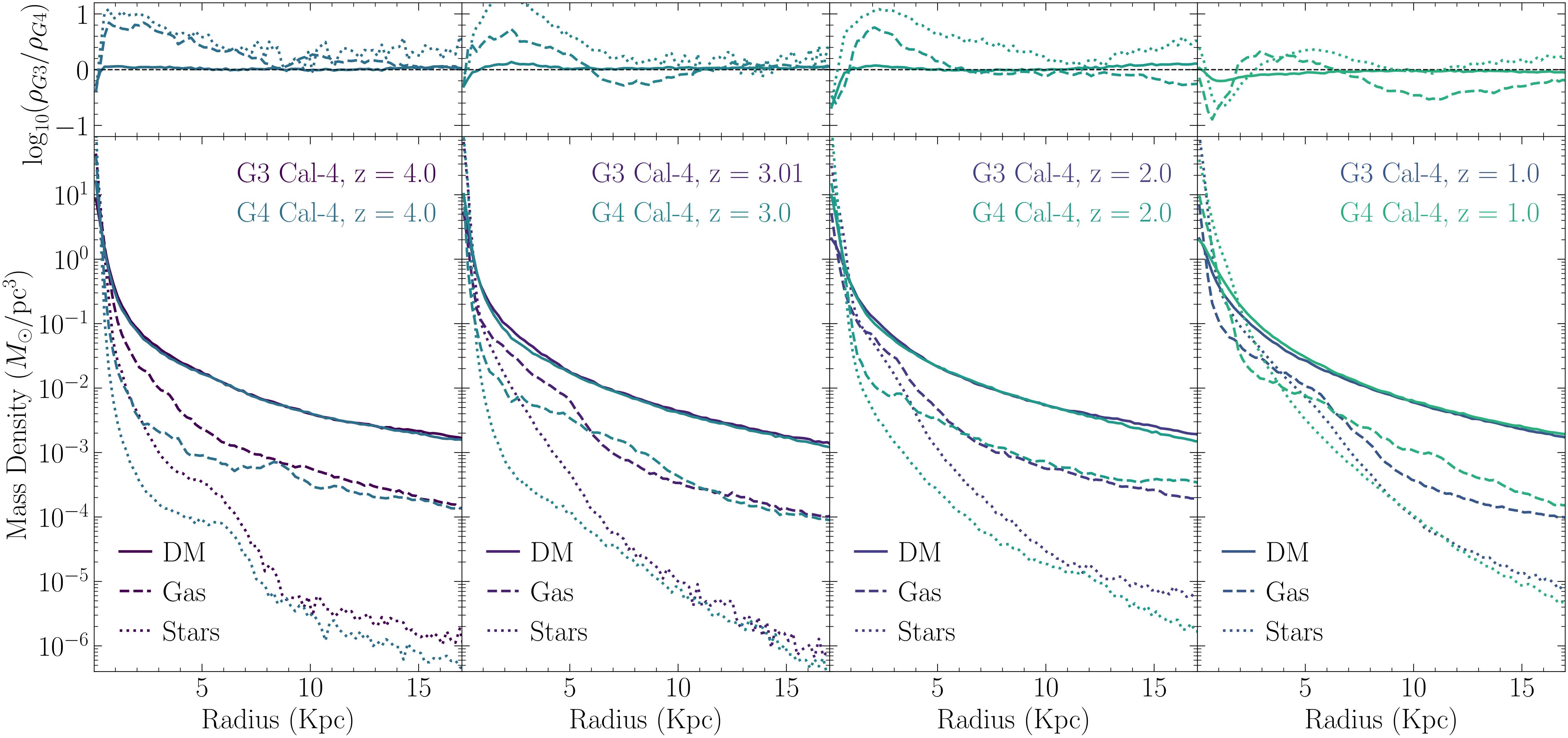}
    }
    \end{adjustwidth}
    \caption{Spherically averaged density profiles for fiducial runs. {(\textbf{a})} %
 Isolated \textit{SFF} at $t=500$ Myr. Dark matter (solid), gas (dashed), and~stars (dotted) for \gthree and \gfour. The~upper subplot in each panel displays the logarithmic density ratio $\log_{10}(\rho_{\rm G3}/\rho_{\rm G4})$ to highlight~deviations.     (\textbf{b}) Same as (\textbf{a}) but for cosmological Cal-4 simulations at $z=4,3,2,1$ (left \mbox{to right}).}\label{13}
\end{figure}

\begin{figure}[H]\begin{adjustwidth}{-\extralength}{0cm}
  \centering
    \subfloat[\centering\label{fig:surf_model}]{
    \includegraphics[width=0.48\linewidth]{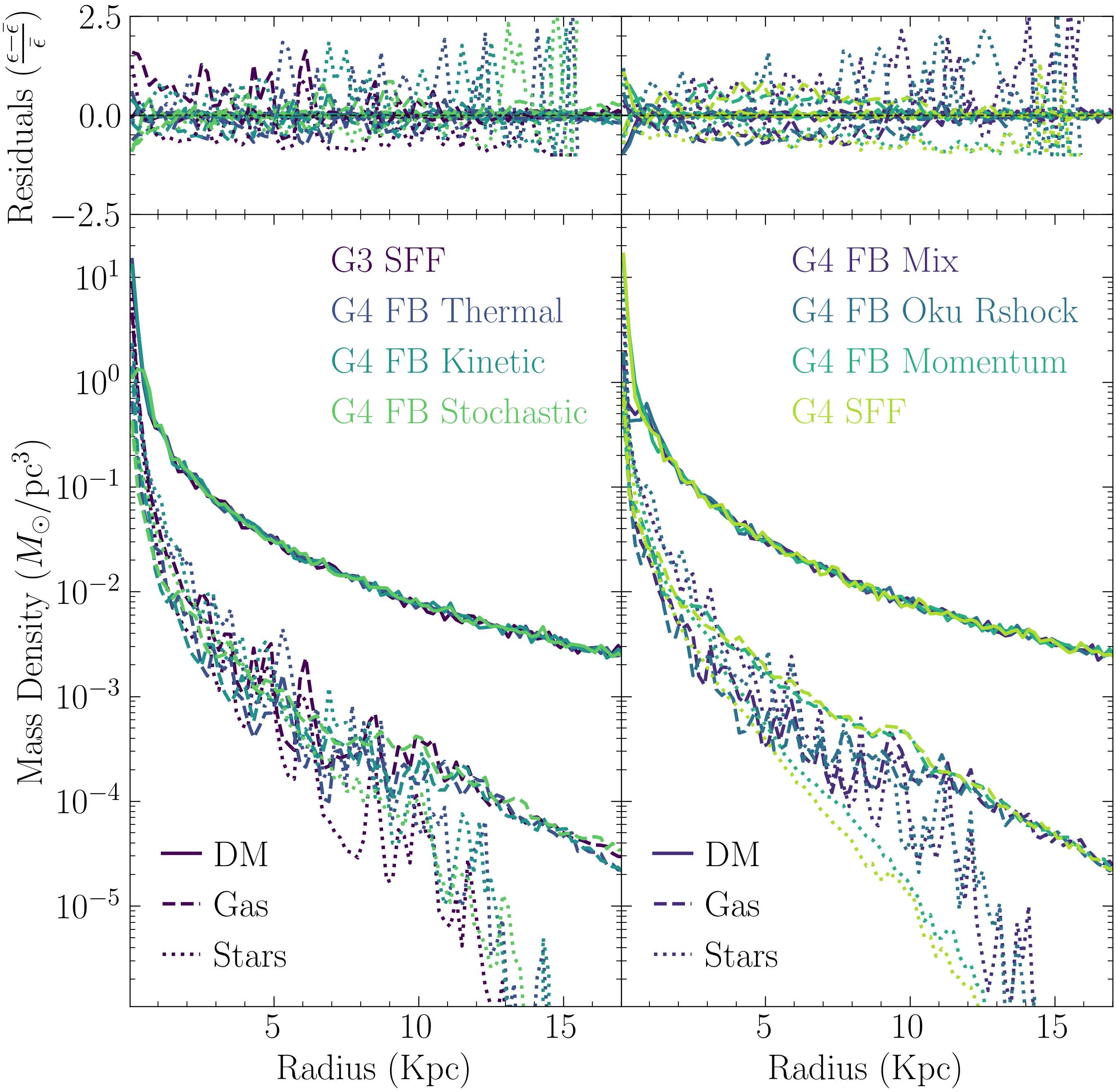}
    }~
    \subfloat[\centering\label{fig:sfr_model}]{
    \includegraphics[width=0.48\linewidth]{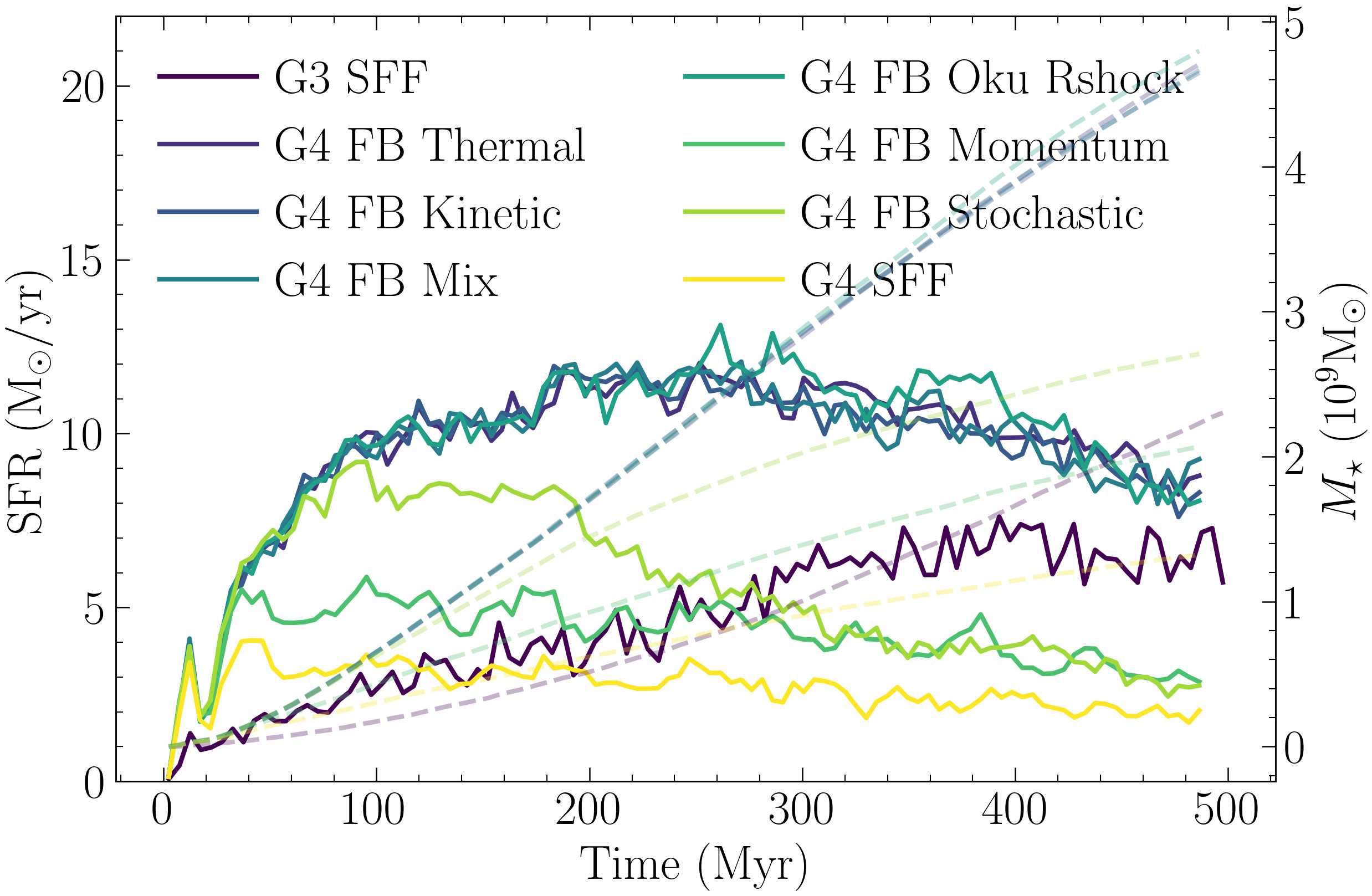}
    }
    \end{adjustwidth}
    \caption{(\textbf{a}) Spherically-averaged mass density profiles for feedback model variations at \mbox{$t=500$ Myr}. Dark matter (solid), gas (dashed), and~stars (dotted) are shown, with~residuals displayed in the top sub-panel to highlight differences. Models are color-coded by complexity (see legend). 
    \mbox{(\textbf{b}) Star} formation rate (solid, left axis) and stellar mass (dashed, right axis) versus time for feedback model variations. The~overall stellar mass correlates with clump counts (Figure~\ref{17}a).}\label{16}
\end{figure}

Moving on to the radial profiles in {Figure}~\ref{13}a (second panel), we can again see the presence of clumps in \gthree as the biggest difference between the two codes. Dark matter profiles now precisely converge (verifying that the disagreement in \textit{NSFF} profiles was a centering issue rather than physical). The~differences arise from profile fluctuations within each code, which are driven by the particularly low mass resolution of dark matter particles in the initial conditions. However, this agreement is not surprising, since these initial conditions already impose a particular density profile (NFW) at the start, so it does not arise naturally as in cosmological runs. Stellar and gas profiles exhibit the largest deviations: \gthree shows order-of-magnitude higher density spikes corresponding to individual clumps, while \gfour follows a smooth power law. Between~the clumps, \gthree's stellar density drops to \gfour's level. Even though the total stellar mass is actually $\sim$$ 90\%$ \textit{higher} in \gthree ($M_\star \sim 2.1 \times 10^9 M_\odot$ vs. $1.2 \times 10^9 M_\odot$ in \gfour, {Figure}~\ref{16}b), it is locked in clumps rather than smoothly distributed. Gas profiles anti-correlate with stellar profiles: \gthree exhibits voids between clumps where gas has been converted to stars, while \gfour maintains a smooth gas~disk.

\vspace{-6pt}

The \gthree simulation we analyzed here is quite old; the Osaka feedback model has since evolved substantially (as we also observe in the Cal-4 run, which was performed much later than this one). Previous studies have implemented new modules and examined their effects on the same AGORA-isolated galaxy; therefore, we will briefly discuss~them.

In \citet{Oku_2022}, the~effect of different feedback implementations in \gthree was studied. These modules were not used for the AGORA simulations; rather, they are the predecessors of the modules implemented in \gfour and used for our set of simulations. As~shown there, the~smoother and thicker disk that is free of clumping instabilities is a direct consequence of the terminal momentum injection module, whereas the outflows result from the stochastic feedback module. We will revisit and expand on these results with \gfour when we discuss our feedback model~variations. 

In another study, \citet{Shimizu_2019} compared different feedback models in \gthree in isolated galaxy simulations prior to the development of the momentum injection model. They obtained a smooth disk using pure thermal feedback and kinetic models. However, their model also includes Early Stellar Feedback (ESFB), radiation pressure from massive stars, AGB and SN-Ia feedback, and a~stochastic thermal model ({this} model is the same as \citet{Dalla_Vecchia_2012}, not the entropy-based one in \gfour and originally presented in \citet{Oku_2022}). In addition, it~turns off cooling during SN feedback for 1$\sim$3 Myr, which prevents gas from rapidly radiating away its energy (avoiding the classic overcooling catastrophe; see the introduction section in \citet{Oku_2022}). Among~these, stochastic heating is the primary reason they obtained a smooth disk without clumps. This discussion will be revisited in detail in Section~\ref{sec:fbvaresults} with a more complete~picture.

\subsubsection{{Cosmological} Zoom-Ins: Fiducial Feedback (Cal-4)}
\label{sec:cziff}

Now, we will explore the effects of feedback in a cosmological setting with Cal-4. In~this case, the~convergence requirement is much more relaxed, with~each run using their preferred feedback. The~only constraint is for the stellar mass of the main halo at $z = 4$ to be within $1$--$5 \times 10^{9} M_{\odot}$, with~the explicit goal of matching the mass given by semiempirical models for the selected halo ($M_{200} = 2 \times 10^{11} M_{\odot}$ at this redshift). In~practice, this means boosting the energy injected by each SN event to match this stellar mass. In~the original Paper III runs, most simulations include this boosting (sometimes for momentum or metallicity too) to achieve convergence. Here, for~\gfour, the~final run has a boost factor of $10$ for a final energy injection of $E_{\rm SN} = 10^{52}$ erg/SN, which reproduces a stellar mass at the upper end of the imposed range. We will briefly explore the effect of different choices of this parameter in Section~\ref{sec:fbstresults}. Additionally, for~a detailed analysis of the disk morphology and evolution in both codes, see AGORA Paper VIII \citep{jung2025}, where \gfour and \gthree results are explicitly compared in their~Appendix.

\begin{figure}[H]
        \centering
        \includegraphics[width=0.75\linewidth]{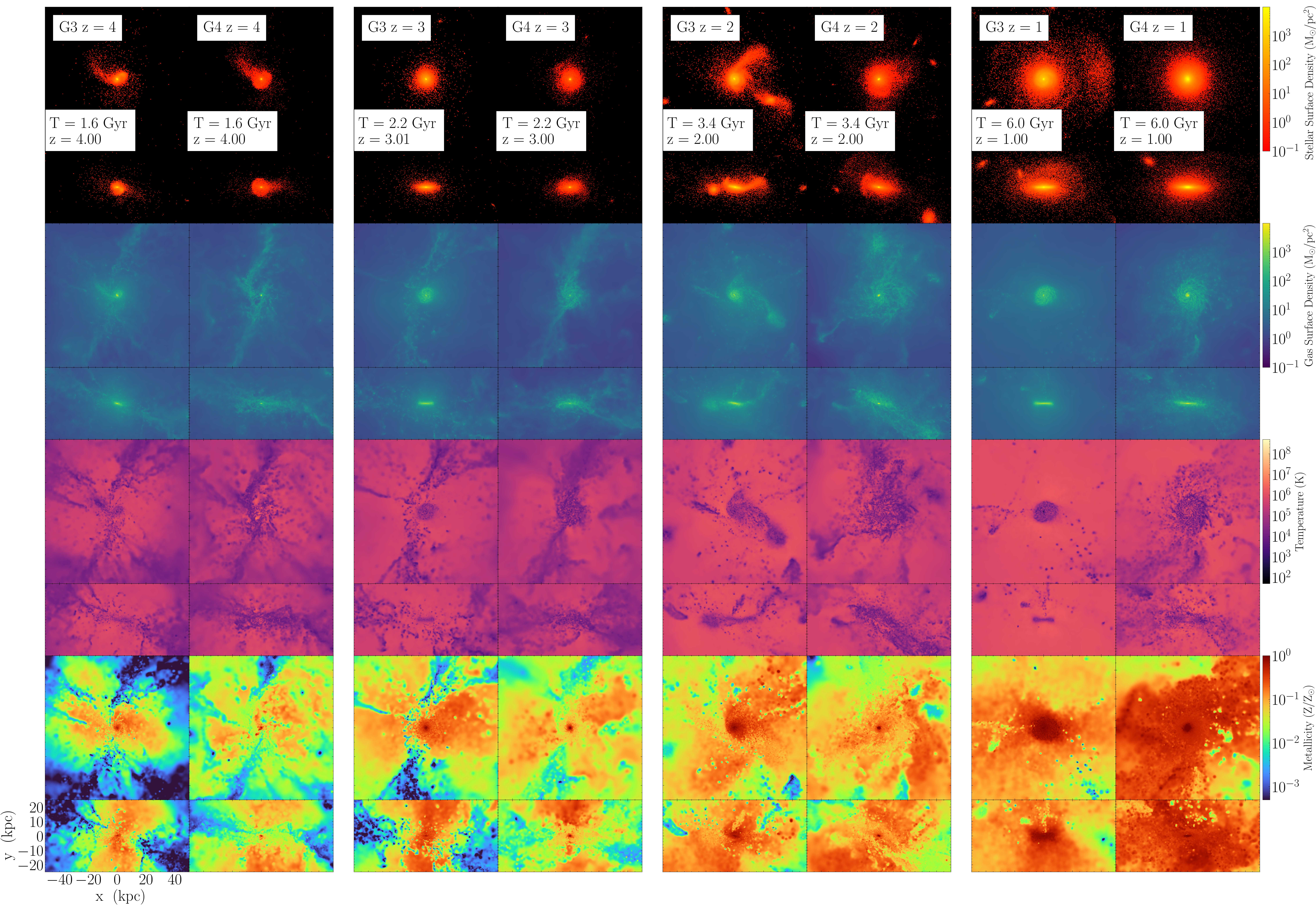}
        \caption{Face-on and edge-on projections at $z=4,3,2,1$ for cosmological Cal-4 runs. Fields are plotted in each row: Stellar density, gas density, temperature, and metallicity. Times are separated by white spaces, with~earliest $z=4$ in the left and most recent $z = 1$ in the right. \gthree is always plotted in the first columns and \gfour in the right. Projections use 100 kpc thick slabs. Both codes produce rotationally supported disks with comparable sizes ($r$ $\sim$$10$ kpc), but~\gthree exhibits stronger spiral arm contrast and a more compact bulge~component.}
        \label{fig:proj_cosmo}
\end{figure}

In Figure~\ref{fig:proj_cosmo}, we repeat the projections seen in the \textit{SFF} case, but now for the cosmological runs at $z = 4, 3, 2, 1$ ({we} stop at $z = 1$, since lower redshift data for \gthree results were not available at the time of analysis). Both codes produce rotationally supported disks with comparable sizes and morphologies, in~stark contrast to the isolated case. Why does \gthree perform better here?

\startlandscape
\finishlandscape

The key difference is \textit{cosmological context}. Unlike the isolated disk, which is an artificial equilibrium configuration, the~cosmological halo assembles hierarchically with continuous gas accretion along cold filaments. This accretion (1) replenishes gas depleted by star formation, preventing runaway fragmentation and~(2) introduces angular momentum coherently aligned with the disk, naturally forming a rotationally supported structure. Additionally, the~\gthree cosmological feedback includes ESFB and AGB (Section~\ref{sec:g3feedback}), which were absent in the isolated~run.

Nevertheless, important differences persist. \gthree exhibits stronger spiral arm contrast (higher density amplitude) and a more extended bulge component (\mbox{$r < 2$ kpc}), while \gfour produces an extremely compact and dense bulge (see {Figure}~\ref{13}b). Gas and temperature distributions are more turbulent and disturbed in \gfour near the main disk. While \gthree sits in a relatively unperturbed and homogenous CGM (particularly at $z=1$), \gfour displays more gas, temperature, and~metal outflows emerging from the disk. This is the cosmological analog of the isolated \textit{SFF} wind differences; stochastic heating in \gfour efficiently expels hot gas to large radii, while \gthree's thermal feedback mostly heats gas in~situ without driving large-scale outflows. At~larger ($r > 100$ kpc) scales (not shown here), the~IGM is heated efficiently by \gfour, while \gthree's remains at the $T$ $\sim$$10^{4}$ K equilibrium. It can also be observed that the cold streams feeding the galaxy are clumpy in \gthree but~diffuse in \gfour. This is something we have seen in previous calibrations due to the implementation of artificial conductivity and~cooling.

Metallicity maps corroborate this: both codes achieve $Z$$\sim$$Z_\odot$ in the central disk, with metal-enriched gas reaching $\sim$$ 30$ kpc; however, \gthree's metals are much more concentrated to the galactic disk (especially at $z = 1$), while \gfour manages to enrich the CGM further with its outflows. The caveat might be raised that the effective metal yields exposed in Section~\ref{sec:g4feedback} were higher for \gfour than other AGORA codes; however, tests with different yield tables presented very similar distributions. Although~the total metallicity in those may be slightly lower, their outflows still reach further than those of \gthree. A~detailed comparison to CGM observations is beyond our scope here, but was previously performed for the AGORA \textit{CosmoRun} suite in Paper VI \citep{Clayton2024}.

The phase diagrams (Figure~\ref{fig:phase_fid}, panels 3--4) highlight two key differences. First, \gthree exhibits a prominent warm--dense phase at $T$ $\sim$$10^{4-6}$ K, \mbox{$n > 10^{-24}$ g cm$^{-3}$} that is absent in \gfour. This arises from the delayed cooling prescription in \gthree, which artificially holds SN-heated gas at these temperatures for \mbox{$\sim$$ 1$--$5$ Myr} (\mbox{Section~\ref{sec:g3feedback}}). GEAR simulations, which also use delayed cooling, exhibit the same feature (Paper III, Figure~15). \gfour instead shows a hot--dense phase at \mbox{$T > 10^7$ K,} \mbox{$n$ $\sim$$10^{-24}$ g cm$^{-3}$}; these are the transient stochastic thermal bubbles (seen too in the previously discussed \textit{SFF} run), which exist only briefly before expanding and cooling, resulting in their much lower mass~fraction.

Second, \gfour develops cooling tails at low temperatures similar to those seen in \gthree's Cal-2/Cal-3 runs, another \textsc{GRACKLE} table interpolation artifact. Switching to non-equilibrium chemistry eliminates these tails (Section~\ref{sec:cal3coolvar}), confirming that they are numerical rather than~physical.

\begin{figure}[H]\begin{adjustwidth}{-\extralength}{0cm}
  \centering
    \subfloat[\centering\label{fig:sfr_str_iso}]{
    \includegraphics[width=0.52\linewidth]{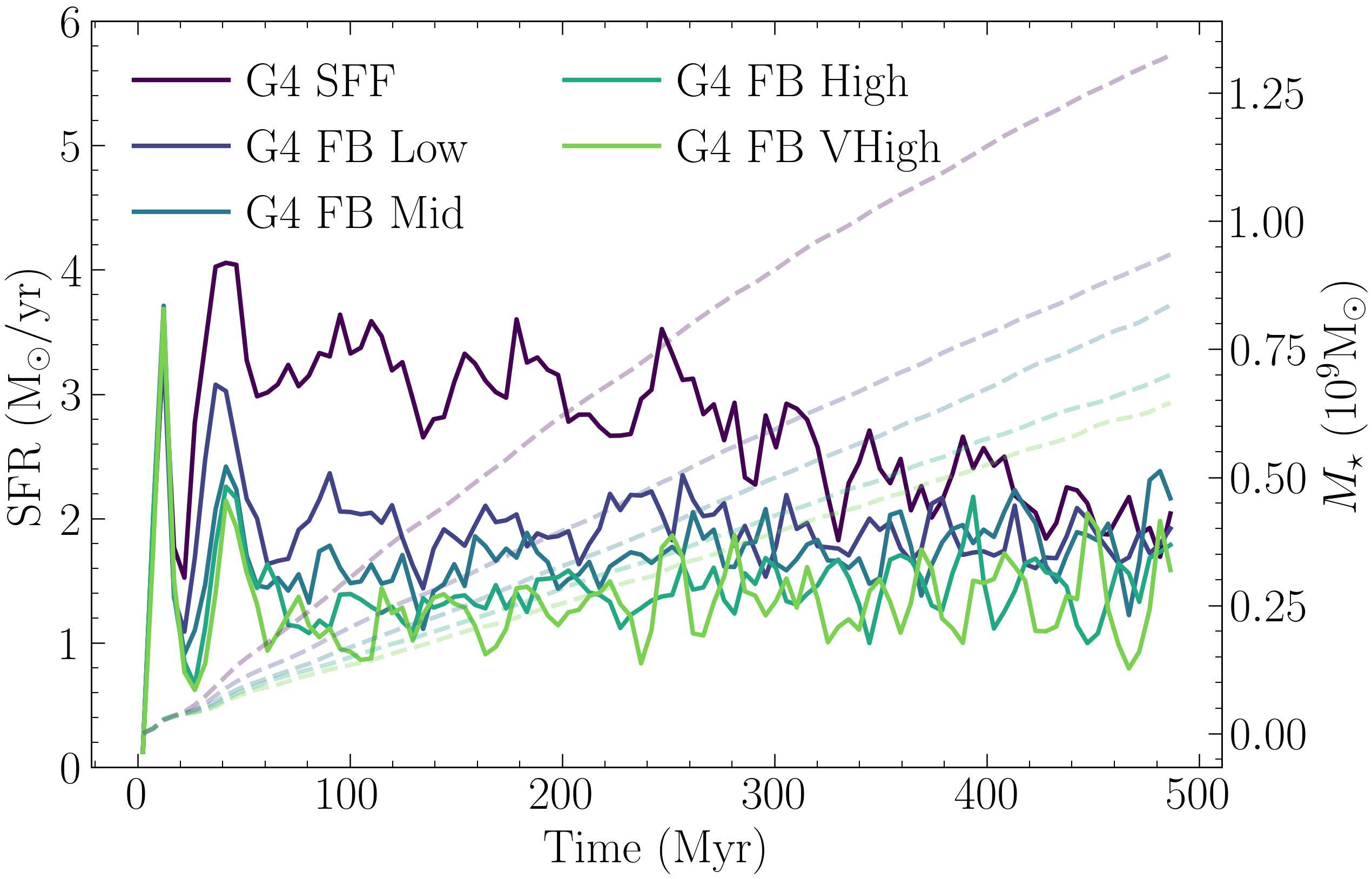}
    }~
    \subfloat[\centering\label{fig:sfr_str}]{
    \includegraphics[width=0.45\linewidth]{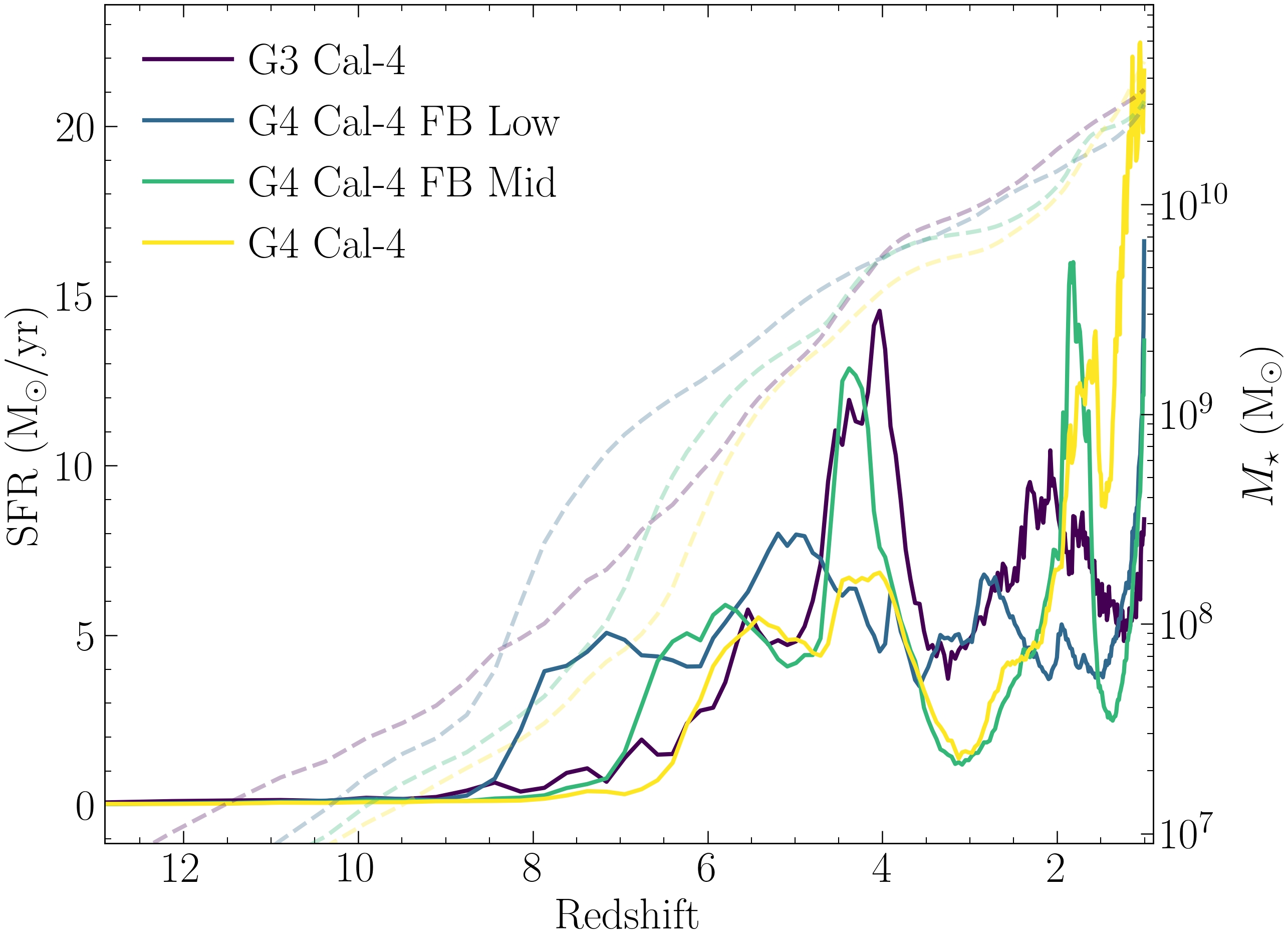}
    }
    \end{adjustwidth}
    \caption{(\textbf{a}) Star formation rate (solid, left axis) and stellar mass (dashed, right axis) with respect to time for \gfour \textit{SFF} feedback strength variations up to 500 Myr. Data plotted for simulations with $E_{\rm SN} = 1.0, 3.0, 5.5, 8.0, 10.0 \times 10^{51}$\,erg. 
    (\textbf{b}) Star formation rate (solid, left axis) and stellar mass (dashed, right axis) versus redshift for \gfour Cal-4 feedback strength variations. Merger events (stellar mass surges near $z \sim$4 and $z \sim$2) occur at different times \mbox{across models}.}\label{19}
\end{figure}

{Figure}~\ref{19}b shows the stellar mass growth history for Cal-4 fiducial runs together with the feedback strength variations. The~total stellar mass (plotted as dashed lines) shows that \gfour has a consistently lower stellar mass than \gthree, except~near $z=1$ where there is a small overtake. Major merger events are visible as SFR spikes near \mbox{$z \sim 4$} and $z \sim 2$. A~detailed analysis of these events will be presented in an upcoming paper of the AGORA collaboration, which will include the \gfour run presented~here.

\subsubsection{{Feedback} Model~Deconstruction}
\label{sec:fbvaresults}

To systematically disentangle which components of the \gfour feedback model drive the differences observed in Section~\ref{sec:fbresults}, we perform isolated disk simulations with progressively enabled features (Section~\ref{sec:fb_variations}): \textit{G4-FB-Thermal} (100\% thermal, closest to \gthree, \textit{G4-FB-Kinetic} (100\% kinetic), \textit{G4-FB-Mix} (50/50 thermal/kinetic split), \textit{G4-FB-Oku-Rshock} (improved shock radius), \textit{G4-FB-Momentum} (momentum injection), \textit{G4-FB-Stochastic} (stochastic heating), and \textit{G4-SFF} (full model with TIGRESS). This ladder of complexity allows us to isolate which physical processes are most important for disk regulation and outflow~generation.

\begin{figure}[H]
        \centering
        \includegraphics[width=0.7\linewidth]{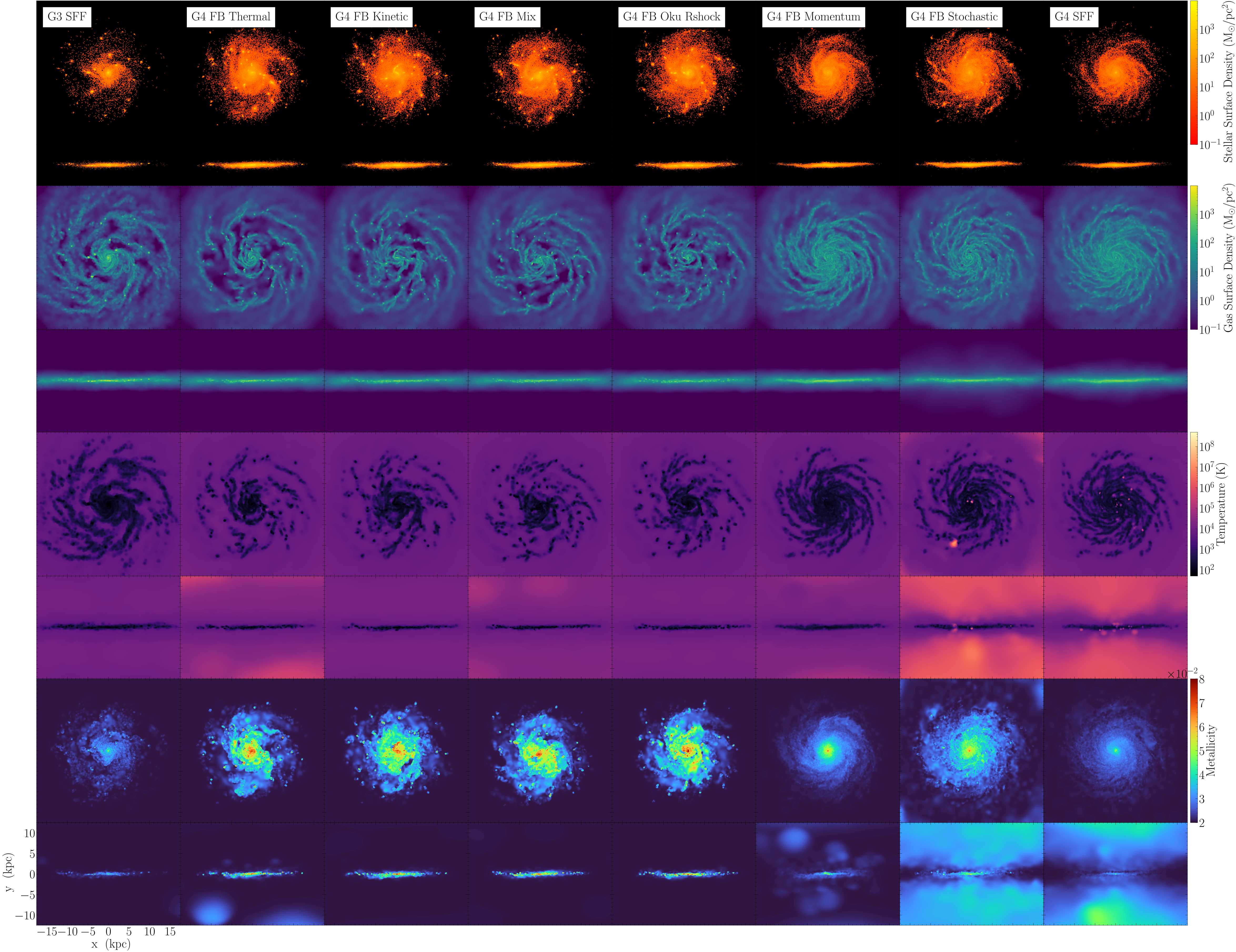}
        \caption{Face-on (rows 1, 3, 5, and 7) and edge-on (rows 2, 4, 6, and 8) projections at $t=500$ Myr for all isolated feedback variations, systematically increasing in model complexity from left to right: \textit{G3-SFF}, \textit{G4-FB-Thermal}, \textit{G4-FB-Kinetic}, \textit{G4-FB-Mix}, \textit{G4-FB-Oku-Rshock}, \textit{G4-FB-Momentum}, \textit{G4-FB-Stochastic}, \textit{G4-SFF}. Rows show (from top): stellar surface density, gas surface density, temperature, and~metallicity. All panels use 35 kpc thick~slabs.}
        \label{fig:proj_model}
\end{figure}

Figure~\ref{fig:proj_model} displays the full suite in a multipanel: face-on and edge-on projections of stellar density, gas density, temperature, and~metallicity for all eight models. The~progression from the old code to the present one is clear, as we describe~below.

The \textit{G4-FB-Thermal} model closely resembles \gthree, consisting of a highly clumpy disk with $\sim$$ 40$ massive clumps ({Figure}~\ref{17}a), thin scale height, no hot halo, and~metallicity confined to the disk. This confirms that pure thermal feedback (even with \gfour's modern SPH improvements) cannot prevent overcooling and clump formation. Large voids can be seen in the gas distribution due to the high thermal energy injection by SN in this case. This also causes some hot gas to be present outside the disk. No clear outflows are present, but a hot halo with~increased metallicity surrounds the~disk~at larger distances than shown here.

\startlandscape
\finishlandscape

\begin{figure}[H]\begin{adjustwidth}{-\extralength}{0cm}
        \centering
        \includegraphics[width=.99\linewidth]{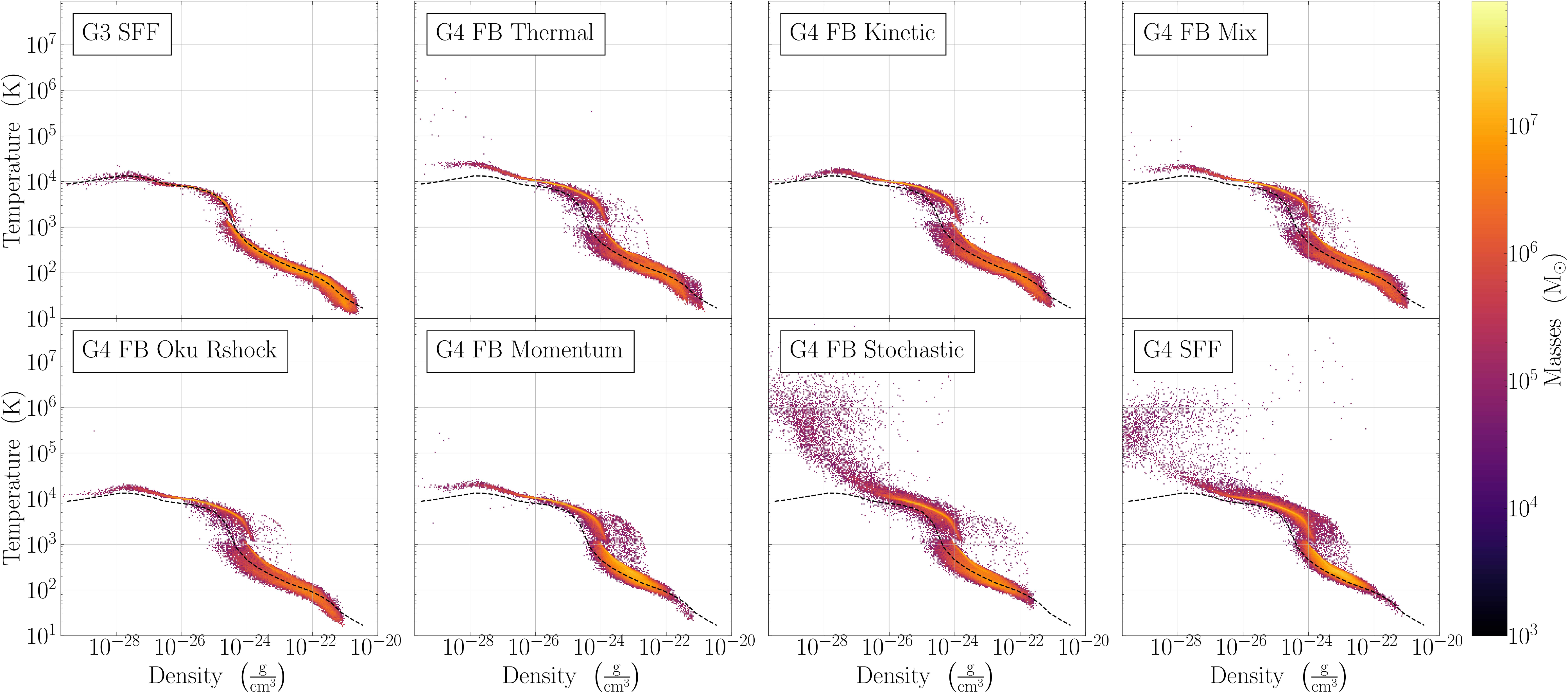}\end{adjustwidth}
        \caption{Gas density--temperature phase diagrams for feedback model variations at \mbox{$t=500$ Myr}, showing gas within 30~kpc of the disk center. Eight panels arranged in $4 \times 2$ grid display (left to right, top to bottom): \textit{G3-SFF}, \textit{G4-FB-Thermal}, \textit{G4-FB-Kinetic}, \textit{G4-FB-Mix}, \textit{G4-FB-Oku-Rshock}, \textit{G4-FB-Momentum}, \textit{G4-FB-Stochastic}, and \textit{G4-SFF}. Colors represent gas mass per~bin.}
        \label{fig:phase_model}
\end{figure}

The \textit{G4-FB-Kinetic}, \textit{G4-FB-Mix}, and \textit{G4-FB-Oku-Rschock} models present very similar morphologies to the first case. However, pure kinetic feedback cannot create a hot halo. This is also present in the gas phase diagrams (Figure~\ref{fig:phase_model}), where only pure thermal and mix models manage to heat gas more than a few $10^4$ K. Curiously, changing the shock radius to the Oku formulation, even with a $50/50$ split, makes heated gas more unlikely. We hypothesize that this is due to the radius being on~average smaller, and consequently affecting fewer particles. Additionally, the~four phase diagrams corresponding to these feedback formulations are very similar. They better resemble \gthree's, with~denser gas now present in the many clumps formed. If~we look at the total stellar mass in Figure~\ref{16}b, all four have the highest stellar mass and SFR compared to the remaining runs, which is a symptom of overcooling. Contrary to the \gthree case, we did not turn off cooling during SN feedback; therefore, gas can rapidly cool and continue star formation undeterred. This is also the reason for the SFR being higher than in the \gthree~run.

The \textit{G4-FB-Momentum} model marks the turning point. The clump count goes down to $\lesssim$5 (Figure~\ref{17}a), stellar mass drops by $\sim$$ 50\%$ to $M_\star$ $\sim$$2 \times 10^9 M_\odot$, and~we start to see metal outflows in the metallicity projections. The~disk is colder and more filamentary, but~still thinner than in the fiducial run. Metals are also distributed evenly across the disk, even without explicit diffusion. The~phase diagram (Figure~\ref{fig:phase_model}, sixth panel) now completely lacks the highest-density gas, instead concentrating more in the less dense but still cold area in phase space, which forms the filamentary structure of the disk. Momentum-driven feedback, even with a limited model only considering SNII remnants, is the key factor for eliminating clumps and distributing metals~homogenously.

The \textit{G4-FB-Stochastic} model (only stochastic thermal feedback, no TIGRESS, no momentum injection) produces the missing elements to reach our fiducial model. While the clumps are not as reduced as with momentum injection (see Figure~\ref{17}a), there is a clear lack of voids, a~much thicker disk, and temperature and metallicity bipolar outflows emerging from it. Phase diagrams show the low-density hot gas indicative of the diffuse CGM we saw in the fiducial \textit{SFF} runs. The~cold dense gas from other runs is also avoided; the material that would have formed denser clumps is instead transferred to the CGM. \rev{This run also presents an intermediate SFR and stellar mass, sitting in between the momentum and various velocity-kick models in Figure~\ref{16}b.}

The \textit{G4-SFF} model (full model: momentum + stochastic + TIGRESS) sits in between the last two cases; outflows are present, but~not as hot and with a lesser mass fraction, and~the disk is thicker than in the pure momentum-driven model, but~not as much as our stochastic~one.

Radial profiles (Figure~\ref{16}a) quantify this progression. Stellar profiles transition from highly peaked (clumps dominate in \textit{Thermal/Kinetic/Mix/Oku Rshock}) to smooth in \textit{Momentum/Stochastic/SFF}, although~with bigger deviations in the stochastic run. Gas profiles show the inverse: clumpy models exhibit voids between clumps, while regulated models maintain smooth profiles. Dark matter profiles converge across all models, with~central deviations present in some due to misidentification of the center, as we saw in the \mbox{\textit{NSFF} run}.

Star formation histories (Figure~\ref{16}b) exhibit some correlation with clump counts: the higher the clump count goes, the~higher the final stellar mass will be. The~only exception is the pure momentum scheme, which shows exceptionally low clump counts. However, we know from \gthree's run that low SFR and a high number of clumps can coexist if cooling is turned off after each SN~event.

Figure~\ref{17}b shows the Kennicutt--Schmidt (KS) relation for all feedback models, constructed from mock observations at 750 pc resolution (Figure~\ref{fig:mock_obs}) matching \citet{Bigiel_2008}. We overplot the \citet{Kn2007} best-fit slope ($\Sigma_{\rm SFR} \propto \Sigma_{\rm gas}^{1.37}$). The~star formation law sets the slope of the KS relation for our simulations (see Section~\ref{sec:commonphys}), namely, $\Sigma_{\rm SFR} \propto \Sigma_{\rm gas}^{1.5}$, giving a slope of 1.5 for all~runs.

The same models that overproduced stars (\textit{Thermal/Kinetic/Mix/Oku-Rshock}) scatter toward higher $\Sigma_{\rm SFR}$ at fixed $\Sigma_{\rm gas}$, with~normalization $\sim$$ 0.5$--$1.5$ dex above observations. This reflects unregulated clump collapse: once gas exceeds the density threshold (\mbox{$n_{\rm H} > 10$ cm$^{-3}$}), it rapidly converts to stars with efficiency $\epsilon_\star = 0.01$ per free-fall time, unchecked by feedback. Momentum-driven and stochastic models (\textit{Momentum/Stochastic/SFF}) are more consistent with the \citet{Kn2007} fit. This is particularly true for the fiducial and pure-momentum models, which exhibit substantial overlap. However, the~stochastic feedback lies between less complex models and the fiducial model, which we attribute to its higher stellar~mass.

The mock observation images (Figure~\ref{fig:mock_obs}) visually demonstrate this. \textit{Thermal/Kinetic/}\linebreak\emph{Mix/Oku-Rshock} models show patchy, clumpy $\Sigma_{\rm SFR}$ distributions with peak\linebreak \mbox{$\Sigma_{\rm SFR}$ $\sim$$10 M_\odot$ yr$^{-1}$ kpc$^{-2}$}, while \textit{Momentum/Stochastic/Fiducial} models display smooth, centrally concentrated $\Sigma_{\rm SFR}$ with peak $\Sigma_{\rm SFR}$ $\sim$$10^{-1} M_\odot$ yr$^{-1}$ kpc$^{-2}$.

Our feedback deconstruction reveals a clear hierarchy: \textit{momentum injection is necessary to prevent clumping and regulate star formation}, while \textit{stochastic heating sustains hot enriched outflows}. Energy partition alone (thermal vs. kinetic) is insufficient. The~TIGRESS wind model can provide quantitative refinement, but is not qualitatively essential: the \mbox{momentum + stochastic} combination captures most of the regulatory~effect.

Similar conclusions have been previously observed in the literature. For~example, \citet{Smith_2018} showed that outflows are resolution dependent, only appearing at very high resolutions ($\sim$20 $M_{\odot}$) in simplified feedback models, with~the exception of their momentum-driven scheme. Their KS relation results and finding of overcooling for simple thermal/kinetic/mix models are also similar to our findings. In~\citet{Hopkins_2012_gal}, it was similarly found that SN heating and momentum injection is very important for generating galactic winds. In~\citet{Agertz13}, a momentum-injection scheme was introduced and compared with previous methods such as delayed cooling,~finding that it could better regulate~SFR.

Overall, momentum-based schemes have been shown to be one of the leading sub-grid techniques for stellar feedback. Coupling it with stochastic thermal heating, as~in our fiducial model, appears to yield the most robust~results.

\subsubsection{{Feedback} Strength~Variations}
\label{sec:fbstresults}

Having isolated the algorithmic components essential for feedback regulation (\mbox{Section~\ref{sec:fbvaresults}}), we now examine the degeneracy between feedback \textit{strength} (parameterized by $E_{\rm SN}$) and feedback \textit{algorithm}. Specifically, we want to know how sensitive \gfour's results are to the choice of $E_{\rm SN}$.

Figure~\ref{19}a shows isolated disk star formation histories for \gfour with $E_{\rm SN} = 1.0, 3.0, 5.5, 8.0, 10.0 \times 10^{51}$ erg. The~relationship is monotonic and intuitive: the higher $E_{\rm SN}$, the~lower the final stellar mass. At~$t = 500$ Myr, $M_\star$ spans $0.65$, $0.7$, $0.8$, $0.9$, to~$1.3 \times 10^9 M_\odot$ (fiducial, lowest energy) for the five energies, a~factor of $\sim$$ 2$ dynamic range from a factor of $10$ energy variation. Diminishing returns seem to set in beyond $E_{\rm SN}$ $\sim$$3 \times 10^{51}$ erg. These runs also show a progressive increase in the mass of the hot halo around the galaxy and a decrease in disk metallicity, which in turn increases the CGM metallicity. However, the~KS relation is not affected and remains very similar across~runs.

The cosmological case (Figure~\ref{19}b) is more complex. At~high redshift ($z > 4$), the~same monotonic relationship holds: $E_{\rm SN} = 5.5 \times 10^{51}$ erg produces the highest SFR and stellar mass, while $E_{\rm SN} = 10.0 \times 10^{51}$ erg (fiducial) produces the lowest. However, the~ordering disappears afterwards. What breaks the monotonicity?

We hypothesize that this is due to \textit{gas recycling} and \textit{merger timing stochasticity}. In~the isolated disk, gas driven out by SNe escapes the galaxy and does not return during the duration of our simulation. In~the cosmological halo, gas launched into the CGM can cool and be re-accreted on longer timescales. Because we run cosmological simulations for \mbox{$\sim$$ 13$ Gyr}, multiple recycling episodes may occur. Stronger feedback initially suppresses star formation (as in the isolated case), but also enriches the CGM with metals. When this metal-rich gas is re-accreted, it cools efficiently and fuels a second burst of star formation at lower redshift. This ``delayed star formation'' effect partially compensates for the initial suppression, reducing the overall dynamic range in the final $M_\star$. 

In any case, the~SFR can be different even for simulations with the same parameters due to stochastic heating, which leads to our second~point.

Major mergers at $z$ $\sim$$4$ and $z$ $\sim$$2$ (visible as SFR spikes in Figure~\ref{19}b) occur at slightly different times across runs (see discussion in Section~\ref{sec:cal3coolvar}), with~satellite orbits differing more as the simulation proceeds due to chaotic effects. Similar stochastic variations were studied in AURIGA simulations \citep{Pakmor_2025}, where identical runs exhibited $8\%$ scatter in $M_\star(z=0)$ and $40\%$ variation in the stellar mass outside the~disk.

These effects are absent in the isolated disk, which has no CGM recycling (gas is lost forever) and no mergers (it is an isolated system), leading to the cleaner monotonic relationship in Figure~\ref{19}a.

\section{Conclusions}
\label{sec:conclusion}

In this work we introduce \gfour, a~modern galaxy formation simulation code, into~the AGORA Project. Preliminary results obtained with the \gfour framework have already been incorporated into two recent AGORA comparison studies (Paper~VIII~\cite{jung2025} and  Paper~X~\cite{kim2025agora}), where they contributed to broader code-comparison efforts. Building on those initial applications, this paper provides a systematic analysis of \gfour through a controlled multi-stage comparison with \gthree, its predecessor. 
By isolating numerical and physical updates in both idealized and cosmological settings, we validate the performance of \gfour and clarify the origins of differences between the two generations of the Osaka code in both isolated and cosmological contexts. 
Our investigation confirms that while changes in hydrodynamics solvers and cooling implementations introduce minor systematic offsets, the~dominant driver of galaxy evolution in these models remains the implementation of stellar~feedback.

Our primary findings can be summarized as follows:

\begin{itemize}
  \item The transition from a purely thermal feedback model (\gthree in isolated test) to a comprehensive model including mechanical momentum injection and stochastic thermal heating (\gfour) fundamentally changes simulation outcomes. The~\gfour model successfully prevents catastrophic gravitational fragmentation of the gas disk, replacing a clumpy and unrealistic morphology with a smooth stable disk featuring prominent spiral arms. This regulation brings the simulated star formation into agreement with the observed Kennicutt--Schmidt relation without requiring artificial delays in cooling.
  \item The momentum-driven and stochastic components of the \gfour feedback model are crucial for launching galactic-scale outflows. These outflows enrich the CGM with metals and establish a hot volume-filling gaseous halo. Our systematic feedback deconstruction reveals that momentum injection is the primary agent for suppressing clumps, while stochastic heating is essential for driving hot outflows.
  \item Following the rigorous AGORA protocol, we successfully calibrated the \gfour cosmological zoom-in simulation to match the stellar-halo mass relation at redshift $z = 4$. Despite the more complicated nature of cosmological assembly, the~core physical lessons from the isolated disk tests hold: the \gfour feedback model produces a more realistic multiphase interstellar medium and drives more significant outflows compared to the older \gthree cosmological run.
  \item Our staged comparison confirms that gravity solvers in both codes produce convergent dark matter structures. Differences in SPH formulation (including, e.g.,~artificial conduction and metal diffusion) and the interface with the \textsc{GRACKLE} cooling library were found to cause small but persistent offsets in gas temperature and clumpiness.
\end{itemize}

\subsection*{{Limitations} %
 and Future~Work}
\label{sec:limitfuture}

Despite its successes, this work has several limitations that point toward future avenues of~research:
\begin{itemize}
  \item The calibration of sub-grid parameters is unsatisfactory. In~particular, energy from SN is the most problematic, and~setting it to $10 \times$ more in \gfour seems somewhat unrealistic. We attribute this to the absence of other regulatory mechanisms, such as AGN and stellar radiative feedback. In~particular, this last one is present in \gthree but absent in our runs, which may warrant the higher boost we used. Additionally, further comparisons with observations are needed to enable a more complete calibration procedure.
  \item As mentioned in our last point, the~simulations used here do not include all physical processes relevant at galaxy-formation scales; magnetic fields, cosmic rays, radiative feedback, AGN, and more would need their own sub-grid models, which were either not activated or not present in \gfour. Comparing these implementations between codes is also a goal of future AGORA projects.
  \item This study was conducted at a single fixed resolution. Momentum-based feedback has been found to be relatively resolution-independent \citep{Smith_2018}, and~isolated tests have been performed at different resolutions using our feedback models \citep{Shimizu_2019,Oku_2022}. However, resolution tests for zoom-in cosmological runs have not been performed with our code, and~recent studies have found systematic differences even among modern sub-grid models in galaxy-scale simulations \citep{Pakmor_2025}.
  \item Deeper analysis of the current cosmological zoom-in simulation. While this work focused on the calibration targets at high redshift, the~cosmological simulation has been evolved to $z = 0$. This can be used for a wide range of analyses, including detailed studies of chemical abundance patterns, the~assembly history and morphology of the low-redshift stellar disk, the~evolution of the CGM, the~effects of individual mergers, and~the properties of the satellite galaxy population. These analyses constitute the core of Papers IV to X of the AGORA collaboration, with~disk analysis already performed in Paper VIII and merger analysis underway in Paper IX, including \gfour.
  \item Feedback variations in the full cosmological runs, like those performed for the isolated run in Section~\ref{sec:fbvaresults}, are a future endeavor that could validate many of the conclusions reached here and deepen our understanding of sub-grid models.
  \item Expanded comparisons with observational data are needed. \rev{While this study establishes numerical stability and calibration within the AGORA framework, the~absence of direct comparisons with multi-waveband observations limits our assessment of the model's physical realism. With~the code now validated, we can generate synthetic observations to enable direct quantitative comparisons with astronomical surveys. This includes testing our model against galaxy scaling relations such as the \mbox{mass--metallicity} relation and~the observed properties of the CGM through mock quasar absorption-line spectroscopy.}
\end{itemize}

The final calibrated cosmological simulation from \gfour has been contributed to the AGORA collaboration's \textit{CosmoRun} suite. By~continuing to refine the physical models and rigorously comparing them against other codes and observational data, we aim to further enhance the fidelity and predictive power of galaxy formation~simulations.

\vspace{6pt}

\authorcontributions{P.G. performed all \gfour simulations in this paper, analyzed their outputs, and wrote the first draft. Y.O. developed the \gfour code. Both Y.O. and K.N. interpreted, supervised, and polished this~work. All authors have read and agreed to the published version of the manuscript.}

\funding{This work is supported by the MEXT/JSPS KAKENHI,
Grant Numbers JP20H00180, JP22K21349, 24H00002,
24H00241, and~25K01032 (K.N.).}

\dataavailability{\gthree simulation data are publicly available through the AGORA data {release} %
 (\url{https://sites.google.com/site/santacruzcomparisonproject/data}, accessed on 25 February 2026) \citep{AGORAdatareleaseRoc}. \gfour fiducial runs will be released together with the previous data in the future, and~variation runs are available upon request to the authors. Analysis scripts are provided in a GitHub {repository} (\url{https://github.com/Ghippax/Paper-Scripts}, accessed on 25 February 2026) for~reproducibility.} 

\acknowledgments{This work used computational resources provided by the SQUID at the D3 Center of the University of Osaka through the HPCI System Research Project (Project
IDs: hp230089, hp240141, hp250119). We thank Santi Roca-Fàbrega for providing the analysis codes used for AGORA runs and helpful comments, and~thank Ji-hoon Kim and Joel Primack for their leadership in the AGORA collaboration. We are grateful to Volker Springel and others for developing the original version of the GADGET-3 and GADGET-4 codes, on~which the \gthree and \gfour codes are based. We dedicate this paper to Joel Primack and Avishai Dekel, both strong proponents of the AGORA project, who passed away during the preparation of this~manuscript.}

\conflictsofinterest{The authors declare no conflicts of~interest.}

\clearpage 
\abbreviations{Abbreviations}{
The following abbreviations are used in this manuscript:
\\

\noindent 
\begin{tabular}{@{}m{1.5cm}<{\raggedright}m{10.5cm}<{\raggedright}}
AGORA & Assembling Galaxies of Resolved Anatomy\\
AGN & Active Galactic Nucleus\\
SPH & Smoothed Particle Hydrodynamics\\
AMR & Adaptive Mesh Refinement\\
CDM & Cold Dark Matter\\
NFW & Navarro--Frenk--White\\
FMM & Fast Multipole Method\\
PM & Particle Mesh\\
IMF & Initial Mass Function\\
ESFB & Early Stellar Feedback\\
SN & Supernovae\\
ISM & Interstellar medium\\
MW & Milky Way\\
SF & Star Formation\\
FB & Feedback\\
SFR & Star Formation Rate\\

\end{tabular}

\noindent
\begin{tabular}{@{}m{1.5cm}<{\raggedright}m{10.5cm}<{\raggedright}}

IGM & Intergalactic Medium\\
CGM & Circumgalactic Medium\\
AGB & Asymptotic Giant Branch\\
KS & Kennicutt--Schmidt\\
NSFF & No Star Formation or Feedback\\
SFF & Star Formation and Feedback\\
\end{tabular}
}

\begin{adjustwidth}{-\extralength}{0cm}

\reftitle{{References} %
}

\PublishersNote{}
\end{adjustwidth}

\begin{thebibliography}{999}
\bibitem[White and Rees(1978)]{White1978}
White, S.D.M.; Rees, M.J.
\newblock Core condensation in heavy halos: A two-stage theory for galaxy
formation and clustering.
\newblock {\em Mon. Not. R. Astron. Soc.} {\bf 1978},
{\em 183},~341--358. [\href{http://doi.org/10.1093/mnras/183.3.341}{CrossRef}]
\bibitem[Springel et~al.(2005)Springel, White, Jenkins, Frenk, Yoshida, Gao,
Navarro, Thacker, Croton, Helly, Peacock, Cole, Thomas, Couchman, Evrard,
Colberg, and Pearce]{Springel_2005_1}
Springel, V.; White, S.D.M.; Jenkins, A.; Frenk, C.S.; Yoshida, N.; Gao, L.;
Navarro, J.; Thacker, R.; Croton, D.; Helly, J.;  et~al.
\newblock Simulations of the formation, evolution and clustering of galaxies
and quasars.
\newblock {\em Nature} {\bf 2005}, {\em 435},~629--636. [\href{http://dx.doi.org/10.1038/nature03597}{CrossRef}] [\href{http://www.ncbi.nlm.nih.gov/pubmed/15931216}{PubMed}]
\bibitem[Klypin et~al.(2011)Klypin, Trujillo-Gomez, and Primack]{Klypin_2011}
Klypin, A.A.; Trujillo-Gomez, S.; Primack, J.
\newblock {Dark} 
Matter Halos in the Standard Cosmological Model: Results from
the Bolshoi Simulation.
\newblock {\em  Astrophys. J.} {\bf 2011}, {\em 740},~102. [\href{http://dx.doi.org/10.1088/0004-637X/740/2/102}{CrossRef}]
\bibitem[Hirashima et~al.(2023)Hirashima, Moriwaki, Fujii, Hirai, Saitoh, and
Makino]{hirashima2023}
Hirashima, K.; Moriwaki, K.; Fujii, M.S.; Hirai, Y.; Saitoh, T.R.; Makino, J.
\newblock 3D-Spatiotemporal Forecasting the Expansion of Supernova Shells Using
Deep Learning toward High-Resolution Galaxy Simulations. \textit{arXiv}  \textbf{2023}, {arXiv:2302.00026}. [\href{http://dx.doi.org/10.1093/mnras/stad2864}{CrossRef}]
\bibitem[Cavelan et~al.(2020)Cavelan, Cabezón, Grabarczyk, and
Ciorba]{Cavelan_2020}
Cavelan, A.; Cabezón, R.M.; Grabarczyk, M.; Ciorba, F.M.
\newblock A Smoothed Particle Hydrodynamics Mini-App for Exascale.
\newblock In \textit{Proceedings of the Platform for Advanced
Scientific Computing Conference (PASC ’20), Geneva Switzerland, 29 June--1 July 2020};  ACM: {New York, NY, USA,} 
2020; pp. 1--11. [\href{http://dx.doi.org/10.1145/3394277.3401855}{CrossRef}]
\bibitem[{Hirashima} et~al.(2025){Hirashima}, {Fujii}, {Saitoh}, {Harada},
{Nomura}, {Yoshikawa}, {Hirai}, {Asano}, {Moriwaki}, {Iwasawa}, {Okamoto},
and {Makino}]{2025arXiv251023330H}
{Hirashima}, K.; {Fujii}, M.S.; {Saitoh}, T.R.; {Harada}, N.; {Nomura}, K.;
{Yoshikawa}, K.; {Hirai}, Y.; {Asano}, T.; {Moriwaki}, K.; {Iwasawa}, M.;
et~al.
\newblock {The First Star-by-star $N$-body/Hydrodynamics Simulation of Our
Galaxy Coupling with a Surrogate Model}.
\newblock {\em arXiv } {\bf 2025}, arXiv:2510.23330. [\href{http://dx.doi.org/10.48550/arXiv.2510.23330}{CrossRef}]
\bibitem[Springel et~al.(2017)Springel, Pakmor, Pillepich, Weinberger, Nelson,
Hernquist, Vogelsberger, Genel, Torrey, Marinacci, and Naiman]{Springel_2017}
Springel, V.; Pakmor, R.; Pillepich, A.; Weinberger, R.; Nelson, D.; Hernquist,
L.; Vogelsberger, M.; Genel, S.; Torrey, P.; Marinacci, F.;  et~al.
\newblock First results from the IllustrisTNG simulations: Matter and galaxy
clustering.
\newblock {\em Mon. Not. R. Astron. Soc.} {\bf 2017},
{\em 475},~676--698. [\href{http://dx.doi.org/10.1093/mnras/stx3304}{CrossRef}]
\bibitem[Schaye et~al.(2014)Schaye, Crain, Bower, Furlong, Schaller, Theuns,
Dalla~Vecchia, Frenk, McCarthy, Helly, Jenkins, Rosas-Guevara, White, Baes,
Booth, Camps, Navarro, Qu, Rahmati, Sawala, Thomas, and
Trayford]{Schaye_2014}
Schaye, J.; Crain, R.A.; Bower, R.G.; Furlong, M.; Schaller, M.; Theuns, T.;
Dalla~Vecchia, C.; Frenk, C.S.; McCarthy, I.G.; Helly, J.C.;  et~al.
\newblock The EAGLE project: Simulating the evolution and assembly of galaxies
and their environments.
\newblock {\em Mon. Not. R. Astron. Soc.} {\bf 2014},
{\em 446},~521--554. [\href{http://dx.doi.org/10.1093/mnras/stu2058}{CrossRef}]
\bibitem[Kaviraj et~al.(2017)Kaviraj, Laigle, Kimm, Devriendt, Dubois, Pichon,
Slyz, Chisari, and Peirani]{Kaviraj_2017}
Kaviraj, S.; Laigle, C.; Kimm, T.; Devriendt, J.E.G.; Dubois, Y.; Pichon, C.;
Slyz, A.; Chisari, E.; Peirani, S.
\newblock The Horizon-AGN simulation: Evolution of galaxy properties over
cosmic time.
\newblock {\em Mon. Not. R. Astron. Soc.} {\bf 2017}, {\textit{467}, 4739--4752}. [\href{http://dx.doi.org/10.1093/mnras/stx126}{CrossRef}]
\bibitem[Oku and Nagamine(2024)]{Oku2024}
Oku, Y.; Nagamine, K.
\newblock Osaka Feedback Model. III. Cosmological Simulation CROCODILE.
\newblock {\em  Astrophys. J.} {\bf 2024}, {\em 975},~183. [\href{http://dx.doi.org/10.3847/1538-4357/ad77d3}{CrossRef}]
\bibitem[Davé et~al.(2019)Davé, Anglés-Alcázar, Narayanan, Li,
Rafieferantsoa, and Appleby]{Dav__2019}
Davé, R.; Anglés-Alcázar, D.; Narayanan, D.; Li, Q.; Rafieferantsoa, M.H.;
Appleby, S.
\newblock simba: Cosmological simulations with black hole growth and feedback.
\newblock {\em Mon. Not. R. Astron. Soc.} {\bf 2019},
{\em 486},~2827--2849. [\href{http://dx.doi.org/10.1093/mnras/stz937}{CrossRef}]
\bibitem[Dolag et~al.(2025)Dolag, Remus, Valenzuela, Kimmig, Seidel, Fortune,
Stoiber, Ivleva, Hoffmann, Biffi, Marini, Popesso, and
Vladutescu-Zopp]{dolag2025emagneticum}
Dolag, K.; Remus, R.S.; Valenzuela, L.M.; Kimmig, L.C.; Seidel, B.; Fortune,
S.; Stoiber, J.; Ivleva, A.; Hoffmann, T.; Biffi, V.;  et~al.
\newblock Encyclopedia Magneticum: Scaling Relations from Cosmic Dawn to
Present Day.  \textit{arXiv} \textbf{2025}, arXiv:2504.01061.  [\href{http://dx.doi.org/10.48550/arXiv.2504.01061}{CrossRef}]
\bibitem[Feldmann et~al.(2023)Feldmann, Quataert, Faucher-Giguère, Hopkins,
Çatmabacak, Kereš, Bassini, Bernardini, Bullock, Cenci, Gensior, Liang,
Moreno, and Wetzel]{Feldmann_2023}
Feldmann, R.; Quataert, E.; Faucher-Giguère, C.A.; Hopkins, P.F.; Çatmabacak,
O.; Kereš, D.; Bassini, L.; Bernardini, M.; Bullock, J.S.; Cenci, E.;
et~al.
\newblock FIREbox: Simulating galaxies at high dynamic range in a cosmological
volume.
\newblock {\em Mon. Not. R. Astron. Soc.} {\bf 2023},
{\em 522},~3831--3860. [\href{http://dx.doi.org/10.1093/mnras/stad1205}{CrossRef}]
\bibitem[Schaye et~al.(2025)Schaye, Chaikin, Schaller, Ploeckinger, Huško,
McGibbon, Trayford, Benítez-Llambay, Correa, Frenk, Richings, Moreno, Bahé,
Borrow, Durrant, Gebek, Helly, Jenkins, Lacey, Ludlow, and
Nobels]{schaye2025colibre}
Schaye, J.; Chaikin, E.; Schaller, M.; Ploeckinger, S.; Huško, F.; McGibbon,
R.; Trayford, J.W.; Benítez-Llambay, A.; Correa, C.; Frenk, C.S.;  et~al.
\newblock The COLIBRE project: Cosmological hydrodynamical simulations of
galaxy formation and evolution. \textit{arXiv} \textbf{2025}, arXiv:2508.21126. [\href{http://dx.doi.org/10.48550/arXiv.2508.21126}{CrossRef}]
\bibitem[Scannapieco et~al.(2012)Scannapieco, Wadepuhl, Parry, Navarro,
Jenkins, Springel, Teyssier, Carlson, Couchman, Crain, Vecchia, Frenk,
Kobayashi, Monaco, Murante, Okamoto, Quinn, Schaye, Stinson, Theuns, Wadsley,
White, and Woods]{Scannapieco_2012}
Scannapieco, C.; Wadepuhl, M.; Parry, O.H.; Navarro, J.F.; Jenkins, A.;
Springel, V.; Teyssier, R.; Carlson, E.; Couchman, H.M.P.; Crain, R.A.;
et~al.
\newblock The Aquila comparison project: The effects of feedback and numerical
methods on simulations of galaxy formation: The Aquila comparison project.
\newblock {\em Mon. Not. R. Astron. Soc.} {\bf 2012},
{\em 423},~1726--1749. [\href{http://dx.doi.org/10.1111/j.1365-2966.2012.20993.x}{CrossRef}]
\bibitem[Wadsley et~al.(2004)Wadsley, Stadel, and Quinn]{Wadsley_2004}
Wadsley, J.; Stadel, J.; Quinn, T.
\newblock Gasoline: A flexible, parallel implementation of TreeSPH.
\newblock {\em New Astron.} {\bf 2004}, {\em 9},~137--158. [\href{http://dx.doi.org/10.1016/j.newast.2003.08.004}{CrossRef}]
\bibitem[Springel(2005)]{Springel_2005}
Springel, V.
\newblock The cosmological simulation code gadget-2.
\newblock {\em Mon. Not. R. Astron. Soc.} {\bf 2005},
{\em 364},~1105--1134. [\href{http://dx.doi.org/10.1111/j.1365-2966.2005.09655.x}{CrossRef}]
\bibitem[Bryan et~al.(2014)Bryan, Norman, O’Shea, Abel, Wise, Turk, Reynolds,
Collins, Wang, Skillman, Smith, Harkness, Bordner, Kim, Kuhlen, Xu, Goldbaum,
Hummels, Kritsuk, Tasker, Skory, Simpson, Hahn, Oishi, So, Zhao, Cen, and
Li]{Bryan_2014}
Bryan, G.L.; Norman, M.L.; O’Shea, B.W.; Abel, T.; Wise, J.H.; Turk, M.J.;
Reynolds, D.R.; Collins, D.C.; Wang, P.; Skillman, S.W.;  et~al.
\newblock ENZO: {An} Adaptive Mesh Refinement Code for Astrophysics.
\newblock {\em  Astrophys. J. Suppl. Ser.} {\bf 2014}, {\em
211},~19. [\href{http://dx.doi.org/10.1088/0067-0049/211/2/19}{CrossRef}]
\bibitem[Kravtsov et~al.(1997)Kravtsov, Klypin, and Khokhlov]{Kravtsov_1997}
Kravtsov, A.V.; Klypin, A.A.; Khokhlov, A.M.
\newblock Adaptive Refinement Tree: A New High‐ResolutionN‐Body Code for
Cosmological Simulations.
\newblock {\em  Astrophys. J. Suppl. Ser.} {\bf 1997}, {\em
111},~73--94. [\href{http://dx.doi.org/10.1086/313015}{CrossRef}]
\bibitem[Teyssier(2002)]{Teyssier_2002}
Teyssier, R.
\newblock Cosmological hydrodynamics with adaptive mesh refinement: A new high
resolution code called RAMSES.
\newblock {\em Astron. Astrophys.} {\bf 2002}, {\em 385},~337--364. [\href{http://dx.doi.org/10.1051/0004-6361:20011817}{CrossRef}]
\bibitem[Weinberger et~al.(2020)Weinberger, Springel, and
Pakmor]{Weinberger_2020}
Weinberger, R.; Springel, V.; Pakmor, R.
\newblock The AREPO Public Code Release.
\newblock {\em  Astrophys. J. Suppl. Ser.} {\bf 2020}, {\em
248},~32. [\href{http://dx.doi.org/10.3847/1538-4365/ab908c}{CrossRef}]
\bibitem[Hopkins(2012)]{Hopkins_2012}
Hopkins, P.F.
\newblock A general class of Lagrangian smoothed particle hydrodynamics methods
and implications for fluid mixing problems.
\newblock {\em Mon. Not. R. Astron. Soc.} {\bf 2012},
{\em 428},~2840--2856. [\href{http://dx.doi.org/10.1093/mnras/sts210}{CrossRef}]
\bibitem[Sawala et~al.(2016)Sawala, Frenk, Fattahi, Navarro, Bower, Crain,
Vecchia, Furlong, Helly, Jenkins, Oman, Schaller, Schaye, Theuns, Trayford,
and White]{Sawala_2016}
Sawala, T.; Frenk, C.S.; Fattahi, A.; Navarro, J.F.; Bower, R.G.; Crain, R.A.;
Vecchia, C.D.; Furlong, M.; Helly, J.C.; Jenkins, A.;  et~al.
\newblock The APOSTLE simulations: Solutions to the Local Group’s cosmic
puzzles.
\newblock {\em Mon. Not. R. Astron. Soc.} {\bf 2016},
{\em 457},~1931--1943. [\href{http://dx.doi.org/10.1093/mnras/stw145}{CrossRef}]
\bibitem[Wetzel et~al.(2016)Wetzel, Hopkins, Kim, Faucher-Giguère, Kereš, and
Quataert]{Wetzel_2016}
Wetzel, A.R.; Hopkins, P.F.; Kim, J.H.; Faucher-Giguère, C.A.; Kereš, D.;
Quataert, E.
\newblock {Reconciling} Dwarf Galaxies with $\Lambda$CDM Cosmology: Simulating a
Realistic Population of Satellites Around a Milky Way--Mass Galaxy.
\newblock {\em  Astrophys. J. Lett.} {\bf 2016}, {\em 827},~L23. [\href{http://dx.doi.org/10.3847/2041-8205/827/2/L23}{CrossRef}]
\bibitem[Grand et~al.(2017)Grand, Gómez, Marinacci, Pakmor, Springel,
Campbell, Frenk, Jenkins, and White]{Grand_2017}
Grand, R.J.J.; Gómez, F.A.; Marinacci, F.; Pakmor, R.; Springel, V.; Campbell,
D.J.R.; Frenk, C.S.; Jenkins, A.; White, S.D.M.
\newblock The Auriga Project: The properties and formation mechanisms of disc
galaxies across cosmic time.
\newblock {\em Mon. Not. R. Astron. Soc.} {\bf 2017},
{\textit{467}, 179--207.}
\newblock [\href{http://dx.doi.org/10.1093/mnras/stx071}{CrossRef}]
\bibitem[Guedes et~al.(2011)Guedes, Callegari, Madau, and Mayer]{Guedes_2011}
Guedes, J.; Callegari, S.; Madau, P.; Mayer, L.
\newblock {Forming} Realistic Late-Type Spirals in a $\Lambda$CDM Universe: The
Eris Simulation.
\newblock {\em  Astrophys. J.} {\bf 2011}, {\em 742},~76. [\href{http://dx.doi.org/10.1088/0004-637X/742/2/76}{CrossRef}]
\bibitem[Wang et~al.(2015)Wang, Dutton, Stinson, Macciò, Penzo, Kang, Keller,
and Wadsley]{Wang_2015}
Wang, L.; Dutton, A.A.; Stinson, G.S.; Macciò, A.V.; Penzo, C.; Kang, X.;
Keller, B.W.; Wadsley, J.
\newblock NIHAO project---I. Reproducing the inefficiency of galaxy formation
across cosmic time with a large sample of cosmological hydrodynamical
simulations.
\newblock {\em Mon. Not. R. Astron. Soc.} {\bf 2015},
{\em 454},~83--94. [\href{http://dx.doi.org/10.1093/mnras/stv1937}{CrossRef}]
\bibitem[Agertz et~al.(2021)Agertz, Renaud, Feltzing, Read, Ryde, Andersson,
Rey, Bensby, and Feuillet]{Agertz_2021}
Agertz, O.; Renaud, F.; Feltzing, S.; Read, J.I.; Ryde, N.; Andersson, E.P.;
Rey, M.P.; Bensby, T.; Feuillet, D.K.
\newblock VINTERGATAN---I. The origins of chemically, kinematically, and
structurally distinct discs in a simulated Milky Way-mass galaxy.
\newblock {\em Mon. Not. R. Astron. Soc.} {\bf 2021},
{\em 503},~5826--5845. [\href{http://dx.doi.org/10.1093/mnras/stab322}{CrossRef}]
\bibitem[Ceverino et~al.(2014)Ceverino, Klypin, Klimek, Trujillo-Gomez,
Churchill, Primack, and Dekel]{Ceverino_2014}
Ceverino, D.; Klypin, A.; Klimek, E.S.; Trujillo-Gomez, S.; Churchill, C.W.;
Primack, J.; Dekel, A.
\newblock Radiative feedback and the low efficiency of galaxy formation in
low-mass haloes at high redshift.
\newblock {\em Mon. Not. R. Astron. Soc.} {\bf 2014},
{\em 442},~1545--1559. [\href{http://dx.doi.org/10.1093/mnras/stu956}{CrossRef}]
\bibitem[Frenk et~al.(1999)Frenk, White, Bode, Bond, Bryan, Cen, Couchman,
Evrard, Gnedin, Jenkins, Khokhlov, Klypin, Navarro, Norman, Ostriker, Owen,
Pearce, Pen, Steinmetz, Thomas, Villumsen, Wadsley, Warren, Xu, and
Yepes]{Frenk_1999}
Frenk, C.S.; White, S.D.M.; Bode, P.; Bond, J.R.; Bryan, G.L.; Cen, R.;
Couchman, H.M.P.; Evrard, A.E.; Gnedin, N.; Jenkins, A.;  et~al.
\newblock The Santa Barbara Cluster Comparison Project: A Comparison of
Cosmological Hydrodynamics Solutions.
\newblock {\em  Astrophys. J.} {\bf 1999}, {\em 525},~554--582. [\href{http://dx.doi.org/10.1086/307908}{CrossRef}]
\bibitem[Sembolini et~al.(2016)Sembolini, Yepes, Pearce, Knebe, Kay, Power,
Cui, Beck, Borgani, Dalla~Vecchia, Davé, Elahi, February, Huang, Hobbs,
Katz, Lau, McCarthy, Murante, Nagai, Nelson, Newton, Perret, Puchwein, Read,
Saro, Schaye, Teyssier, and Thacker]{Sembolini_2016}
Sembolini, F.; Yepes, G.; Pearce, F.R.; Knebe, A.; Kay, S.T.; Power, C.; Cui,
W.; Beck, A.M.; Borgani, S.; Dalla~Vecchia, C.;  et~al.
\newblock nIFTy galaxy cluster simulations---I. Dark matter and non-radiative
models.
\newblock {\em Mon. Not. R. Astron. Soc.} {\bf 2016},
{\em 457},~4063--4080. [\href{http://dx.doi.org/10.1093/mnras/stw250}{CrossRef}]
\bibitem[Cui et~al.(2018)Cui, Knebe, Yepes, Pearce, Power, Dave, Arth, Borgani,
Dolag, Elahi, Mostoghiu, Murante, Rasia, Stoppacher, Vega-Ferrero, Wang,
Yang, Benson, Cora, Croton, Sinha, Stevens, Vega-Martínez, Arthur, Baldi,
Cañas, Cialone, Cunnama, De Petris, Durando, Ettori, Gottlöber, Nuza, Old,
Pilipenko, Sorce, and Welker]{Cui_2018}
Cui, W.; Knebe, A.; Yepes, G.; Pearce, F.; Power, C.; Dave, R.; Arth, A.;
Borgani, S.; Dolag, K.; Elahi, P.;  et~al.
\newblock The Three Hundred project: A large catalogue of theoretically
modelled galaxy clusters for cosmological and astrophysical applications.
\newblock {\em Mon. Not. R. Astron. Soc.} {\bf 2018},
{\em 480},~2898--2915. [\href{http://dx.doi.org/10.1093/mnras/sty2111}{CrossRef}]
\bibitem[Hu et~al.(2023)Hu, Smith, Teyssier, Bryan, Verbeke, Emerick,
Somerville, Burkhart, Li, Forbes, and Starkenburg]{Hu_2023Lag}
Hu, C.Y.; Smith, M.C.; Teyssier, R.; Bryan, G.L.; Verbeke, R.; Emerick, A.;
Somerville, R.S.; Burkhart, B.; Li, Y.; Forbes, J.C.;  et~al.
\newblock Code Comparison in Galaxy-scale Simulations with Resolved Supernova
Feedback: Lagrangian versus Eulerian Methods.
\newblock {\em  Astrophys. J.} {\bf 2023}, {\em 950},~132. [\href{http://dx.doi.org/10.3847/1538-4357/accf9e}{CrossRef}]
\bibitem[Smith et~al.(2018)Smith, Sijacki, and Shen]{Smith_2018}
Smith, M.C.; Sijacki, D.; Shen, S.
\newblock Supernova feedback in numerical simulations of galaxy formation:
Separating physics from numerics.
\newblock {\em Mon. Not. R. Astron. Soc.} {\bf 2018},
{\em 478},~302--331. [\href{http://dx.doi.org/10.1093/mnras/sty994}{CrossRef}]
\bibitem[Stewart et~al.(2017)Stewart, Maller, Oñorbe, Bullock, Joung,
Devriendt, Ceverino, Kereš, Hopkins, and Faucher-Giguère]{Stewart_2017}
Stewart, K.R.; Maller, A.H.; Oñorbe, J.; Bullock, J.S.; Joung, M.R.;
Devriendt, J.; Ceverino, D.; Kereš, D.; Hopkins, P.F.; Faucher-Giguère,
C.A.
\newblock High Angular Momentum Halo Gas: A Feedback and Code-independent
Prediction of LCDM.
\newblock {\em  Astrophys. J.} {\bf 2017}, {\em 843},~47. [\href{http://dx.doi.org/10.3847/1538-4357/aa6dff}{CrossRef}]
\bibitem[{Hopkins} et~al.(2018){Hopkins}, {Wetzel}, {Kere{\v{s}}},
{Faucher-Gigu{\`e}re}, {Quataert}, {Boylan-Kolchin}, {Murray}, {Hayward},
{Garrison-Kimmel}, {Hummels}, {Feldmann}, {Torrey}, {Ma},
{Angl{\'e}s-Alc{\'a}zar}, {Su}, {Orr}, {Schmitz}, {Escala}, {Sanderson},
{Grudi{\'c}}, {Hafen}, {Kim}, {Fitts}, {Bullock}, {Wheeler}, {Chan},
{Elbert}, and {Narayanan}]{Fire2_2018}
{Hopkins}, P.F.; {Wetzel}, A.; {Kere{\v{s}}}, D.; {Faucher-Gigu{\`e}re}, C.A.;
{Quataert}, E.; {Boylan-Kolchin}, M.; {Murray}, N.; {Hayward}, C.C.;
{Garrison-Kimmel}, S.; {Hummels}, C.;  et~al.
\newblock {FIRE-2 simulations: Physics versus numerics in galaxy formation}.
\newblock {\em \mnras} {\bf 2018}, {\em 480},~800--{863.} 
\newblock [\href{http://dx.doi.org/10.1093/mnras/sty1690}{CrossRef}]
\bibitem[Springel et~al.(2005)Springel, Di~Matteo, and
Hernquist]{Springel_2005fdb}
Springel, V.; Di~Matteo, T.; Hernquist, L.
\newblock Modelling feedback from stars and black holes in galaxy mergers.
\newblock {\em Mon. Not. R. Astron. Soc.} {\bf 2005},
{\em 361},~776--794. [\href{http://dx.doi.org/10.1111/j.1365-2966.2005.09238.x}{CrossRef}]
\bibitem[Hahn and Abel(2011)]{Hahn_2011}
Hahn, O.; Abel, T.
\newblock Multi-scale initial conditions for cosmological simulations:
Multi-scale initial conditions.
\newblock {\em Mon. Not. R. Astron. Soc.} {\bf 2011},
{\em 415},~2101--2121. [\href{http://dx.doi.org/10.1111/j.1365-2966.2011.18820.x}{CrossRef}]
\bibitem[Turk et~al.(2010)Turk, Smith, Oishi, Skory, Skillman, Abel, and
Norman]{Turk_2010}
Turk, M.J.; Smith, B.D.; Oishi, J.S.; Skory, S.; Skillman, S.W.; Abel, T.;
Norman, M.L.
\newblock {yt:} A Multi-Code Analysis Toolkit for Astrophysical Simulation Data.
\newblock {\em  Astrophys. J. Suppl. Ser.} {\bf 2010}, {\em
192},~9. [\href{http://dx.doi.org/10.1088/0067-0049/192/1/9}{CrossRef}]
\bibitem[Kim et~al.(2013)Kim, Abel, Agertz, Bryan, Ceverino, Christensen,
Conroy, Dekel, Gnedin, Goldbaum, Guedes, Hahn, Hobbs, Hopkins, Hummels,
Iannuzzi, Keres, Klypin, Kravtsov, Krumholz, Kuhlen, Leitner, Madau, Mayer,
Moody, Nagamine, Norman, Onorbe, O’Shea, Pillepich, Primack, Quinn, Read,
Robertson, Rocha, Rudd, Shen, Smith, Szalay, Teyssier, Thompson, Todoroki,
Turk, Wadsley, Wise, Zolotov, and Adi]{Kim_2013}
Kim, J.H.; Abel, T.; Agertz, O.; Bryan, G.L.; Ceverino, D.; Christensen, C.;
Conroy, C.; Dekel, A.; Gnedin, N.Y.; Goldbaum, N.J.;  et~al.
\newblock {The} Agora High-Resolution Galaxy Simulations Comparison Project.
\newblock {\em  Astrophys. J. Suppl. Ser.} {\bf 2013}, {\em
210},~14. [\href{http://dx.doi.org/10.1088/0067-0049/210/1/14}{CrossRef}]
\bibitem[Kim et~al.(2016)Kim, Agertz, Teyssier, Butler, Ceverino, Choi,
Feldmann, Keller, Lupi, Quinn, Revaz, Wallace, Gnedin, Leitner, Shen, Smith,
Thompson, Turk, Abel, Arraki, Benincasa, Chakrabarti, DeGraf, Dekel,
Goldbaum, Hopkins, Hummels, Klypin, Li, Madau, Mandelker, Mayer, Nagamine,
Nickerson, O’Shea, Primack, Roca-Fàbrega, Semenov, Shimizu, Simpson,
Todoroki, Wadsley, and Wise]{Kim_2016}
Kim, J.H.; Agertz, O.; Teyssier, R.; Butler, M.J.; Ceverino, D.; Choi, J.H.;
Feldmann, R.; Keller, B.W.; Lupi, A.; Quinn, T.;  et~al.
\newblock The Agora High-Resolution Galaxy Simulations Comparison Project. II.
Isolated Disk Test.
\newblock {\em  Astrophys. J.} {\bf 2016}, {\em 833},~202. [\href{http://dx.doi.org/10.3847/1538-4357/833/2/202}{CrossRef}]
\bibitem[Roca-Fàbrega et~al.(2021)Roca-Fàbrega, Kim, Hausammann, Nagamine,
Lupi, Powell, Shimizu, Ceverino, Primack, Quinn, Revaz, Velázquez, Abel,
Buehlmann, Dekel, Dong, Hahn, Hummels, Kim, Smith, Strawn, Teyssier, and
Turk]{Roca_F_brega_2021}
Roca-Fàbrega, S.; Kim, J.H.; Hausammann, L.; Nagamine, K.; Lupi, A.; Powell,
J.W.; Shimizu, I.; Ceverino, D.; Primack, J.R.; Quinn, T.R.;  et~al.
\newblock The AGORA High-resolution Galaxy Simulations Comparison Project. III.
Cosmological Zoom-in Simulation of a Milky Way--mass Halo.
\newblock {\em  Astrophys. J.} {\bf 2021}, {\em 917},~64. [\href{http://dx.doi.org/10.3847/1538-4357/ac088a}{CrossRef}]
\bibitem[Roca-Fàbrega et~al.(2024)Roca-Fàbrega, hoon Kim, Primack, Jung,
Genina, Hausammann, Kim, Lupi, Nagamine, Powell, Revaz, Shimizu, Strawn,
Velázquez, Abel, Ceverino, Dong, Quinn, jin Shin, Segovia-Otero, Agertz,
Barrow, Cadiou, Dekel, Hummels, Oh, Teyssier, and
Collaboration]{rocafabrega2024}
Roca-Fàbrega, S.; hoon Kim, J.; Primack, J.R.; Jung, M.; Genina, A.;
Hausammann, L.; Kim, H.; Lupi, A.; Nagamine, K.; Powell, J.W.;  et~al.
\newblock The AGORA High-resolution Galaxy Simulations Comparison Project IV:
Halo and Galaxy Mass Assembly in a Cosmological Zoom-in Simulation at
$z\le2$.  \textit{arXiv} \textbf{2024}, {arXiv:2402.06202}. [\href{http://dx.doi.org/10.3847/1538-4357/ad43de}{CrossRef}]
\bibitem[Jung et~al.(2024)Jung, Roca-Fàbrega, Kim, Genina, Hausammann, Kim,
Lupi, Nagamine, Powell, Revaz, Shimizu, Velázquez, Ceverino, Primack, Quinn,
Strawn, Abel, Dekel, Dong, Oh, and Teyssier]{Jung_2024}
Jung, M.; Roca-Fàbrega, S.; Kim, J.H.; Genina, A.; Hausammann, L.; Kim, H.;
Lupi, A.; Nagamine, K.; Powell, J.W.; Revaz, Y.;  et~al.
\newblock The AGORA High-resolution Galaxy Simulations Comparison Project. V.
Satellite Galaxy Populations in a Cosmological Zoom-in Simulation of a Milky
Way--Mass Halo.
\newblock {\em  Astrophys. J.} {\bf 2024}, {\em 964},~123. [\href{http://dx.doi.org/10.3847/1538-4357/ad245b}{CrossRef}]
\bibitem[{Strawn} et~al.(2024){Strawn}, {Roca-F{\`a}brega}, {Primack}, {Kim},
{Genina}, {Hausammann}, {Kim}, {Lupi}, {Nagamine}, {Powell}, {Revaz},
{Shimizu}, {Vel{\'a}zquez}, {Abel}, {Ceverino}, {Dong}, {Jung}, {Quinn},
{Shin}, {Barrow}, {Dekel}, {Oh}, {Mandelker}, {Teyssier}, {Hummels}, {Maji},
{Man}, {Mayerhofer}, and {The Agora Collaboration}]{Clayton2024}
{Strawn}, C.; {Roca-F{\`a}brega}, S.; {Primack}, J.R.; {Kim}, J.H.; {Genina},
A.; {Hausammann}, L.; {Kim}, H.; {Lupi}, A.; {Nagamine}, K.; {Powell}, J.W.;
et~al.
\newblock {The AGORA High-resolution Galaxy Simulations Comparison Project. VI.
Similarities and Differences in the Circumgalactic Medium}.
\newblock {\em \apj} {\bf 2024}, {\em 962},~{29.} \newblock [\href{http://dx.doi.org/10.3847/1538-4357/ad12cb}{CrossRef}]
\bibitem[Rodríguez-Cardoso et~al.(2025)Rodríguez-Cardoso, Roca-Fàbrega,
Jung, Nguyen, Kim, Primack, Agertz, Barrow, Gallego, Nagamine, Powell, Revaz,
Velázquez, Genina, Kim, Lupi, Abel, Cen, Ceverino, Dekel, Oh, and
Quinn]{Rodr_guez_Cardoso_2025}
Rodríguez-Cardoso, R.; Roca-Fàbrega, S.; Jung, M.; Nguyen, T.H.; Kim, J.H.;
Primack, J.; Agertz, O.; Barrow, K.S.S.; Gallego, J.; Nagamine, K.;  et~al.
\newblock The AGORA High-Resolution Galaxy Simulations Comparison Project: VII.
Satellite quenching in zoom-in simulation of a Milky Way-mass halo.
\newblock {\em Astron. Astrophys.} {\bf 2025}, {\em 698},~A303. [\href{http://dx.doi.org/10.1051/0004-6361/202453639}{CrossRef}]
\bibitem[Jung et~al.(2025)Jung, hoon Kim, Nguyen, Rodriguez-Cardoso,
Roca-Fàbrega, Primack, Barrow, Genina, Kim, Nagamine, Powell, Revaz,
Velázquez, Lupi, Shimizu, Abel, Agertz, Cen, Ceverino, Dekel, Jeong, Mayer,
Oh, Quinn, Song, and for The AGORA~Collaboration]{jung2025}
Jung, M.; hoon Kim, J.; Nguyen, T.H.; Rodriguez-Cardoso, R.; Roca-Fàbrega, S.;
Primack, J.R.; Barrow, K.; Genina, A.; Kim, H.; Nagamine, K.;  et~al.
\newblock The {\it AGORA} High-resolution Galaxy Simulations Comparison
Project. VIII: Disk Formation and Evolution of Simulated Milky Way Mass
Galaxy Progenitors at $1<z<5$.  \textit{arXiv} \textbf{2025}, {arXiv:2505.05720}. [\href{http://dx.doi.org/10.3847/1538-4357/ae112d}{CrossRef}]
\bibitem[Springel et~al.(2021)Springel, Pakmor, Zier, and
Reinecke]{Springel_2021}
Springel, V.; Pakmor, R.; Zier, O.; Reinecke, M.
\newblock Simulating cosmic structure formation with the \textsc{gadget}-4
code.
\newblock {\em Mon. Not. R. Astron. Soc.} {\bf 2021},
{\em 506},~2871--2949. [\href{http://dx.doi.org/10.1093/mnras/stab1855}{CrossRef}]
\bibitem[Shimizu et~al.(2019)Shimizu, Todoroki, Yajima, and
Nagamine]{Shimizu_2019}
Shimizu, I.; Todoroki, K.; Yajima, H.; Nagamine, K.
\newblock Osaka feedback model: Isolated disc galaxy simulations.
\newblock {\em Mon. Not. R. Astron. Soc.} {\bf 2019},
{\em 484},~2632--2655. [\href{http://dx.doi.org/10.1093/mnras/stz098}{CrossRef}]
\bibitem[Oku et~al.(2022)Oku, Tomida, Nagamine, Shimizu, and Cen]{Oku_2022}
Oku, Y.; Tomida, K.; Nagamine, K.; Shimizu, I.; Cen, R.
\newblock Osaka Feedback Model. II. Modeling Supernova Feedback Based on
High-resolution Simulations.
\newblock {\em  Astrophys. J. Suppl. Ser.} {\bf 2022}, {\em
262},~9. [\href{http://dx.doi.org/10.3847/1538-4365/ac77ff}{CrossRef}]
\bibitem[{Navarro} et~al.(1997){Navarro}, {Frenk}, and {White}]{1997NFW}
{Navarro}, J.F.; {Frenk}, C.S.; {White}, S.D.M.
\newblock {A Universal Density Profile from Hierarchical Clustering}.
\newblock {\em \apj} {\bf 1997}, {\em 490},~493--{508.}
\newblock [\href{http://dx.doi.org/10.1086/304888}{CrossRef}]
\bibitem[{Hernquist}(1990)]{1990hernquist}
{Hernquist}, L.
\newblock {An Analytical Model for Spherical Galaxies and Bulges}.
\newblock {\em \apj} {\bf 1990}, {\em 356},~359. [\href{http://dx.doi.org/10.1086/168845}{CrossRef}]
\bibitem[{Komatsu} et~al.(2011){Komatsu}, {Smith}, {Dunkley}, {Bennett},
{Gold}, {Hinshaw}, {Jarosik}, {Larson}, {Nolta}, {Page}, {Spergel},
{Halpern}, {Hill}, {Kogut}, {Limon}, {Meyer}, {Odegard}, {Tucker}, {Weiland},
{Wollack}, and {Wright}]{komatsuic}
{Komatsu}, E.; {Smith}, K.M.; {Dunkley}, J.; {Bennett}, C.L.; {Gold}, B.;
{Hinshaw}, G.; {Jarosik}, N.; {Larson}, D.; {Nolta}, M.R.; {Page}, L.;
et~al.
\newblock {Seven-year Wilkinson Microwave Anisotropy Probe (WMAP) Observations:
Cosmological Interpretation}.
\newblock {\em \apjs} {\bf 2011}, {\em 192},~{18.}
\newblock [\href{http://dx.doi.org/10.1088/0067-0049/192/2/18}{CrossRef}]
\bibitem[Hinshaw et~al.(2013)Hinshaw, Larson, Komatsu, Spergel, Bennett,
Dunkley, Nolta, Halpern, Hill, Odegard, Page, Smith, Weiland, Gold, Jarosik,
Kogut, Limon, Meyer, Tucker, Wollack, and Wright]{Hinshaw_2013}
Hinshaw, G.; Larson, D.; Komatsu, E.; Spergel, D.N.; Bennett, C.L.; Dunkley,
J.; Nolta, M.R.; Halpern, M.; Hill, R.S.; Odegard, N.;  et~al.
\newblock {Nine}-Year Wilkinson Microwave Anisotropy Probe (WMAP) Observations:
Cosmological Parameter Results.
\newblock {\em  Astrophys. J. Suppl. Ser.} {\bf 2013}, {\em
208},~19. [\href{http://dx.doi.org/10.1088/0067-0049/208/2/19}{CrossRef}]
\bibitem[Springel et~al.(2001)Springel, Yoshida, and White]{Springel_2001G1}
Springel, V.; Yoshida, N.; White, S.D.
\newblock GADGET: A code for collisionless and gasdynamical cosmological
simulations.
\newblock {\em New Astron.} {\bf 2001}, {\em 6},~79--117. [\href{http://dx.doi.org/10.1016/S1384-1076(01)00042-2}{CrossRef}]
\bibitem[Dehnen and Aly(2012)]{Dehnen_2012}
Dehnen, W.; Aly, H.
\newblock Improving convergence in smoothed particle hydrodynamics simulations
without pairing instability: SPH without pairing instability.
\newblock {\em Mon. Not. R. Astron. Soc.} {\bf 2012},
{\em 425},~1068--1082. [\href{http://dx.doi.org/10.1111/j.1365-2966.2012.21439.x}{CrossRef}]
\bibitem[Saitoh and Makino(2013)]{Saitoh_2013}
Saitoh, T.R.; Makino, J.
\newblock {A} Density-Independent Formulation of Smoothed Particle Hydrodynamics.
\newblock {\em  Astrophys. J.} {\bf 2013}, {\em 768},~44. [\href{http://dx.doi.org/10.1088/0004-637X/768/1/44}{CrossRef}]
\bibitem[Frontiere et~al.(2017)Frontiere, Raskin, and Owen]{Frontiere_2017}
Frontiere, N.; Raskin, C.D.; Owen, J.M.
\newblock CRKSPH---A Conservative Reproducing Kernel Smoothed Particle
Hydrodynamics Scheme.
\newblock {\em J. Comput. Phys.} {\bf 2017}, {\em
332},~160--209. [\href{http://dx.doi.org/10.1016/j.jcp.2016.12.004}{CrossRef}]
\bibitem[Rosswog(2020)]{Rosswog_2020}
Rosswog, S.
\newblock The Lagrangian hydrodynamics code magma2.
\newblock {\em Mon. Not. R. Astron. Soc.} {\bf 2020},
{\em 498},~4230--4255. [\href{http://dx.doi.org/10.1093/mnras/staa2591}{CrossRef}]
\bibitem[Price(2008)]{Price_2008}
Price, D.J.
\newblock Modelling discontinuities and Kelvin--Helmholtz instabilities in SPH.
\newblock {\em J. Comput. Phys.} {\bf 2008}, {\em
227},~10040--10057. [\href{http://dx.doi.org/10.1016/j.jcp.2008.08.011}{CrossRef}]
\bibitem[Borrow et~al.(2021)Borrow, Schaller, Bower, and Schaye]{Borrow_2021}
Borrow, J.; Schaller, M.; Bower, R.G.; Schaye, J.
\newblock \textsc{Sphenix}: Smoothed particle hydrodynamics for the next
generation of galaxy formation simulations.
\newblock {\em Mon. Not. R. Astron. Soc.} {\bf 2021},
{\em 511},~2367--2389. [\href{http://dx.doi.org/10.1093/mnras/stab3166}{CrossRef}]
\bibitem[Morris and Monaghan(1997)]{Morris_1997}
Morris, J.; Monaghan, J.
\newblock A Switch to Reduce SPH Viscosity.
\newblock {\em J. Comput. Phys.} {\bf 1997}, {\em 136},~41--50. [\href{http://dx.doi.org/10.1006/jcph.1997.5690}{CrossRef}]
\bibitem[{Morris}(1996)]{Morris1996}
{Morris}, J.P.
\newblock {A study of the stability properties of smooth particle
hydrodynamics}.
\newblock {\em \pasa} {\bf 1996}, {\em 13},~97--102. [\href{http://dx.doi.org/10.1017/s1323358000020610}{CrossRef}]
\bibitem[Saitoh and Makino(2009)]{Saitoh_2009}
Saitoh, T.R.; Makino, J.
\newblock {A} Necessary Condition for Individual Time Steps in SPH Simulations.
\newblock {\em  Astrophys. J.} {\bf 2009}, {\em 697},~L99--L102. [\href{http://dx.doi.org/10.1088/0004-637X/697/2/L99}{CrossRef}]
\bibitem[Smith et~al.(2016)Smith, Bryan, Glover, Goldbaum, Turk, Regan, Wise,
Schive, Abel, Emerick, O’Shea, Anninos, Hummels, and Khochfar]{Smith_2016}
Smith, B.D.; Bryan, G.L.; Glover, S.C.O.; Goldbaum, N.J.; Turk, M.J.; Regan,
J.; Wise, J.H.; Schive, H.Y.; Abel, T.; Emerick, A.;  et~al.
\newblock grackle: A chemistry and cooling library for astrophysics.
\newblock {\em Mon. Not. R. Astron. Soc.} {\bf 2016},
{\em 466},~2217--2234. [\href{http://dx.doi.org/10.1093/mnras/stw3291}{CrossRef}]
\bibitem[Ferland et~al.(2013)Ferland, Porter, van Hoof, Williams, Abel, Lykins,
Shaw, Henney, and Stancil]{ferland20132013releasecloudy}
Ferland, G.J.; Porter, R.L.; van Hoof, P.A.M.; Williams, R.J.R.; Abel, N.P.;
Lykins, M.L.; Shaw, G.; Henney, W.J.; Stancil, P.C.
\newblock The 2013 Release of Cloudy.  \textit{arXiv} \textbf{2013}, {arXiv:1302.4485}. [\href{http://dx.doi.org/10.48550/arXiv.1302.4485}{CrossRef}]
\bibitem[{Haardt} and {Madau}(2012)]{Haardtable}
{Haardt}, F.; {Madau}, P.
\newblock {Radiative Transfer in a Clumpy Universe. IV. New Synthesis Models of
the Cosmic UV/X-Ray Background}.
\newblock {\em \apj} {\bf 2012}, {\em 746},~125. [\href{http://dx.doi.org/10.1088/0004-637X/746/2/125}{CrossRef}]
\bibitem[{Schmidt}(1959)]{Schmidt1959}
{Schmidt}, M.
\newblock {The Rate of Star Formation.}
\newblock {\em \apj} {\bf 1959}, {\em 129},~243. [\href{http://dx.doi.org/10.1086/146614}{CrossRef}]
\bibitem[Chabrier(2003)]{Chabrier_2003}
Chabrier, G.
\newblock Galactic Stellar and Substellar Initial Mass Function.
\newblock {\em Publ. Astron. Soc. Pac.} {\bf
2003}, {\em 115},~763--795. [\href{http://dx.doi.org/10.1086/376392}{CrossRef}]
\bibitem[Saitoh(2017)]{Saitoh_2017}
Saitoh, T.R.
\newblock Chemical Evolution Library for Galaxy Formation Simulation.
\newblock {\em  Astron. J.} {\bf 2017}, {\em 153},~85. [\href{http://dx.doi.org/10.3847/1538-3881/153/2/85}{CrossRef}]
\bibitem[{Truelove} et~al.(1997){Truelove}, {Klein}, {McKee}, {Holliman},
{Howell}, and {Greenough}]{truelove}
{Truelove}, J.K.; {Klein}, R.I.; {McKee}, C.F.; {Holliman}, J.H., II; {Howell},
L.H.; {Greenough}, J.A.
\newblock {The Jeans Condition: A New Constraint on Spatial Resolution in
Simulations of Isothermal Self-gravitational Hydrodynamics}.
\newblock {\em \apjl} {\bf 1997}, {\em 489},~L179--L183. [\href{http://dx.doi.org/10.1086/310975}{CrossRef}]
\bibitem[{Chevalier}(1974)]{Chevalier74}
{Chevalier}, R.A.
\newblock {The Evolution of Supernova Remnants. Spherically Symmetric Models}.
\newblock {\em \apj} {\bf 1974}, {\em 188},~501--516. [\href{http://dx.doi.org/10.1086/152740}{CrossRef}]
\bibitem[Kim et~al.(2020)Kim, Ostriker, Somerville, Bryan, Fielding, Forbes,
Hayward, Hernquist, and Pandya]{Kim_2020}
Kim, C.G.; Ostriker, E.C.; Somerville, R.S.; Bryan, G.L.; Fielding, D.B.;
Forbes, J.C.; Hayward, C.C.; Hernquist, L.; Pandya, V.
\newblock First Results from SMAUG: Characterization of Multiphase Galactic
Outflows from a Suite of Local Star-forming Galactic Disk Simulations.
\newblock {\em  Astrophys. J.} {\bf 2020}, {\em 900},~61. [\href{http://dx.doi.org/10.3847/1538-4357/aba962}{CrossRef}]
\bibitem[Aoyama et~al.(2016)Aoyama, Hou, Shimizu, Hirashita, Todoroki, Choi,
and Nagamine]{Aoyama_2016}
Aoyama, S.; Hou, K.C.; Shimizu, I.; Hirashita, H.; Todoroki, K.; Choi, J.H.;
Nagamine, K.
\newblock Galaxy simulation with dust formation and destruction.
\newblock {\em Mon. Not. R. Astron. Soc.} {\bf 2016},
{\em 466},~105--121. [\href{http://dx.doi.org/10.1093/mnras/stw3061}{CrossRef}]
\bibitem[{Smagorinsky}(1963)]{Smagorinsky}
{Smagorinsky}, J.
\newblock {General Circulation Experiments with the Primitive Equations}.
\newblock {\em Mon. Weather Rev.} {\bf 1963}, {\em 91},~99. [\href{http://dx.doi.org/10.1175/1520-0493(1963)091<0099:GCEWTP>2.3.CO;2}{CrossRef}]
\bibitem[Dalla~Vecchia and Schaye(2012)]{Dalla_Vecchia_2012}
Dalla~Vecchia, C.; Schaye, J.
\newblock Simulating galactic outflows with thermal supernova feedback:
Galactic outflows with thermal SN feedback.
\newblock {\em Mon. Not. R. Astron. Soc.} {\bf 2012},
{\em 426},~140--158. [\href{http://dx.doi.org/10.1111/j.1365-2966.2012.21704.x}{CrossRef}]
\bibitem[{Portinari} et~al.(1998){Portinari}, {Chiosi}, and
{Bressan}]{Portinari}
{Portinari}, L.; {Chiosi}, C.; {Bressan}, A.
\newblock {Galactic chemical enrichment with new metallicity dependent stellar
yields}.
\newblock {\em \aap} {\bf 1998}, {\em 334},~505--{539.}
\newblock [\href{http://dx.doi.org/10.48550/arXiv.astro-ph/9711337}{CrossRef}]
\bibitem[{Nomoto} et~al.(2013){Nomoto}, {Kobayashi}, and {Tominaga}]{Nomoto}
{Nomoto}, K.; {Kobayashi}, C.; {Tominaga}, N.
\newblock {Nucleosynthesis in Stars and the Chemical Enrichment of Galaxies}.
\newblock {\em \araa} {\bf 2013}, {\em 51},~457--509. [\href{http://dx.doi.org/10.1146/annurev-astro-082812-140956}{CrossRef}]
\bibitem[{Aoyama} et~al.(2020){Aoyama}, {Hirashita}, and
{Nagamine}]{Aoyama2020}
{Aoyama}, S.; {Hirashita}, H.; {Nagamine}, K.
\newblock {Galaxy simulation with the evolution of grain size distribution}.
\newblock {\em \mnras} {\bf 2020}, {\em 491},~3844--{3859.}
\newblock [\href{http://dx.doi.org/10.1093/mnras/stz3253}{CrossRef}]
\bibitem[Romano et~al.(2022)Romano, Nagamine, and Hirashita]{Romano_2022}
Romano, L.E.C.; Nagamine, K.; Hirashita, H.
\newblock The co-evolution of molecular hydrogen and the grain size
distribution in an isolated galaxy.
\newblock {\em Mon. Not. R. Astron. Soc.} {\bf 2022},
{\em 514},~1461--1476. [\href{http://dx.doi.org/10.1093/mnras/stac1386}{CrossRef}]
\bibitem[{Sutherland} and {Dopita}(1993)]{1993Sutherland}
{Sutherland}, R.S.; {Dopita}, M.A.
\newblock {Cooling Functions for Low-Density Astrophysical Plasmas}.
\newblock {\em \apjs} {\bf 1993}, {\em 88},~253. [\href{http://dx.doi.org/10.1086/191823}{CrossRef}]
\bibitem[Bryan and Norman(1998)]{Bryan_1998}
Bryan, G.L.; Norman, M.L.
\newblock Statistical Properties of X‐Ray Clusters: Analytic and Numerical
Comparisons.
\newblock {\em  Astrophys. J.} {\bf 1998}, {\em 495},~80--99. [\href{http://dx.doi.org/10.1086/305262}{CrossRef}]
\bibitem[Hobbs et~al.(2013)Hobbs, Read, Power, and Cole]{Hobbs_2013}
Hobbs, A.; Read, J.; Power, C.; Cole, D.
\newblock Thermal instabilities in cooling galactic coronae: Fuelling star
formation in galactic discs.
\newblock {\em Mon. Not. R. Astron. Soc.} {\bf 2013},
{\em 434},~1849--1868. [\href{http://dx.doi.org/10.1093/mnras/stt977}{CrossRef}]
\bibitem[Hu et~al.(2014)Hu, Naab, Walch, Moster, and Oser]{Hu_2014}
Hu, C.Y.; Naab, T.; Walch, S.; Moster, B.P.; Oser, L.
\newblock SPHGal: Smoothed particle hydrodynamics with improved accuracy for
galaxy simulations.
\newblock {\em Mon. Not. R. Astron. Soc.} {\bf 2014},
{\em 443},~1173--1191. [\href{http://dx.doi.org/10.1093/mnras/stu1187}{CrossRef}]
\bibitem[Pakmor et~al.(2025)Pakmor, Bieri, Fragkoudi, Gómez, Grand, Simpson,
Talbot, van de Voort, and Werhahn]{Pakmor_2025}
Pakmor, R.; Bieri, R.; Fragkoudi, F.; Gómez, F.A.; Grand, R.J.J.; Simpson,
C.M.; Talbot, R.Y.; van de Voort, F.; Werhahn, M.
\newblock Quantifying the intrinsic variability due to randomness of the Auriga
galaxy formation model.
\newblock {\em Mon. Not. R. Astron. Soc.} {\bf 2025},
{\em 543},~1761--1774. [\href{http://dx.doi.org/10.1093/mnras/staf1542}{CrossRef}]
\bibitem[Eisenstein and Hut(1998)]{Eisenstein_1998}
Eisenstein, D.J.; Hut, P.
\newblock HOP: A New Group‐finding Algorithm forN‐Body Simulations.
\newblock {\em  Astrophys. J.} {\bf 1998}, {\em 498},~137--142. [\href{http://dx.doi.org/10.1086/305535}{CrossRef}]
\bibitem[Bigiel et~al.(2008)Bigiel, Leroy, Walter, Brinks, de~Blok, Madore, and
Thornley]{Bigiel_2008}
Bigiel, F.; Leroy, A.; Walter, F.; Brinks, E.; de~Blok, W.J.G.; Madore, B.;
Thornley, M.D.
\newblock {The} Star Formation Law in Nearby Galaxies on sub-kpc Scales.
\newblock {\em  Astron. J.} {\bf 2008}, {\em 136},~2846--2871. [\href{http://dx.doi.org/10.1088/0004-6256/136/6/2846}{CrossRef}]
\bibitem[Kennicutt et~al.(2007)Kennicutt, Calzetti, Walter, Helou, Hollenbach,
Armus, Bendo, Dale, Draine, Engelbracht, Gordon, Prescott, Regan, Thornley,
Bot, Brinks, de~Blok, de~Mello, Meyer, Moustakas, Murphy, Sheth, and
Smith]{Kn2007}
Kennicutt, R.C., Jr.; Calzetti, D.; Walter, F.; Helou, G.; Hollenbach, D.J.;
Armus, L.; Bendo, G.; Dale, D.A.; Draine, B.T.; Engelbracht, C.W.;  et~al.
\newblock Star Formation in NGC 5194 (M51a). II. The Spatially Resolved Star
Formation Law.
\newblock {\em  Astrophys. J.} {\bf 2007}, {\em 671},~333--348. [\href{http://dx.doi.org/10.1086/522300}{CrossRef}]
\bibitem[{Nakashima} et~al.(2018){Nakashima}, {Inoue}, {Yamasaki}, {Sofue},
{Kataoka}, and {Sakai}]{Nakashima2018}
{Nakashima}, S.; {Inoue}, Y.; {Yamasaki}, N.; {Sofue}, Y.; {Kataoka}, J.;
{Sakai}, K.
\newblock {Spatial Distribution of the Milky Way Hot Gaseous Halo Constrained
by Suzaku X-Ray Observations}.
\newblock {\em \apj} {\bf 2018}, {\em 862},~{34.}
\newblock [\href{http://dx.doi.org/10.3847/1538-4357/aacceb}{CrossRef}]
\bibitem[Hopkins et~al.(2012)Hopkins, Quataert, and Murray]{Hopkins_2012_gal}
Hopkins, P.F.; Quataert, E.; Murray, N.
\newblock Stellar feedback in galaxies and the origin of galaxy-scale winds:
Stellar feedback and galactic winds.
\newblock {\em Mon. Not. R. Astron. Soc.} {\bf 2012},
{\em 421},~3522--3537. [\href{http://dx.doi.org/10.1111/j.1365-2966.2012.20593.x}{CrossRef}]
\bibitem[{Agertz} et~al.(2013){Agertz}, {Kravtsov}, {Leitner}, and
{Gnedin}]{Agertz13}
{Agertz}, O.; {Kravtsov}, A.V.; {Leitner}, S.N.; {Gnedin}, N.Y.
\newblock {Toward a Complete Accounting of Energy and Momentum from Stellar
Feedback in Galaxy Formation Simulations}.
\newblock {\em \apj} {\bf 2013}, {\em 770},~{25.}
\newblock [\href{http://dx.doi.org/10.1088/0004-637X/770/1/25}{CrossRef}]
\bibitem[Kim et~al.(2025)Kim, hoon Kim, Jung, Roca-Fàbrega, Ceverino, Granizo,
Nagamine, Primack, Velázquez, Barrow, Feldmann, Fukushima, Mayer, Oh,
Powell, Abel, Jeong, Lupi, Oku, Quinn, Revaz, Rodríguez-Cardoso, Shimizu,
and Teyssier]{kim2025agora}
Kim, H.; hoon Kim, J.; Jung, M.; Roca-Fàbrega, S.; Ceverino, D.; Granizo, P.;
Nagamine, K.; Primack, J.R.; Velázquez, H.; Barrow, K.S.S.;  et~al.
\newblock The AGORA High-resolution Galaxy Simulations Comparison Project. X:
Formation and Evolution of Galaxies at the High-redshift Frontier. \textit{arXiv} \textbf{2025}, {arXiv:2511.04435}. [\href{http://dx.doi.org/10.48550/arXiv.2511.04435}{CrossRef}]
\bibitem[{Roca-F{\`a}brega} et~al.(2024){Roca-F{\`a}brega}, {Kim}, {Primack},
{Genina}, {Jung}, {Lupi}, {Nagamine}, {Powell}, {Quinn}, {Revaz}, {Shimizu},
{Vel{\'a}zquez}, and {the AGORA Collaboration}]{AGORAdatareleaseRoc}
{Roca-F{\`a}brega}, S.; {Kim}, J.H.; {Primack}, J.R.; {Genina}, A.; {Jung}, M.;
{Lupi}, A.; {Nagamine}, K.; {Powell}, J.W.; {Quinn}, T.R.; {Revaz}, Y.;
et~al.
\newblock {The AGORA high-resolution galaxy simulations comparison project:
CosmoRun data release}.
\newblock {\em arXiv} {\bf 2024}, arXiv:2408.00432. [\href{http://dx.doi.org/10.48550/arXiv.2408.00432}{CrossRef}]
\end{thebibliography}
\end{document}